\PassOptionsToPackage{english}{babel}
\documentclass[reprint,aps,prx,twocolumn,amssymb,amsfonts, superscriptaddress,longbibliography]{revtex4-2} 

\usepackage{chemformula} 
\usepackage[T1]{fontenc}
\usepackage{bbold}
\usepackage{bm}
\usepackage{subfigure}
\usepackage{graphicx}

\newcommand{\dd}{\text{d}}
\newcommand{\ie}{\textit{i.e. }}
\newcommand{\eg}{\textit{e.g. }}
\newcommand{\qqa}[0]{\bm{\mathsf{q}}^{\alpha}}
\newcommand{\qqf}[0]{\bm{\mathsf{q}}^{f}}
\newcommand{\qqm}[0]{\bm{\mathsf{q}}^{m}}
\newcommand{\qqh}[0]{\hat{\bm{\mathsf{q}}}}
\newcommand{\hha}[0]{\bm{\mathsf{H}}^{\alpha}}
\newcommand{\hhf}[0]{\bm{\mathsf{H}}^{f}}
\newcommand{\hhm}[0]{\bm{\mathsf{H}}^{m}}
\newcommand{\hhh}[0]{\hat{\bm{\mathsf{H}}}}

\newcommand{\mmf}[0]{\bm{\mathsf{M}}^{f}}
\newcommand{\mmm}[0]{\bm{\mathsf{M}}^{m}}
\newcommand{\mmh}[0]{\hat{\bm{\mathsf{M}}}}
\newcommand{\rra}[0]{\rho_{\alpha}}
\newcommand{\rrf}[0]{\rho_{f}}
\newcommand{\rrm}[0]{\rho_{m}}

\setcitestyle{numbers,square}

\begin{document}

\title{Unravelling looping efficiency of stochastic Cosserat polymers}
\author{Giulio Corazza}\email{giulio.corazza@epfl.ch}
\affiliation{Laboratory for Computation and Visualization in Mathematics and Mechanics (LCVMM), Institute of Mathematics, Ecole Polytechnique F{\'e}d{\'e}rale de Lausanne (EPFL), CH-1015 Lausanne, Switzerland}
\author{Raushan Singh}
\affiliation{Laboratory for Computation and Visualization in Mathematics and Mechanics (LCVMM), Institute of Mathematics, Ecole Polytechnique F{\'e}d{\'e}rale de Lausanne (EPFL), CH-1015 Lausanne, Switzerland}
\date{\today}

\begin{abstract}
Understanding looping probabilities, including the particular case of ring-closure or cyclization, of fluctuating polymers (\eg DNA) is important in many applications in molecular biology and chemistry. In a continuum limit the configuration of a polymer is a curve in the group SE(3) of rigid body displacements, whose energy can be modelled via the Cosserat theory of elastic rods. Cosserat rods are a more detailed version of the classic wormlike-chain (WLC) model, which we show to be more appropriate in short-length scale, or stiff, regimes, where the contributions of extension and shear deformations are not negligible and lead to noteworthy high values for the cyclization probabilities (or J-factors). Characterizing the stochastic fluctuations about minimizers of the energy by means of Laplace expansions in a (real) path integral formulation, we develop efficient analytical approximations for the two cases of full looping, in which both end-to-end relative translation and rotation are prescribed, and of marginal looping probabilities, where only end-to-end translation is prescribed. For isotropic Cosserat rods, certain looping boundary value problems admit non-isolated families of critical points of the energy due to an associated continuous symmetry. For the first time, taking inspiration from (imaginary) path integral techniques, a quantum mechanical probabilistic treatment of Goldstone modes in statistical rod mechanics sheds light on J-factor computations for isotropic rods in the semi-classical context. All the results are achieved exploiting appropriate Jacobi fields arising from Gaussian path integrals, and show good agreement when compared with intense Monte Carlo simulations for the target examples.
\end{abstract}
\maketitle

\section{Introduction}

Nowadays it is widely known that polymers involved in biological and chemical processes are anything but static objects. In fact, they are subject to stochastic forcing from the external environment that lead to complex conformational fluctuations. One of the fundamental phenomena which is understood to perform a variety of roles is polymer looping, occurring when two sites separated by several monomers, and therefore considered far from each other, come in close proximity. A basic observation is that the interacting sites alone do not characterize the phenomenon of looping, but rather it is the whole polymeric chain that rearranges itself for this to occur. As a consequence, the length and mechanical properties of the chain, together with the thermodynamic surrounding conditions are finely tuning the likelihood of such events. There are many reasons to study this topic, which have led to a considerable literature. For instance,  looping is involved in the regulation of gene expression by mediating the binding/unbinding of DNA to proteins \cite{LOOP1, LOOP2, LOOP3}, such as the classic example of the Lac operon \cite{LAC1, LAC2}. In addition, DNA packaging (chromatin formation) \cite{PAC1}, replication and recombination \cite{PAC2, LOOP1} depend on the ability of the polymer to deform into loop configurations, as do other cellular processes. Proteins exhibit intrachain loops for organizing the folding of their polypeptide chains \cite{PROT1}, \eg antibodies use loops to bind a wide variety of potential antigens \cite{PROT2}. When dealing with a closed loop, it is usually appropriate to refer to cyclization or ring closure. In this regard, the production of DNA minicircles is being investigated for their possible therapeutic applications \cite{MINI}. Even in the context of nanotechnologies, ring closure studies have been performed for carbon nanotubes subject to thermal fluctuations \cite{NANO} and wormlike micelles \cite{MIC}.

From the modelling point of view, it is appropriate to look back at some of the historical milestones that underpin our work. In 1949, Kratky and Porod \cite{WLC} introduced the wormlike-chain (WLC) model for describing the conformations of stiff polymer chains. Soon after, the complete determination of the polymeric structure of DNA guided scientists towards the application of WLC-type models in the context of DNA statistical mechanics, allowing probabilistic predictions of relevant quantities of interest. Historically, the computations have been performed in terms of Fokker-Plank equations \cite{DAN, DIFF}, but also exploiting the point of view of path integrals \cite{PIC, FR1, FR2}, a technique inherited from Wiener's work \cite{W1, W2} and quantum mechanics \cite{BookFeynman}. These ideas were largely investigated by Yamakawa \cite{YAM0, YAM00, YAM1, YAM2, YAM3, YAM4, YAM6, YAM7}, who in particular considered the problem of computing ring-closure probabilities, now ubiquitous in molecular biology \cite{BIO3, BIO4, BIO5, BIO6}. Nowadays, for a homogeneous chain, the exact statistical mechanical theory of both the WLC and the helical WLC (with twist) is known \cite{WL1, WL2, WL3}, and the topic has been rigorously phrased over the special euclidean group SE(3) \cite{CHIR2}.

In parallel, back in the early years of the 20th century, the Cosserat brothers Eugène and Fran\c cois formulated Kirchhoff’s rod theory using what are now known as directors \cite{COS}. However, the difficulties arising from the generality of the model, which includes the WLC as a particular constrained case, hindered its application to stochastic chains. Only quite  recently, targeting a more realistic description of DNA, the mentioned framework has been partially or fully exploited both within new analytical studies \cite{MMK, ZC, NP, LUDT, LUD, CHIR1} and intense Monte Carlo (MC) simulations \cite{ALEX1, MCDNA, ALEX2, MCD}, the latter being only a partial solution because of time and cost.


In this article we aim to fill the gap between user-friendly but simplistic models (WLC) on one hand, and accurate but expensive simulations (MC) on the other one, still maintaining the analytical aspect which allows to draw conclusions of physical interest. This is achieved using \cite{LUDT, LUD} as a starting point for bridging the two historical lines of research, \ie exploiting efficient (real) path integral techniques in the semi-classical approximation \cite{2020, PAP1, BookChaichian, BookSchulman, BookWiegel, MOR} (or Laplace method \cite{PIT}), and working within the special Cosserat theory of rods in $SE(3)$. Namely, for studying the end-to-end relative displacements of a fluctuating polymer at thermodynamic equilibrium with a heat bath, we describe the configurations of the chain in a continuum limit by means of framed curves over the special euclidean group. Thus, from an assumed Boltzmann distribution on rod configurations, a conditional probability can be expressed as the ratio of a Boltzmann weighted integral over all paths satisfying the desired end conditions, to the analogous weighted integral over all admissible paths (partition function). The resulting path integrals are finally approximated via a quadratic (semi-classical) expansion about a minimal energy configuration, for which the crucial assumption is that the energy required to deform the system is large with respect to the temperature of the heat bath. This means computing probabilities for length scales of some persistence lengths or less, which turns out to be of great relevance in biology.

Although the present study is general and is applicable to various end-to-end statistics, we focus on the computation of ring-closure or cyclization probabilities for elastic rods, targeting three significant aspects. The first is the possibility of systematically distinguishing between the statistics provided by end positions alone (marginal looping) and the ones provided including also end orientations (full looping) \cite{2020}, for Kirchhoff as well as for Cosserat rods. We emphasise that although Kirchhoff rod theory \cite{AN1} generalises both Euler's elastica theory to model deformations in three-dimensions, and the WLC model allowing arbitrary bending, twisting and intrinsic shapes of the rod, it does not allow extension or shearing of the rod centerline. This is indeed a prerogative of the Cosserat, more general framework, where the centerline displacement and the cross-section’s rotation are considered as independent variables. We show that these additional degrees of freedom are crucial in the analysis of polymer chains in short-length scale, or stiff, regimes, both in the full and marginal cases, where the system exploits extension and shear deformations for minimizing the overall elastic energy, in the face of an increasingly penalizing bending contribution. This allows the cyclization probability density to take high values even when the WLC model (and Kirchhoff) is vanishing exponentially.

The second is addressing the ``perfect problem'' in the semi-classical context, where the symmetry of isotropy gives rise to a ``Goldstone mode'' \cite{G1} leading to a singular path integral, and requires a special treatment by suitably adapting (imaginary) quantum mechanical methods \cite{FADPOP, GHO, COL1, POLYA, JAR, BER} and functional determinant theories \cite{FORM, MCK, FAL}, which are novel in such a generality in the context of elastic rod. For simple models, an analysis in this direction is present in \cite{GUER}. The concepts of isotropy and non-isotropy can be roughly related to a circular shape rather than an elliptical shape for the cross-section of the rod, and the two cases have two different mathematical descriptions in terms of Gaussian path integrals, which we discuss in detail in the course of this article. In particular, the effect of non-isotropy for semi-flexible chain statistics has been addressed from a path integral point of view in \cite{LUD} for the planar case and in \cite{LUDT} for the three-dimensional case (and will be here taken up and simplified), but without resolving the singularity arising in the isotropic limit.

\begin{figure*}
\includegraphics[width=0.85\textwidth]{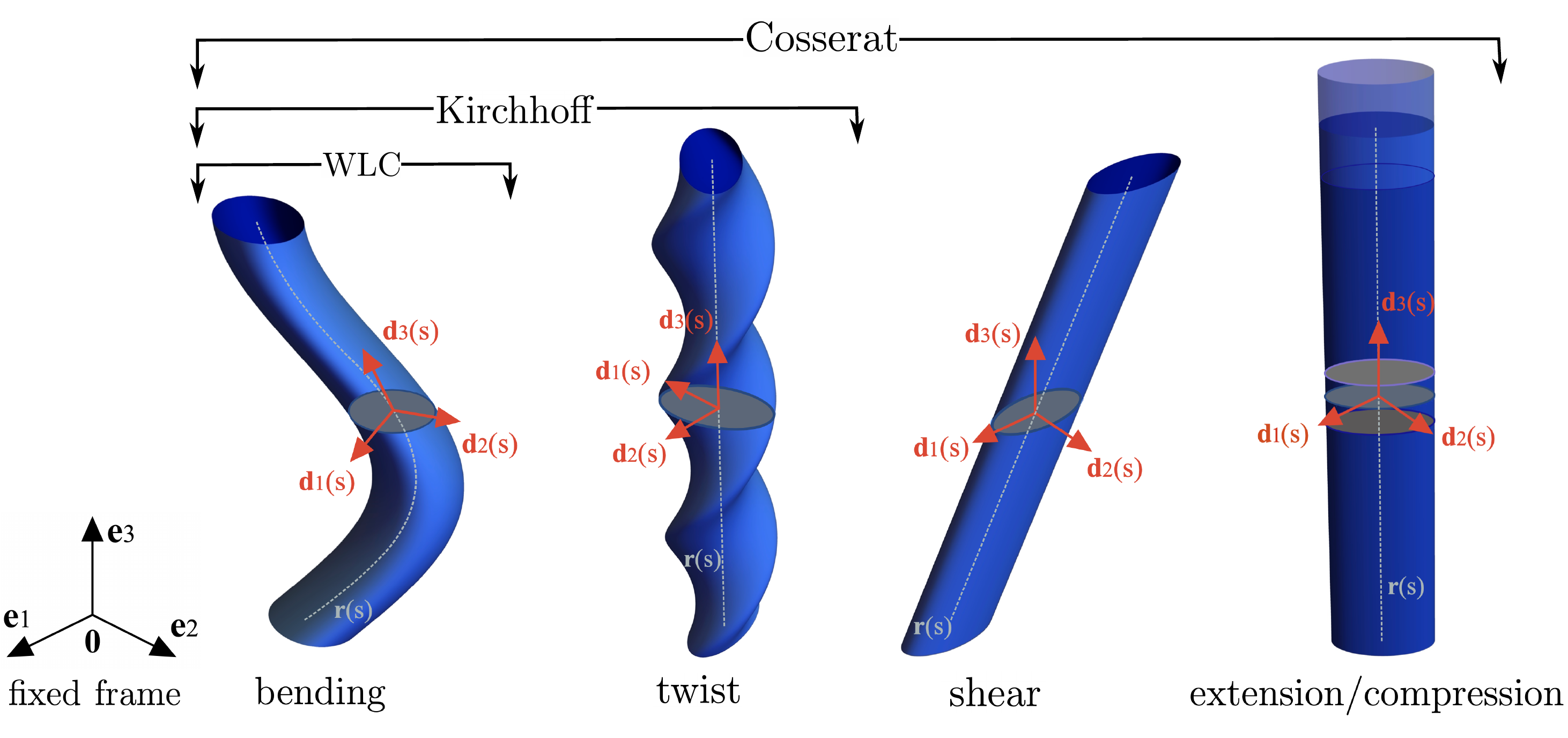}
\caption{Schematic representation of a Cosserat rod with an elliptical cross-section (non-isotropic), where bending, twist, shear and extension/compression are allowed deformations. For Kirchhoff rods only bending and twist are permissible. The standard WLC model only takes bending deformations into account and the cross-section is also assumed to be circular (isotropic).}
\label{fig14}
\end{figure*}

The last significant aspect included in the present work is deriving approximated solution formulas that can always be easily evaluated through straightforward numerical solution of certain systems of Hamiltonian ODE, which in some particularly simple cases can even be evaluated completely explicitly. Versions of the solution formulas, involving evaluation of Jacobi fields at different equilibria and subject to different initial conditions (ICs), are obtained for the two cases of full and marginal ring-closure probabilities. The efficiency aspect in computing looping probabilities, maintaining the same accuracy of MC in the biologically important range less than 1-2 persistence lengths, is fundamental. This is because MC simulation is increasingly intractable due to the difficulty of obtaining sufficiently good sampling with decreasing polymer length, which is the limit where the approximation is increasingly accurate. Contrariwise our approximations are increasingly inaccurate in longer length regimes where good MC sampling is easily achieved. Remarkably, the qualitative behaviour of the probability densities coming from Laplace approximation and from MC sampling are the same regardless of the length scale.

We stress that the stiffness parameters expressing the physical properties of the polymer are allowed to vary along the material parameter of the curve, leading to a non-uniform rod which, in the context of DNA, would represent sequence-dependent variations. In addition, the model allows coupling between bend, twist, stretch and shear, as well as a non-straight intrinsic shape. Notwithstanding the latter generality, we prefer to illustrate our method with some basic examples of uniform and intrinsically straight rods and comparing it with a suitable MC algorithm, in order to highlight the contributions provided by the different choices of cyclization boundary conditions (BCs) in the presence of isotropy or non-isotropy, and to investigate the effect of shear and extension when moving from Kirchhoff to Cosserat rods. Finally, the results will be exposed under the hypothesis of linear elasticity, even tough the theory applies to more general energy functionals.

The structure of the article is as follows. In section \ref{s2} we give an overview of the statics of special Cosserat rods, with particular emphasis on equilibria and stability for the boundary value problems (BVPs) involved, and we further establish the relations with simpler models. In particular, the Hamiltonian formulation of the Euler-Lagrange and Jacobi equations provides a common theoretical framework for both Kirchhoff and Cosserat rods. In section \ref{s3} we set out a preview of the examples that will be considered in the course of the article, namely in the context of linear elasticity. Here we focus on the physical properties that characterise shearable and extensible/compressible polymers and explain how these degrees of freedom improve the understanding of the problem. Therefore, we study the minimizers of the energy, distinguishing between the non-isotropic and isotropic cases. The role of the continuous variational symmetries of isotropy and uniformity is explained. Before describing the computational setting in detail, we devote a section (\ref{s4}) for summarising the general formulas that we obtain for estimating end-to-end probabilities of fluctuating elastic rods as a proxy for interpreting the behaviour of polymers in a thermal bath. Then we introduce the path integral formulation of the problem in section \ref{s5}, prescribing an appropriate parametrisation of the rotation group and giving the functional representations of full and marginal looping probability densities. Afterwards, the explicit approximated formulas for such densities are derived, initially in the case of isolated minimizers and thereafter in presence of non-isolation, for which a special theoretical analysis is performed. Moreover, in section \ref{s6}, we provide a MC algorithm for stochastic elastic rods, exploited to benchmark our results. The examples are finally investigated from the point of view of cyclization probabilities in section \ref{s7}, with special focus on shear and extension contributions for Cosserat rods in the short-length scale regimes. Further discussion and conclusions follow.

\section{Background on elastic rod equilibria and their stability}\label{s2}
A comprehensive overview of the theory of elastic rods in the context of continuum mechanics can be found in \cite{ANT}. In particular, we follow the specific notation and Hamiltonian formulations introduced in \cite{HAM}. Briefly, a configuration of a Cosserat rod is a framed curve $\bm{q}(s)=(\bm{R}(s),\bm{r}(s))$ $\in SE(3)$ for each $s\in[0,L]$, which may be bent, twisted, stretched or sheared. The vector $\bm{r}(s)\in\mathbb{R}^3$ and the matrix $\bm{R}(s)\in SO(3)$ model respectively the rod centerline and the orientation of the material in the rod cross-section via a triad of orthonormal directors $\lbrace\bm{d}_i(s)\rbrace_{i=1,2,3}$ attached to the rod centerline, with respect to a fixed frame $\lbrace\bm{e}_i\rbrace_{i=1,2,3}$. As a matter of notation, the columns of the matrix $\bm{R}(s)$ in coordinates are given by the components of the vectors ${\bm{d}_j(s)}$ in the fixed frame $\lbrace\bm{e}_i\rbrace$, namely $\bm{R}_{i,j}(s)=\bm{e}_i\cdot\bm{d}_j(s),\,\,i,j=1,..,3$. In Fig.\ref{fig14} we show a schematic representation of the the degrees of freedom allowed within the special Cosserat theory of rods in relation to other simpler models that will be outlined in the course of this section.

Strains are defined as $\bm{u}(s)$, $\bm{v}(s)$ where $\bm{d}_i'=\bm{u}\times\bm{d}_i$, $\bm{r}'=\bm{v}$, with $\bm{u}$ the Darboux vector and the prime denoting the derivative with respect to $s$. Sans-serif font is used to denote components in the director basis (\eg $\mathsf{u}_i=\bm{u}\cdot\bm{d}_i$), and we write  $\bm{\mathsf{u}}=(\mathsf{u}_1,\mathsf{u}_2,\mathsf{u}_3)$, $\bm{\mathsf{v}}=(\mathsf{v}_1,\mathsf{v}_2,\mathsf{v}_3)$, etc. Physically, $\mathsf{u}_1$ and $\mathsf{u}_2$ represent the bending strains and $\mathsf{u}_3$ the twist strain. Analogously, $\mathsf{v}_1$ and $\mathsf{v}_2$ are associated with transverse shearing, whereas $\mathsf{v}_3$ with stretching or compression of the rod. In compact form, we have $\bm{\mathsf{u}}^{\times}(s)=\bm{R}(s)^T \bm{R}'(s)$, $\bm{\mathsf{v}}(s)=\bm{R}(s)^T \bm{r}'(s)$, where $\bm{\mathsf{u}}^{\times}$ is the skew-symmetric matrix or cross product matrix of $\bm{\mathsf{u}}$ having $(1,2)$, $(1,3)$ and $(2,3)$ entries respectively equal to $-\mathsf{u}_3$, $\mathsf{u}_2$ and $-\mathsf{u}_1$.

The stresses $\bm{m}(s)$ and $\bm{n}(s)$ are defined as the resultant moment and force arising from averages of the stress field acting across the material cross-section at $\bm{r}(s)$. In the absence of any distributed loading, at equilibrium the stresses satisfy the balance laws ${\bm{{n}}}'=\bm{0}$, ${\bm{{m}}}'+\bm{{r}}'\times\bm{{n}}=\bm{0}$. Equilibrium configurations can be found once constitutive relations are introduced, which we do in a way that facilitates the recovery of the inextensible, unshearable limit tipically adopted in polymer physics.

Namely, we consider a pair of functions $W,\,W^*:\mathbb{R}^3\times\mathbb{R}^3\times [0,L]\rightarrow\mathbb{R}$ that (for each $s\in[0,L]$) are strictly convex, dual functions under Legendre transform in their first two arguments, and with $\bm{0}\in\mathbb{R}^6$ their unique global minimum. If $\hat{\bm{\mathsf{u}}}(s)$ and $\hat{\bm{\mathsf{v}}}(s)$ are the strains of the unique energy minimizing configuration $\hat{\bm{q}}(s)$, then, $\forall\epsilon>0$ we introduce the Hamiltonian function $H=W^*(\bm{\mathsf{m}},\epsilon\bm{\mathsf{n}};s)+\bm{\mathsf{m}}\cdot\hat{\bm{\mathsf{u}}}+\bm{\mathsf{n}}\cdot\hat{\bm{\mathsf{v}}}$, and the constitutive relations are $\bm{\mathsf{u}}={\partial H}/{\partial \bm{\mathsf{m}}}=W_1^*(\bm{\mathsf{m}},\epsilon\bm{\mathsf{n}};s)+\hat{\bm{\mathsf{u}}}$, $\bm{\mathsf{v}}={\partial H}/{\partial\bm{\mathsf{n}}}=\epsilon W_2^*(\bm{\mathsf{m}},\epsilon\bm{\mathsf{n};s})+\hat{\bm{\mathsf{v}}}$, which can be inverted to obtain $\bm{\mathsf{m}}=W_1(\bm{\mathsf{u}}-\hat{\bm{\mathsf{u}}},\frac{\bm{\mathsf{v}}-\hat{\bm{\mathsf{v}}}}{\epsilon};s)$, $\bm{\mathsf{n}}=\frac{1}{\epsilon}W_2(\bm{\mathsf{u}}-\hat{\bm{\mathsf{u}}},\frac{\bm{\mathsf{v}}-\hat{\bm{\mathsf{v}}}}{\epsilon};s)$, where the Lagrangian $W(\bm{\mathsf{u}}-\hat{\bm{\mathsf{u}}},\frac{\bm{\mathsf{v}}-\hat{\bm{\mathsf{v}}}}{\epsilon};s)$ defines the elastic potential energy of the system as $\int_0^L{W(\bm{\mathsf{u}}-\hat{\bm{\mathsf{u}}},\frac{\bm{\mathsf{v}}-\hat{\bm{\mathsf{v}}}}{\epsilon};s)}\,\dd s$. Note the use of the subscripts to denote partial derivatives with respect to the first or second argument. The standard case of linear constitutive relations arises when $W^*(\bm{\mathsf{x}};s)=\frac{1}{2}\bm{\mathsf{x}}\cdot\bm{\mathcal{R}}(s)\bm{\mathsf{x}}$ and $W(\bm{\mathsf{y}};s)=\frac{1}{2}\bm{\mathsf{y}}\cdot\bm{\mathcal{P}}(s)\bm{\mathsf{y}}$ for $\bm{\mathsf{x}}$,$\,\bm{\mathsf{y}}\,\in\mathbb{R}^6$, where $\mathbb{R}^{6\times 6}\ni\bm{\mathcal{P}}^{-1}(s)=\bm{\mathcal{R}}(s)=\bm{\mathcal{R}}(s)^T>0$, with $\bm{\mathcal{P}}(s)$ a general non-uniform stiffness matrix and $\bm{\mathcal{R}}(s)$ the corresponding compliance matrix. For each $\epsilon>0$ and given $W$, $W^*$, we arrive at a well-defined Cosserat rod theory, where, \textit{e.g.}, the full potential energy of the system might include end-loading terms of the form $\bm{\lambda}\cdot(\bm{r}(L)-\bm{r}(0))$, $\bm{\lambda}\in\mathbb{R}^3$. 

The point of the above formulation is that the Hamiltonian and associated constitutive relations behave smoothly in the limit $\epsilon\rightarrow 0$, which imply the unshearability and inextensibility constraint on the strains ${\bm{\mathsf{v}}}(s)=\hat{\bm{\mathsf{v}}}(s)$, where $\hat{\bm{\mathsf{v}}}(s)$ are prescribed. This is precisely a Kirchhoff rod model, abbreviated as $(\mathtt{K})$, in contrast to $(\mathtt{C})$ for Cosserat. However, the $\epsilon\rightarrow 0$ limit of the $(\mathtt{C})$ Lagrangian is not smooth; rather the potential energy density for the $(\mathtt{K})$ rod is the Legendre transform of $W^*(\bm{\mathsf{m}},\bm{0};s)+\bm{\mathsf{m}}\cdot\hat{\bm{\mathsf{u}}}+\bm{\mathsf{n}}\cdot\hat{\bm{\mathsf{v}}}$ w.r.t. $\bm{\mathsf{m}}\in\mathbb{R}^3$, or $W^{(\mathtt{K})}({\bm{\mathsf{u}}}-\hat{\bm{\mathsf{u}}};s)-{\bm{\mathsf{n}}}\cdot\hat{\bm{\mathsf{v}}}$. In the case of linear elasticity for a $(\mathtt{C})$ rod with 
$\bm{\mathcal{P}}(s)=\begin{small}\begin{pmatrix}
       \bm{\mathcal{K}} & \bm{\mathcal{B}}  \\
       \bm{\mathcal{B}}^T & \bm{\mathcal{A}}
     \end{pmatrix}\end{small}$
and $\bm{\mathcal{K}}(s)$, $\bm{\mathcal{B}}(s)$, $\bm{\mathcal{A}}(s)$ in $\mathbb{R}^{3\times 3}$, the $(1,1)$ block of the compliance matrix is $\bm{\mathcal{R}}_{1,1}=(\bm{\mathcal{K}}-\bm{\mathcal{B}}\bm{\mathcal{A}}^{-1}\bm{\mathcal{B}}^T)^{-1}$ and $W^{(\mathtt{K})}({\bm{\mathsf{u}}}-\hat{\bm{\mathsf{u}}};s)=\frac{1}{2}(\bm{\mathsf{u}}-\hat{\bm{\mathsf{u}}})\cdot\bm{\mathcal{K}}^{(\mathtt{K})}(s)(\bm{\mathsf{u}}-\hat{\bm{\mathsf{u}}})$, with $\bm{\mathcal{K}}^{(\mathtt{K})}={\bm{\mathcal{K}}^{(\mathtt{K})}}^T=\bm{\mathcal{R}}_{1,1}^{-1}>0$.

Uniform helical WLC models are recovered in the case of a uniform $(\mathtt{K})$ rod when $\hat{\bm{\mathsf{u}}}(s)$, $\hat{\bm{\mathsf{v}}}(s)$ and $\bm{\mathcal{K}}^{(\mathtt{K})}(s)$ are all taken to be constant. (For any uniform rod, $(\mathtt{C})$ or $(\mathtt{K})$, the Hamiltonian function is constant along equilibria). Linearly elastic $(\mathtt{K})$ rods are (transversely) isotropic when $\bm{\mathcal{K}}^{(\mathtt{K})}(s)=\text{diag}(k_1(s),k_2(s),k_3(s))$ with $k_1=k_2$ and $\hat{{\mathsf{u}}}_1=\hat{{\mathsf{u}}}_2=\hat{{\mathsf{v}}}_1=\hat{{\mathsf{v}}}_2=0$. Then $\mathsf{m}_3$ is constant on equilibria, and $W^{(\mathtt{K})}=\frac{1}{2}[k_1\kappa^2+k_3(\mathsf{u}_3-\hat{\mathsf{u}}_3)^2]$ reduces to a function of the square geometrical curvature $\kappa(s)$ of the curve (where it should be noted that $\mathsf{u}_3(s)$ is still the twist of the $\lbrace\bm{d}_i\rbrace$ frame which is not directly related to the geometrical torsion of the Frenet framing of the rod centerline). The WLC model arises when $k_1(s)$ is constant and the twist moment $\mathsf{m}_3$ vanishes.

There is an extensive literature concerning the study of equilibria of a given elastic rod. Numerically this involves the solution of a two-point BVP, which can reasonably now be regarded as a straightforward well-understood procedure. Often coordinates on $SO(3)$ are introduced and the resulting system of second-order Euler Lagrange equations associated with the potential energy is solved numerically. We adopt an Euler parameters (or quaternions) parametrization of $SO(3)$, but solve the associated first-order canonical Hamiltonian system subject to appropriate (self-adjoint) two-point BCs, so that the inextensible, unshearable $(\mathtt{K})$ rod is a simple smooth limit of the extensible, shearable $(\mathtt{C})$ case. 

In this article we are primarily interested in the two specific BVPs, denoted respectively by $(\mathtt{f})$ and $(\mathtt{m})$:
\begin{equation}\label{f}
(\mathtt{f})\quad\bm{r}(0)=\bm{0}\;,\,\,\bm{R}(0)=\mathbb{1}\;,\,\,\bm{r}(L)=\bm{r}_L\;,\,\,\bm{R}(L)=\bm{R}_L\;,
\end{equation}
\begin{equation}\label{m}
(\mathtt{m})\quad\bm{r}(0)=\bm{0}\;,\,\,\,\bm{R}(0)=\mathbb{1}\;,\,\,\,\bm{r}(L)=\bm{r}_L\;,\,\,\,\bm{m}(L)=\bm{0}\;.
\end{equation}
The BVP $(\mathtt{f})$ arises in modelling looping in $SE(3)$ including the particular case of cyclization where $\bm{r}_L=\bm{0}$ and $\bm{R}_L=\mathbb{1}$. The BVP $(\mathtt{m})$ arises in modelling looping in $\mathbb{R}^3$, where the value of $\bm{R}_L$ is a variable left free, over which one marginalises. In general, for rod two-point BVPs, equilibria with given BCs are non-unique. For isotropic or uniform rods, and for specific choices of $\bm{r}_L$ and $\bm{R}_L$ in $(\mathtt{f})$ and $(\mathtt{m})$, equilibria can arise in continuous isoenergetic families \cite{RING}, a case of primary interest here. 

As we assume hyper-elastic constitutive relations with 
\begin{equation}\label{energy}
E(\bm{q})=\int_0^L{W(\bm{\mathsf{u}}-\hat{\bm{\mathsf{u}}},{\bm{\mathsf{v}}-\hat{\bm{\mathsf{v}}}};s)}\,\dd s\;,
\end{equation}
stability of rod equilibria can reasonably be discussed dependent on whether an equilibrium is a local minimum of the associated potential energy variational principle. For $(\mathtt{C})$ rods classification of which equilibria are local minima has a standard and straightforward solution. The second variation $\delta^2E$ is a quadratic functional of the perturbation field $\bm{\mathsf{h}}=(\delta\bm{\mathsf{c}},\delta\bm{\mathsf{t}})$, where the sans-serif font $\bm{\mathsf{q}}(s)=(\bm{\mathsf{c}}(s),\bm{\mathsf{t}}(s))\in\mathbb{R}^6$ is a given parametrisation of $SE(3)$ for the configuration variable in the director basis which will be specified later in the article, and reads as
\begin{equation}\label{sec}
\delta ^2 E=\int_0^L{\left({\bm{\mathsf{h}}'}\cdot\bm{\mathsf{P}}\bm{\mathsf{h}}'+2{\bm{\mathsf{h}}'}\cdot{\bm{\mathsf{C}}}\bm{\mathsf{h}}+\bm{\mathsf{h}}\cdot{\bm{\mathsf{Q}}}\bm{\mathsf{h}}\right)}\,\dd s\;,
\end{equation}
where $\bm{\mathsf{P}}(s)$, ${\bm{\mathsf{C}}}(s)$ and ${\bm{\mathsf{Q}}}(s)$ are coefficient matrices in $\mathbb{R}^{6\times 6}$ computed at any equilibrium. The Jacobi equations are the (second-order) system of Euler-Lagrange equations for Eq.~(\ref{sec}), or equivalently the linearisation of the original Euler-Lagrange equations for the potential energy variational principle. One then solves a $6\times 6$ matrix valued system, namely an initial value problem for the Jacobi equations with ICs coinciding with the ones given later in the article when computing probability densities from Jacobi fields (shooting towards $s=0$, where in both $(\mathtt{f})$ and $(\mathtt{m})$ Dirichlet BCs are present; the case with Neumann BCs at both ends is more delicate \cite{NEU}). Provided that the determinant of the matrix solution does not vanish in $[0,L)$, then there is no conjugate point and the equilibrium is a local minimum \cite{JHMStab,ISO,HEL}. 

\begin{figure*}
\subfigure[]{\includegraphics[width=.321\textwidth]{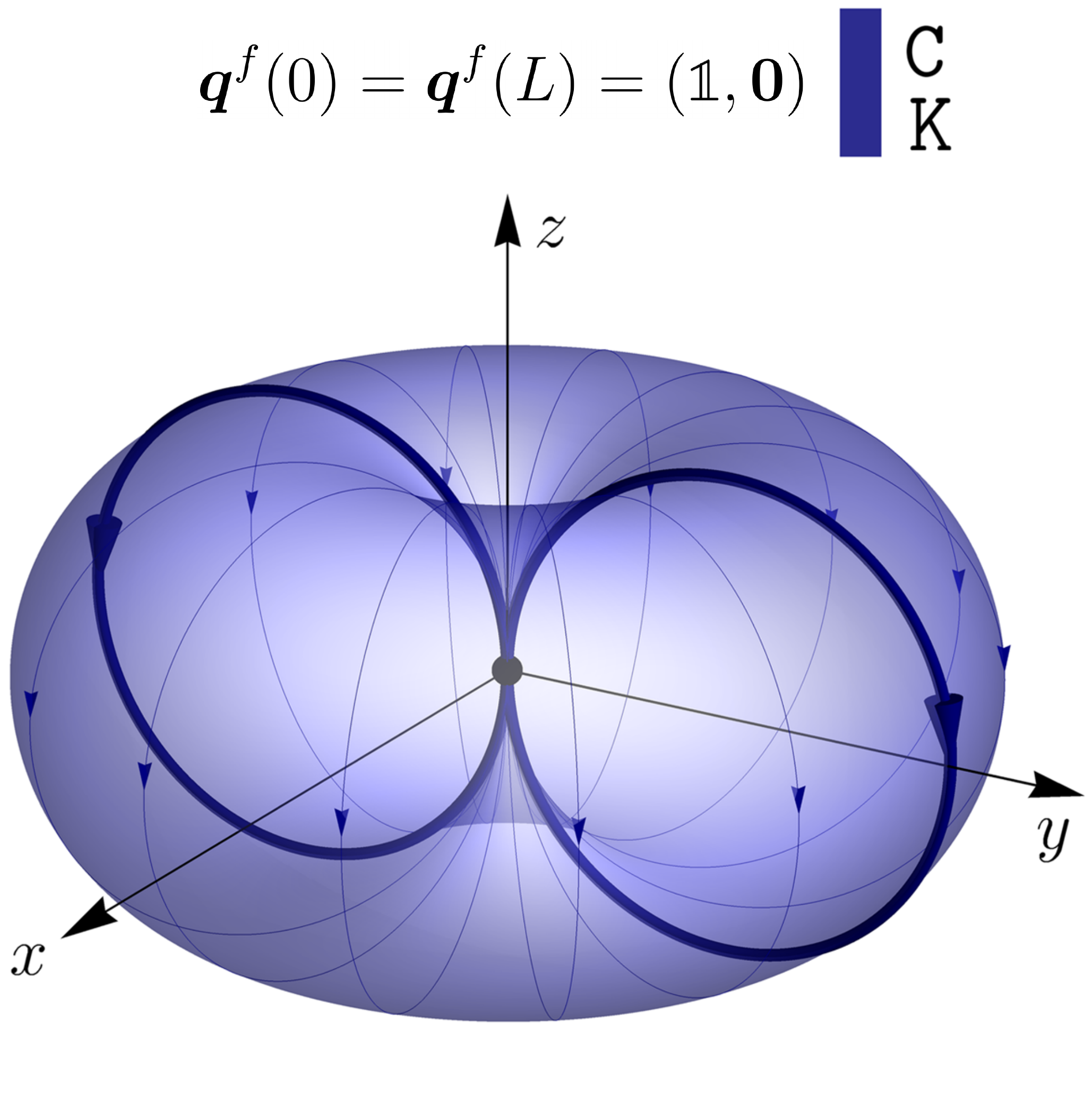}}\qquad\qquad
\subfigure[]{\includegraphics[width=.32\textwidth]{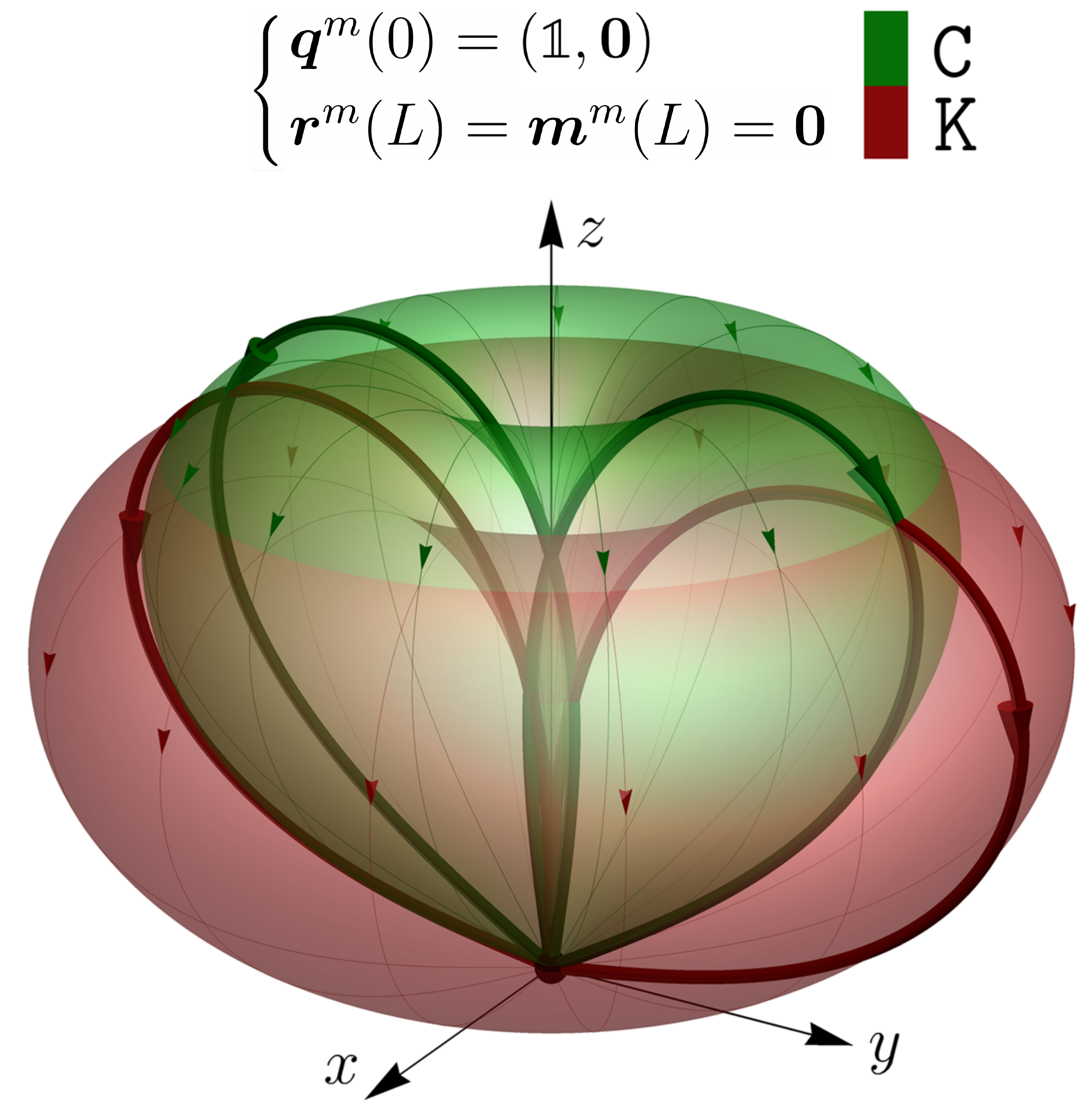}}
\caption{The thick lines represent the pairs of isolated minima for the non-isotropic case; the manifolds of minimizers for the isotropic case are displayed accordingly. In panel (a) the solutions for the $(\mathtt{f})$ case are the same for $(\mathtt{K})$ and $(\mathtt{C})$ rods. In panel (b) we underline the effect of shear and extension for the $(\mathtt{m})$ case, which modifies the red solutions $(\mathtt{K})$ into the green ones $(\mathtt{C})$.}
\label{fig1}
\end{figure*}

As described fully in \cite{JHMT}, the constrained case of $(\mathtt{K})$ is more subtle and a theory dating back to Bolza for isoperimetrically constrained calculus of variations must be applied \cite{BOLZA}. However, the Hamiltonian version of the Jacobi equations for rods (just like the Hamiltonian version of the Euler-Lagrange equilibrium equations) has a smooth limit as $\epsilon\rightarrow 0$, and the limit corresponds to the Hamiltonian formulation of the Bolza conjugate point conditions as described in \cite{ISO}. The Jacobi equations in first-order Hamiltonian form are written as
\begin{equation}\label{JacHam}
\begin{pmatrix}
      {{\bm{\mathsf{H}}'}}  \\
     {{\bm{\mathsf{M}}'}}
     \end{pmatrix}
       =\bm{J}{{\bm{\mathsf{E}}}}\begin{pmatrix}
       {{\bm{\mathsf{H}}}}  \\
     {{\bm{\mathsf{M}}}}
     \end{pmatrix}\;,
\end{equation}
with the Hamiltonian skew-symmetric matrix $\bm{J}=\begin{pmatrix} \mathbb{0} & \mathbb{1}\\ -\mathbb{1} & \mathbb{0} \end{pmatrix}\in\mathbb{R}^{12\times 12}$, $\bm{\mathsf{E}}(s)$ the symmetric matrix driving the system which will be detailed later on, and $\bm{\mathsf{M}}(s)\in\mathbb{R}^{6\times 6}$ the conjugate variable of the Jacobi fields $\bm{\mathsf{H}}(s)$ under the Legendre transform.

In the following, we assume the existence and stability of the minimizers of the elastic energy Eq.~(\ref{energy}) $\bm{q}^f$ and $\bm{q}^m$ satisfying the BCs $(\mathtt{f})$ in Eq.~(\ref{f}) and $(\mathtt{m})$ in Eq.~(\ref{m}) respectively. Note that the intrinsic configuration of the rod $\hat{\bm{q}}$ is itself a minimizer (global) satisfying 
\begin{equation}\label{hat}
\bm{r}(0)=\bm{0}\;,\,\,\,\bm{R}(0)=\mathbb{1}\;,\,\,\,\bm{n}(L)=\bm{m}(L)=\bm{0}\;.
\end{equation}
Stability of equilibria is not the focus of this article, but we will show that the volume of certain Jacobi fields, \textit{i.e.}, the actual (positive) value of a Jacobi determinant, plays a central role in the evaluation formula for the quadratic path integrals that arise in our Laplace approximations to looping probabilities. 

The connection between Jacobi fields and quadratic imaginary path integrals is well known in the case that the coefficient matrix ${\bm{\mathsf{C}}}(s)$ in the cross-terms in Eq.~(\ref{sec}) vanishes (or is symmetric and so can be integrated away). By contrast, for elastic rods a non-symmetric ${\bm{\mathsf{C}}}(s)$ is typically present and the approach of Papadopoulos \cite{PAP1} is required to evaluate the quadratic path integrals, and as described in \cite{LUDT, LUD} a further Riccati transformation for the Papadopoulos solution formula is necessary to recover a Jacobi fields expression. Moreover, in \cite{2020} the latter studies are generalised for different choices of BCs on the paths, in particular for dealing with the partition function and solving the marginalised problem. 

The main contributions of this article are to demonstrate that the approach of \cite{LUDT, LUD} for conditional probability densities can be extended in two ways. First, isolated equilibria to BVP $(\mathtt{m})$ can be treated, in addition to the case of isolated equilibria to BVP $(\mathtt{f})$, and second, the case of non-isolated equilibria of both BVPs $(\mathtt{f})$ and $(\mathtt{m})$ (as arises for isotropic rods) can be handled by appropriately generalising a particular regularization procedure \cite{MCK, FAL} within Forman's theorem in the field of functional determinants \cite{FORM}. Furthermore, the underlying physical phenomena arising from the different cases are discussed and explained within some guiding examples. For a polymer, the questions we are trying to answer would be interpreted as follows: what is a good estimate of the probability of the end monomers coming into contact with each other? How is the latter value changing if we impose an orientation constraint on the binding site? How does the shape of the cross-section (isotropic or non-isotropic) affect the statistics? And finally, what happens if we deviate from the standard inextensible and unshearable model and incorporate shear and extension as possible deformations?

\section{A preview of the examples considered}\label{s3}
The method developed in the present article will be applied, as a fundamental example, to a linearly elastic, uniform, with diagonal stiffness matrix, intrinsically straight and untwisted rod ($\bm{\mathcal{P}}(s)=\bm{\mathcal{P}}=$ diag $\lbrace k_1,k_2,k_3,a_1,a_2,a_3\rbrace$, $\hat{\bm{\mathsf{u}}}=\bm{0}$, $\hat{\bm{\mathsf{v}}}=(0,0,1)$). Neither intrinsic shear nor extension is present. Since we are primarily interested in ring-closure or cyclization probabilities, we look for minimizers of the energy satisfying the BCs reported in Eq.~(\ref{f}), Eq.~(\ref{m}) with $\bm{r}_L=\bm{0}$ and $\bm{R}_L=\mathbb{1}$. 

\begin{figure*}
\centering
\subfigure[]{
\includegraphics[width=.374\textwidth]{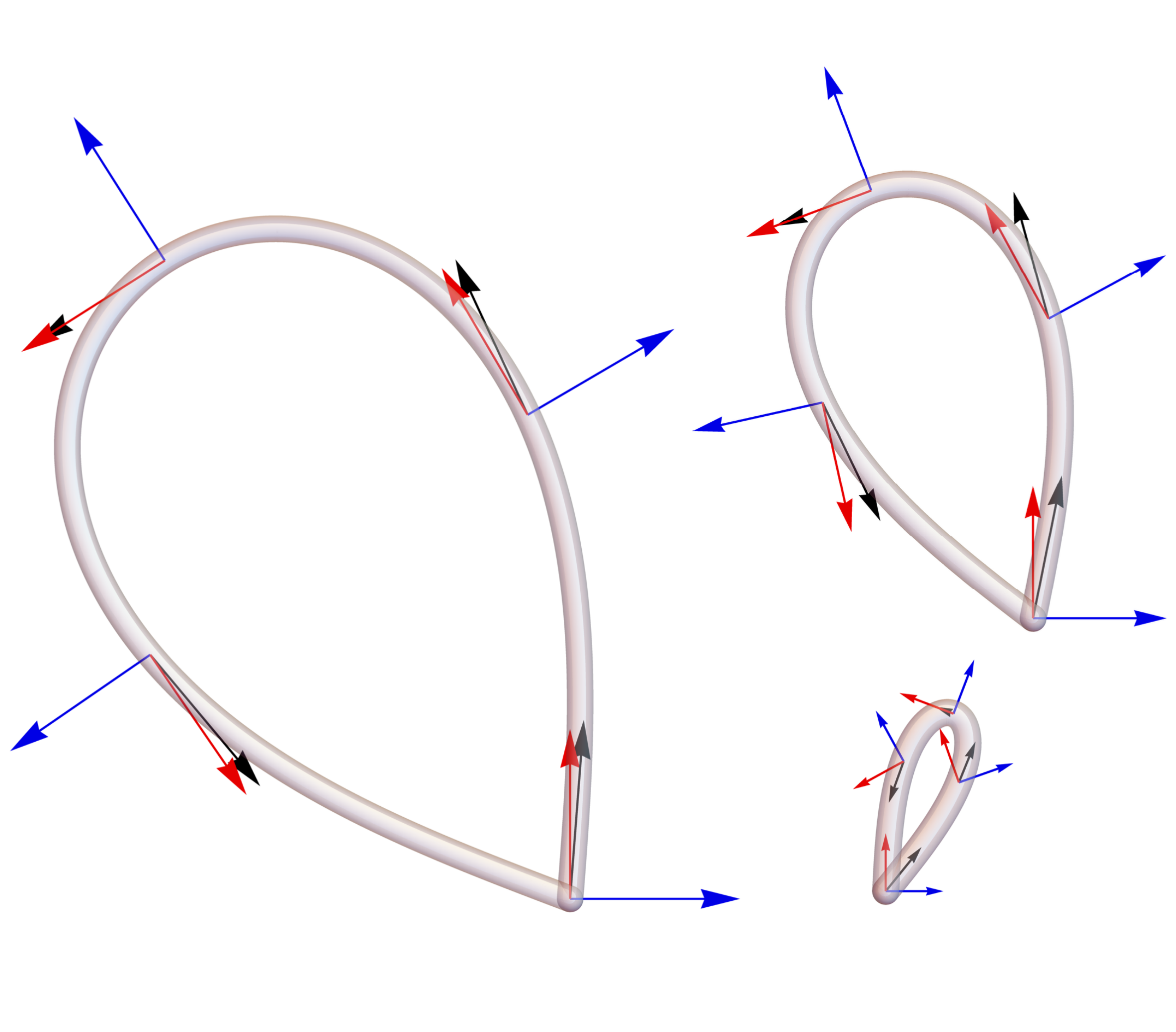}\qquad\qquad
}
\subfigure[]{
\includegraphics[width=.37\textwidth]{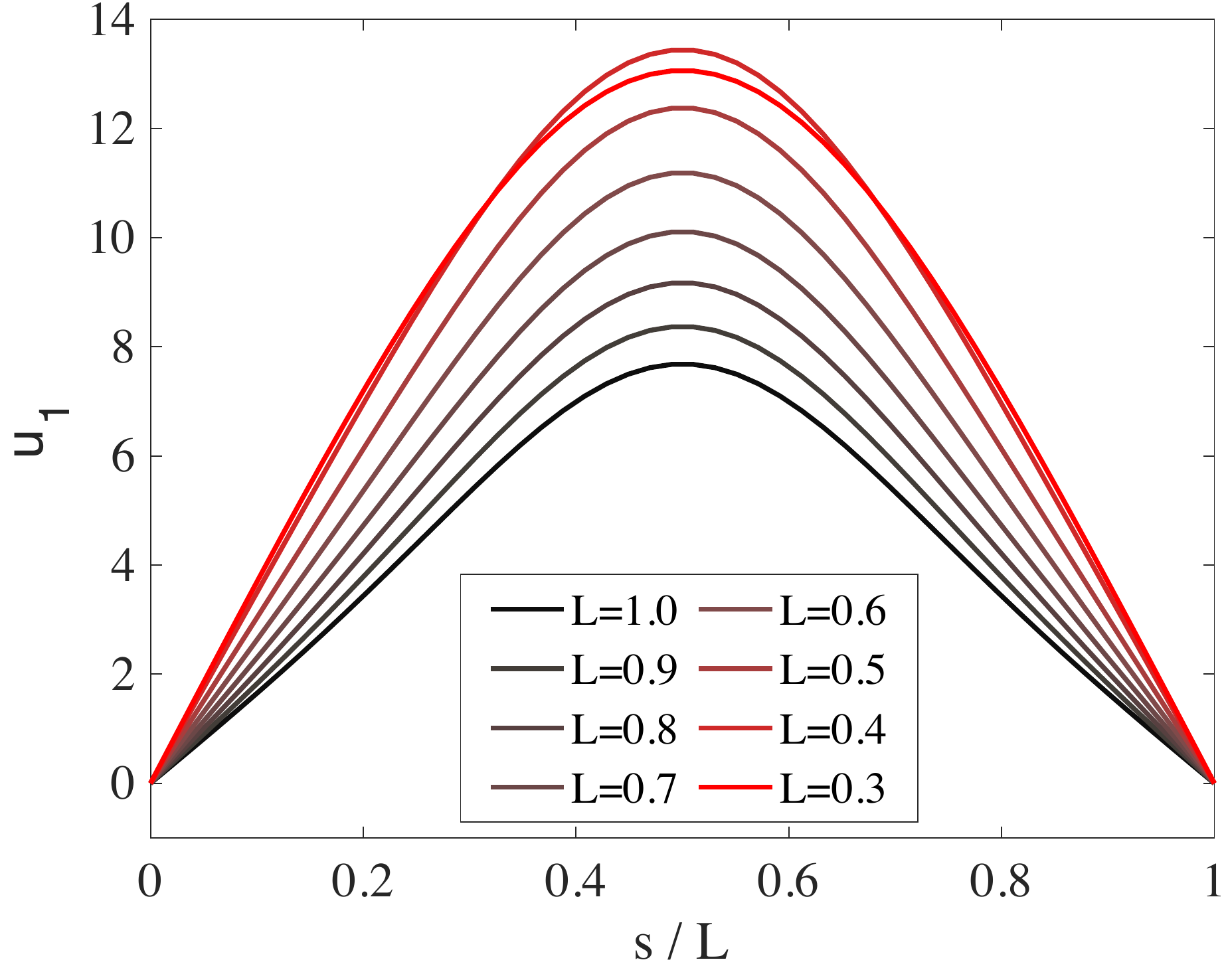}
}

\subfigure[]{
\includegraphics[width=.37\textwidth]{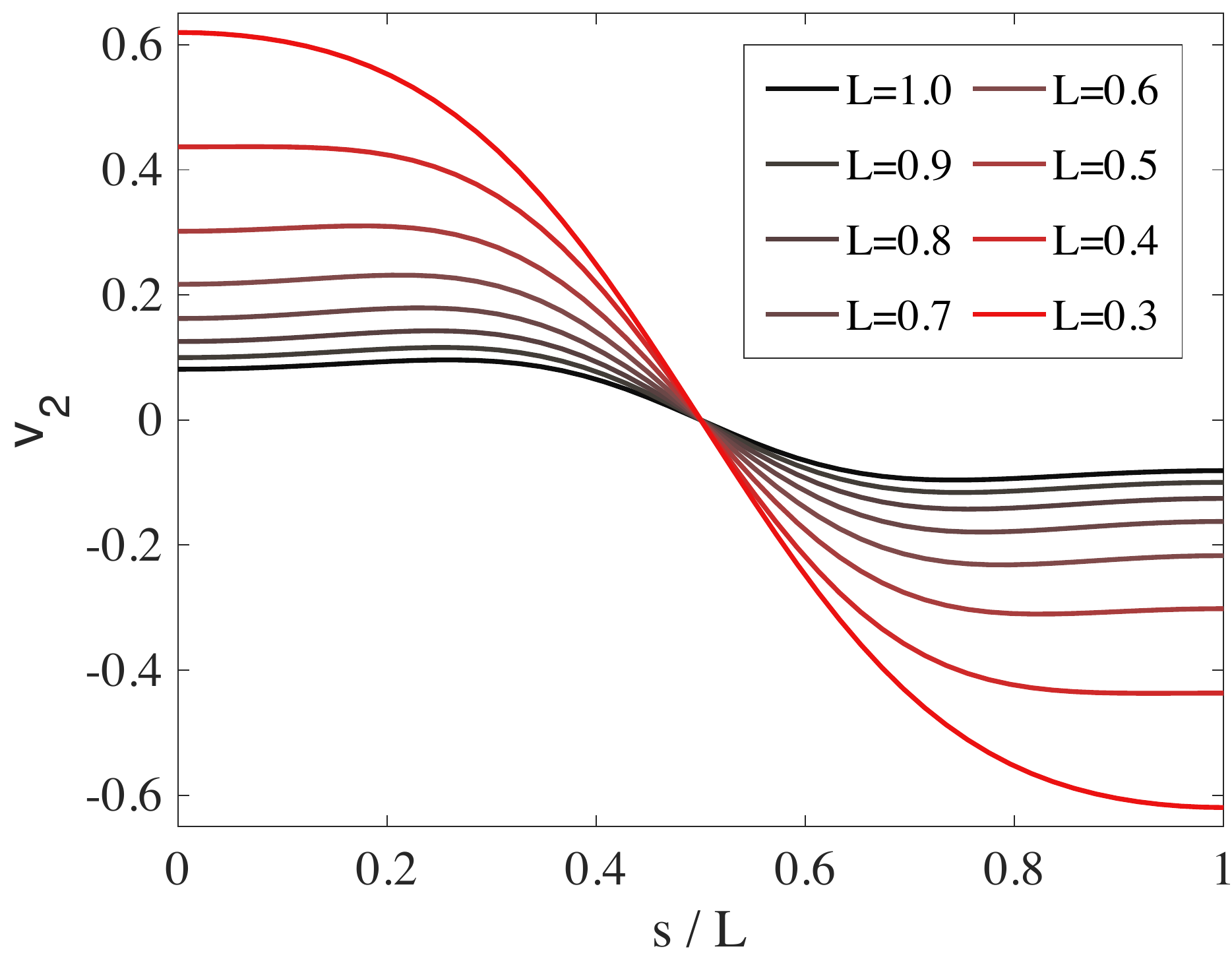}\qquad\qquad\,
}
\subfigure[]{
\includegraphics[width=.373\textwidth]{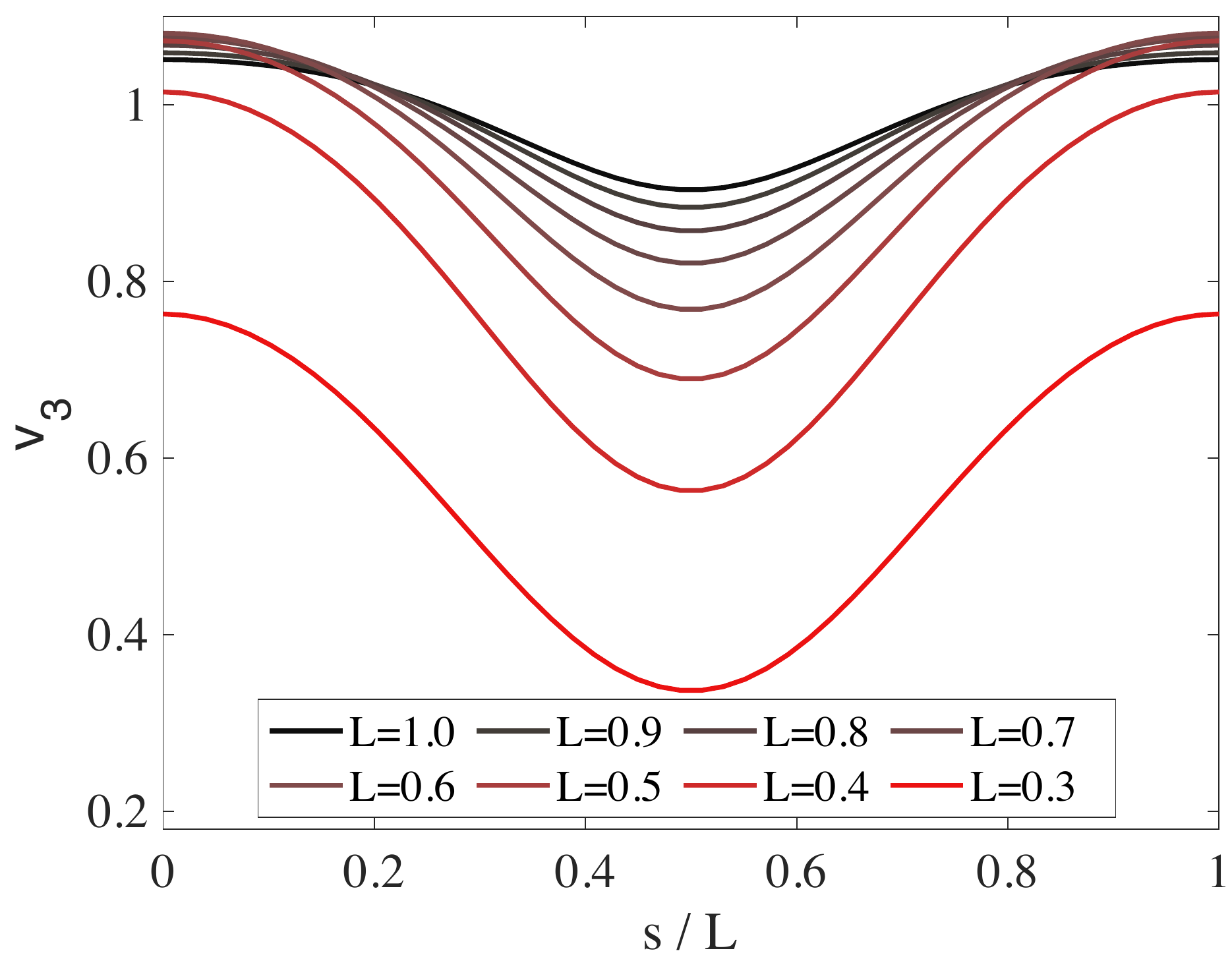}
}
\caption{Analysis of the $(\mathtt{m})$ cyclization problem for a non-isotropic $(\mathtt{C})$ rod with $k_1=0.5$, $k_2=5$, $k_3=10$ and $a_1=a_2=a_3=100$. In panel (a) we report the shapes of the teardrop minimizers for $L=1$, $L=0.6$ and $L=0.3$. The tangent $\bm{r}'(s)$ and the vectors $\bm{d}_2(s)$, $\bm{d}_3(s)$ of the moving frame are displayed in black, blue and red respectively. In panels (b), (c) and (d) we plot the bending, shear and stretch components $\mathsf{u}_1^m(s)$, $\mathsf{v}_2^m(s)$ and $\mathsf{v}_3^m(s)$ respectively for a uniformly spaced set of undeformed lengths ranging from $L=1$ to $L=0.3$. The values are computed numerically.}
\label{fig1518}
\end{figure*}

First, we consider a non-isotropic rod ($k_1\neq k_2$), further assuming w.o.l.o.g. that $k_1<k_2$. For the case of full looping $(\mathtt{f})$, there exist two circular, untwisted, isolated minima $\bm{{q}}^f$ lying on the $y-z$ plane characterised by $\bm{\mathsf{u}}^f=(\pm{2\pi}/{L},0,0)$ and $\bm{\mathsf{v}}^f=(0,0,1)$. In particular, the one having non-positive $y$ coordinate is given by $\bm{r}^f(s)=\frac{L}{2\pi}(0,\cos{({2\pi s}/{L})}-1,\sin{({2\pi s}/{L})})$ and the rotation matrix $\bm{R}^f(s)$ is a counter-clockwise planar rotation about the $x$ axis of an angle $\varphi^f(s)={2\pi s}/{L}$, $s\in[0,L]$. Consequently, $\bm{\mathsf{u}}^f=({2\pi}/{L},0,0)$, $\bm{\mathsf{m}}^f=({2\pi k_1}/{L},0,0)$, $\bm{\mathsf{v}}^f=(0,0,1)$, $\bm{\mathsf{n}}^f=\bm{0}$ and the energy is simply computed as $E(\bm{{q}}^f)={2\pi^2 k_1}/{L}$. We observe that these solutions are special for the fact of being the same both for $(\mathtt{K})$ and $(\mathtt{C})$ rods, which is not the case in general. By contrast, there are no simple analytical expressions for the two planar and untwisted teardrop shaped isolated minimizers $\bm{{q}}^m$ involved in the marginal looping problem $(\mathtt{m})$, and elliptic functions or numerics must be used. For example, in the $(\mathtt{K})$ case, the rotation angle $\varphi^m(s)$ can be derived using elliptic functions in terms of the constant unknown force $\bm{n}^m=(0,n2,n3)$ \cite{AN1, AN2, AN3}. The qualitative shapes of the minimal energy configurations are reported in Fig.\ref{fig1}. It is important to underline that for the $(\mathtt{m})$ problem the solutions for $(\mathtt{K})$ and $(\mathtt{C})$ rods are different, since the latter are characterised by $\bm{\mathsf{v}}^m(s)=(0,\mathsf{v}_2^m(s)\neq 0,\mathsf{v}_3^m(s)\neq 1)$. More precisely, in Fig.\ref{fig1518} we provide a specific numerical analysis for the $(\mathtt{C})$ teardrop solution varying the undeformed length of the rod $L$. We recall that the projection of the tangent $\bm{r}'$ on the director $\bm{d}_2$ is the component $\mathsf{v}_2$ of the shear strain, whereas the projection of the tangent on the director $\bm{d}_3$ is the component $\mathsf{v}_3$ of the stretch.  We observe that the bending and shear components $\mathsf{u}_1^m(s)$, $\mathsf{v}_2^m(s)$ are overall increasing (in the sense of departing from zero) when decreasing $L$, while the stretch $\mathsf{v}_3^m(s)$ decreases and increases (in the sense of departing from one) respectively in the interior and at the boundaries of the interval $[0,L]$. Namely, bending reaches its maximum at $s=L/2$ and vanishes at the boundaries; there is no shear at $s=L/2$ and it is maximized symmetrically within the intervals $[0,L/2)$ and $(L/2,L]$; compression is maximum for $s=L/2$ and slight extension can be observed close to the boundaries. Critical behaviours occur for small values of $L$, where compression dominates and bending starts to decrease: this will be clearer in the following stability analysis. To be precise, among the equilibria satisfying the BCs for the $(\mathtt{f})$ and $(\mathtt{m})$ cases, there are also equilibria with figure eight centerlines, but in the present study their contributions will be neglected because of their higher elastic energy.

\begin{figure*}
\centering
\subfigure[]{
\includegraphics[width=.374\textwidth]{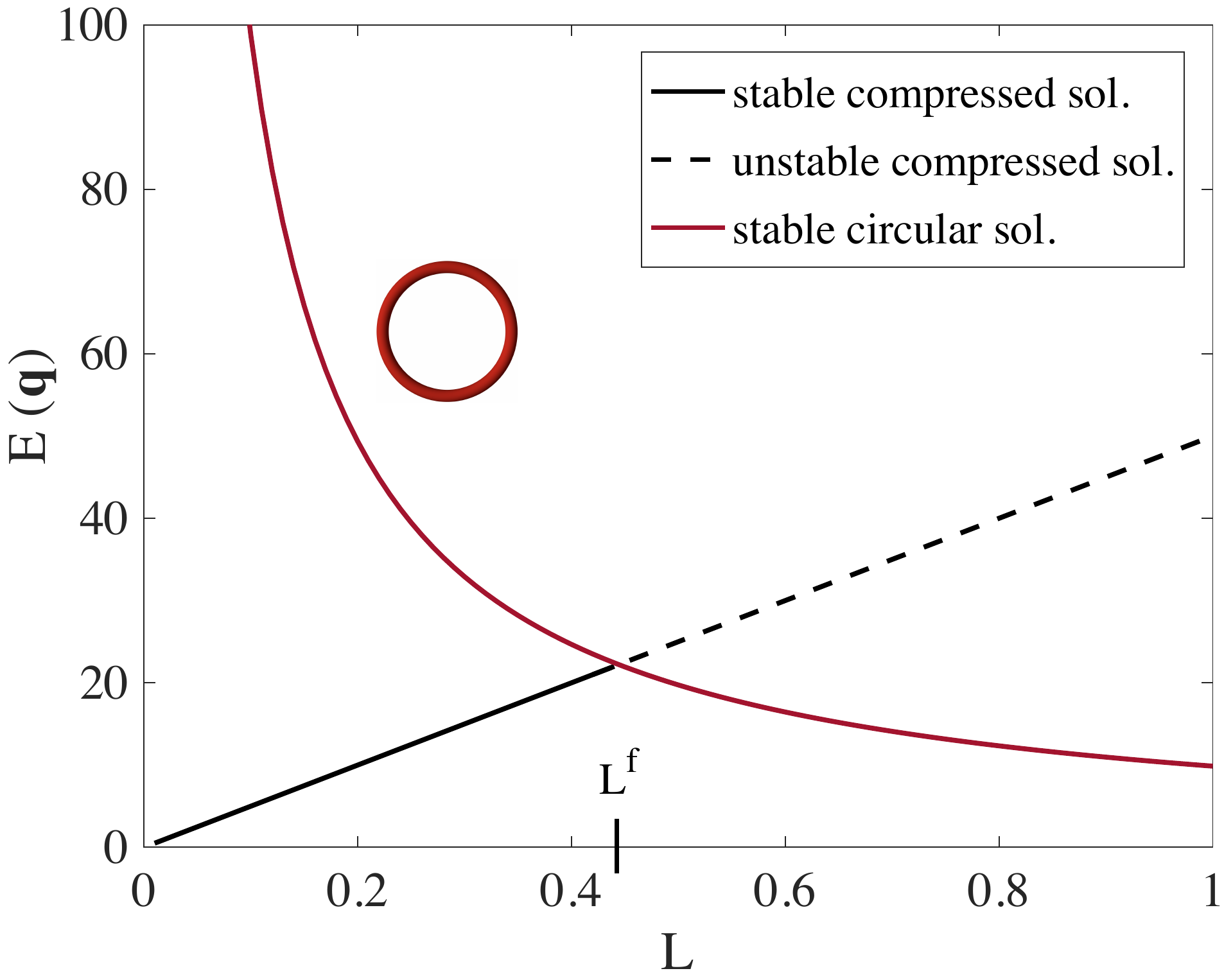}\qquad\qquad
}
\subfigure[]{
\includegraphics[width=.37\textwidth]{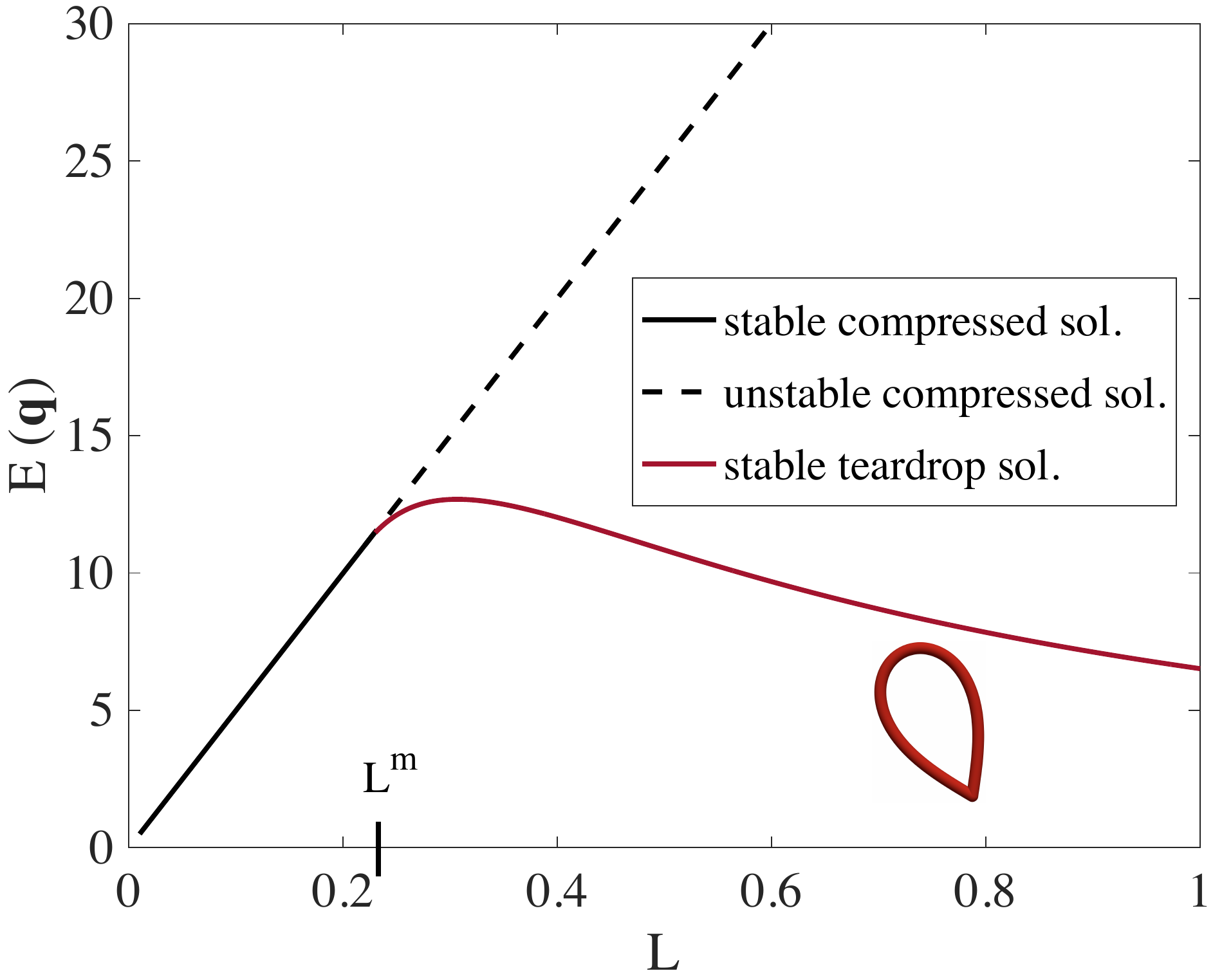}
}

\subfigure[]{
\includegraphics[width=.377\textwidth]{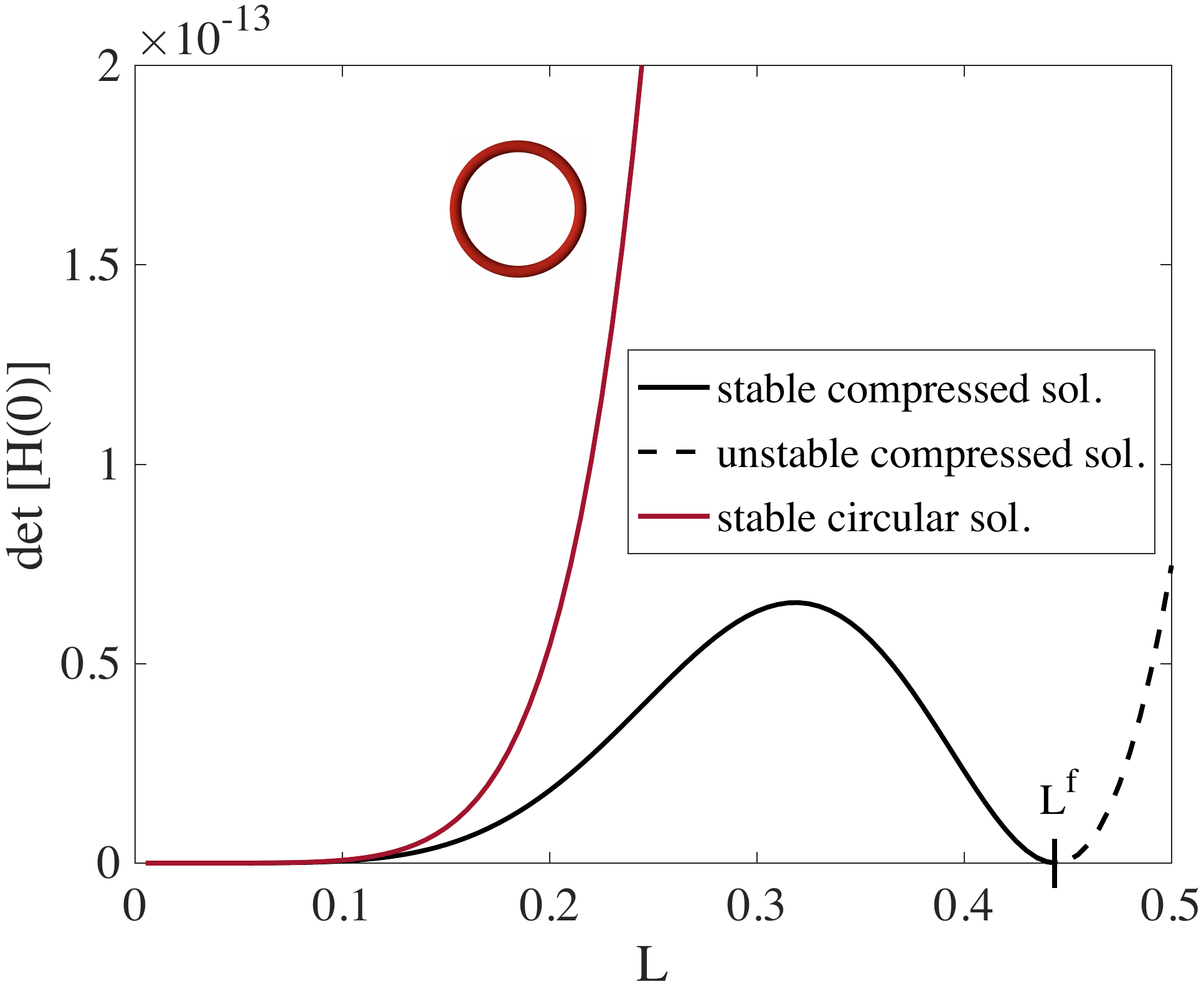}\qquad\qquad
}
\subfigure[]{
\includegraphics[width=.37\textwidth]{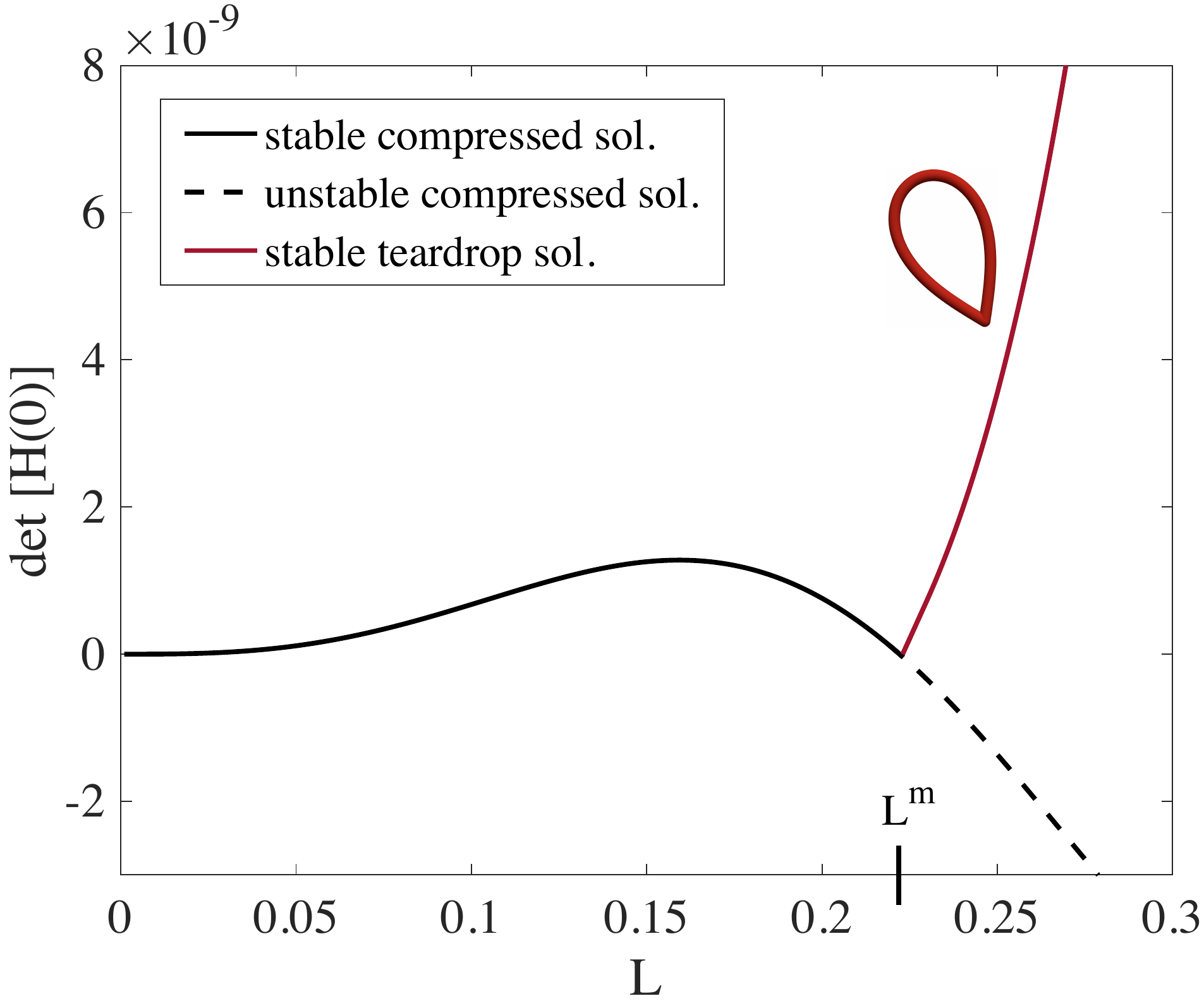}
}
\caption{Stability analysis for a non-isotropic $(\mathtt{C})$ rod with $k_1=0.5$, $k_2=5$, $k_3=10$ and $a_1=a_2=a_3=100$. Continuous lines represent quantities associated to stable solutions, dashed lines to unstable ones. In panels (a) and (b) the energies for the circular and teardrop equilibria are displayed in red, together with the compressed solution, in black, which becomes unstable after the bifurcation point $L^f$ or $L^m$. In panels (c) and (d) we report the values of $\det{[\bm{\mathsf{H}}(0)]}$  computed on the associated solutions for the $(\mathtt{f})$ and $(\mathtt{m})$ cases respectively, with conjugate points arising when the curves hit zero.}
\label{fig2345}
\end{figure*}

We continue the presentation with a brief stability analysis, showing that the circle and teardrop solutions are stable, with exceptions for the $(\mathtt{C})$ rod in the limit of the undeformed length $L$ going to zero, where bifurcations occur. For $(\mathtt{C})$ rods, cyclization problems $(\mathtt{f})$ and $(\mathtt{m})$ always admit a ``compressed'' trivial solution $\bm{q}^c$, characterised by $\bm{r}^c=\bm{0}$, $\bm{R}^c=\bm{\mathbb{1}}$, $\bm{\mathsf{u}}^c=\bm{0}$, $\bm{\mathsf{m}}^c=\bm{0}$, $\bm{\mathsf{v}}^c=\bm{0}$, $\bm{\mathsf{n}}^c=(0,0,-a_3)$ with energy $E(\bm{{q}}^c)={a_3 L}/{2}$, which starts to play an important role (this is not mentioned in \cite{LUD}). In summary, for the full $(\mathtt{C})$ case it exists $L^f>0$ such that the latter solution becomes stable and has lower energy than the circular minimizer $\bm{q}^f$ if $0<L<L^f$. In this regime the system will be mainly driven by the compressed solution (even if the circle remains stable). Moreover, for the marginal $(\mathtt{C})$ case, it exists $L^m>0$ such that the stable teardrop solution $\bm{q}^m$ ceases to exist in the interval $0<L<L^m$, merging with the compressed solution which becomes stable. In both the cases, the above observations will have a strong impact on the trend of the estimated cyclization probability densities, that is confirmed by MC simulations. More precisely, analysing the determinant of the associated Jacobi fields Eq.~(\ref{JacHam}) (with ICs and matrix $\bm{\mathsf{E}}(s)$ given later in Eq.~(\ref{inFM}) and Eq.~(\ref{E1}), Eq.~(\ref{E2}), Eq.~(\ref{E3})) by means of conjugate point theory, we observe that the compressed solution is stable (\ie a minimizer of the energy) in the range $0<L<L^f$ for the $(\mathtt{f})$ case, and in $0<L<L^m$ for the $(\mathtt{m})$ case, where $L^f={2\pi}/{a_3}\min{(\sqrt{k_1\,a_2},\,\,\sqrt{k_2\,a_1})}$ and $L^m=L^f/2$. Moreover, as already mentioned, for full looping $(\mathtt{f})$ there exist also circular solutions $\bm{q}^f$, which are stable for all $L>0$, with energy ${2\pi^2 k_1}/{L}$. (This is true except for $k_1>k_3$, $L<2\pi\sqrt{(k_1-k_3)/{a_1}}$, but in the present article we will not treat such an instability of the circular solution). Note that if $k1\leq k2$ and $a_1=a_2=a_3$, then $E(\bm{q}^c)=E(\bm{q}^f)$ at $L^f={2\pi}\sqrt{k_1/a_3}$ and $E(\bm{q}^c)<E(\bm{q}^f)$  for $0<L<L^f$. For marginal looping $(\mathtt{m})$, the teardrop solution $\bm{q}^m$ is not present in the interval $0<L<L^m$, transforming into the compressed solution which becomes stable. We show the bifurcation diagrams in Fig.\ref{fig2345} for a non-isotropic $(\mathtt{C})$ rod. Observe that $E(\bm{q}^m)$ does not explode for small lengths, but instead reaches a maximum and decreases towards $E(\bm{q}^c)$. By contrast, for a $(\mathtt{K})$ rod the circular and teardrop solutions exist and are stable for all $L>0$, with energy diverging approaching $L=0$, and no compressed solution is present.

In addition to the above statements, the isotropic case requires a more detailed analysis for the presence of a continuous symmetry. Namely, for a general linearly elastic (transversely) isotropic $(\mathtt{C})$ rod defined by $\bm{\mathcal{P}}(s)=\text{diag}(k_1(s),k_2(s),k_3(s),a_1(s),a_2(s),a_3(s))$ with $k_1=k_2$, $a_1=a_2$ and $\hat{{\mathsf{u}}}_1=\hat{{\mathsf{u}}}_2=\hat{{\mathsf{v}}}_1=\hat{{\mathsf{v}}}_2=0$, it is known \cite{RING} that for cyclization BCs $(\mathtt{f})$ in Eq.~(\ref{f}) and $(\mathtt{m})$ in Eq.~(\ref{m}) the equilibria are non-isolated and form a manifold obtained, starting from a known solution, by a rigid rotation of the rod of an angle $\theta$ about the $z$ axis and a subsequent rotation of the framing by an angle $-\theta$ about $\bm{d}_3(s)$, for $\theta\in[0,2\pi)$ (register symmetry). As a consequence, in our particular examples, once selected \eg the non-isotropic solution lying in the $y-z$ plane, $y\leq 0$ and characterized by the configuration $(\bm{R}(s),\,\,\bm{r}(s))$, $s\in[0,L]$, then we get an entire family of minimizers $\bm{R}(s;\theta)=\bm{Q}_{\theta}\bm{R}(s)\bm{Q}^T_{\theta}$, $\bm{r}(s;\theta)=\bm{Q}_{\theta}\bm{r}(s)$, where $\bm{Q}_{\theta}$ is defined as the counter-clockwise planar rotation matrix about the $z$ axis of an angle $\theta\in[0,2\pi)$ (Fig.\ref{fig1}). As a side note for the $(\mathtt{f})$ example, being the circular solutions the same for $(\mathtt{K})$ and $(\mathtt{C})$ rods, the isotropy symmetry arises even if $a_1\neq a_2$.

Furthermore, for a general linearly elastic uniform rod, for which the stiffness matrix $\bm{\mathcal{P}}$ and the intrinsic strains $\hat{\bm{\mathsf{u}}}$, $\hat{\bm{\mathsf{v}}}$ are independent of $s$, another continuous symmetry is present for the cyclization BCs $(\mathtt{f})$ in Eq.~(\ref{f}). In fact, starting from a known solution characterized by the configuration $(\bm{R}(s),\,\,\bm{r}(s))$, $s\in[0,L]$, it is possible to obtain a family of equilibria parametrised by $s^*\in[0,L)$ in the following way: select $s^*\in[0,L)$, rigidly translate the rod by $-\bm{r}(s^*)$, reparametrise the rod using the parameter $t\in[0,L]$ such that $s=t+s^*$ $(\text{mod}\,\,L)$, rigidly rotate the rod about the origin by means of $\bm{R}^T(s^*)$. However, in our uniform examples, the symmetry of uniformity is not playing any role, due to the circular centerline of the minimizers which is a fixed point of the transformation and, in the marginal case, to the impossibility of satisfying the condition $\bm{m}^m(L)=\bm{0}$ after the application of the symmetry.

In the present article we will deal with only one symmetry parameter, namely $\theta\in[0,2\pi)$ associated to isotropic rods, where the presence of a family of minimizers translates into a zero mode $\bm{\mathsf{\psi}}^{\alpha}(s;\theta)$ ($\alpha$ standing both for $f$ and $m$) of the self-adjoint operator $\bm{\mathsf{S}}^{\alpha}$ associated to the second variation Eq.~(\ref{sec}), as will be discussed in due course. Therefore, the stability analysis reported in Fig.\ref{fig2345} is totally analogous for the isotropic case, except from the fact that an entire family of minimizers is involved and a conjugate point is always present due to the zero mode. Furthermore, the theory can be applied to the uniformity symmetry alone and generalised to cases in which isotropy and uniformity allow the coexistence of two non-degenerate symmetry parameters $(\theta,\,s^*)$ generating a manifold of equilibria isomorphic to a torus, as it is the case of figure eight minimizers with $(\mathtt{f})$ cyclization BCs. Finally, note that in the following theory there is no assumption either of uniformity of the rod, nor, in general, of a straight intrinsic shape.

\section{Statement of the problem and general results}\label{s4}
In this section we describe the problem at the heart of this paper and present the general formulas that we derive in the context of end-to-end probabilities for fluctuating elastic rods, valid both in the $(\mathtt{C})$ and $(\mathtt{K})$ cases. The proof and the application of these results will follow in separate sections. Thus we consider an elastic rod at thermodynamic equilibrium with a heat bath in absence of external forces, assuming w.o.l.o.g. that $\bm{q}(0)=\bm{q}_0=(\mathbb{1},\bm{0})$. Then, given a prescribed $\bm{q}_L=(\bm{R}_L,\bm{r}_L)\in SE(3)$, we formulate the problem of computing a conditional probability density function (pdf) for the other end of the rod to satisfy at $s=L$ either $\bm{q}(L)=\bm{q}_L$, or the weaker condition $\bm{r}(L)=\bm{r}_L$. The first case gives rise to a conditional pdf $(\mathtt{f})$ over the space $SE(3)$ denoted by $\rrf(\bm{q}_L,L|\bm{q}_0,0)$, whereas the second one represents the $\mathbb{R}^3$-valued marginal $(\mathtt{m})$ over the final rotation variable, with no displacement constraint on $\bm{R}(L)$, that will be denoted by $\rrm(\bm{r}_L,L|\bm{q}_0,0)$. The following results are given for the case of linear elasticity, although the theory developed in the article is general.

We show that an approximate form of the conditional probability density in the case of an isolated minimizer $\bm{q}^{\alpha}(s)$ of the elastic energy Eq.~(\ref{energy}) (with respect to the associated BVPs $(\mathtt{f})$ and $(\mathtt{m})$) reads as
\begin{equation}\label{fin}
\rra\approx 
\left(\frac{\beta}{2\pi}\right)^{x(\alpha)}
\frac{e^{-\beta E\left(\bm{q}^{\alpha}\right)}}{\sqrt{\det{[\hha(0)]}}}\;,
\end{equation}
with $x(f)=3$, $x(m)=3/2$, and we are interested in the cyclization values $\rrf(\bm{q}_0,L|\bm{q}_0,0)$, $\rrm(\bm{0},L|\bm{q}_0,0)$. We denote by $\hha(s)$ the Jacobi fields computed at $\bm{q}^{\alpha}$, solutions of the associated Jacobi equations Eq.~(\ref{JacHam}) with $\bm{\mathsf{E}}(s)$ reported in Eq.~(\ref{E1}), Eq.~(\ref{E2}), Eq.~(\ref{E3}) and ICs given at $s=L$ as
\begin{equation}\label{inFM}
\begin{split}
\hhf(L)=\mathbb{0}\;,\,\,\mmf(L)=-\mathbb{1}\;;\quad\qquad\qquad\\
\hhm(L)=
\begin{pmatrix}
     \mathbb{1}_{3\times 3} & \mathbb{0}_{3\times 3}\\
      \mathbb{0}_{3\times 3} & \mathbb{0}_{3\times 3}
       \end{pmatrix}\;,\,\,\mmm(L)=\begin{pmatrix}
    \mathbb{0}_{3\times 3} & \mathbb{0}_{3\times 3}\\
      \mathbb{0}_{3\times 3} & -\mathbb{1}_{3\times 3}
       \end{pmatrix}\;.
\end{split}
\end{equation}

We further show that an approximate form of the conditional probability density in the case of non-isolated minimizers $\bm{q}^{\alpha}(s;\theta)$, obtained by means of a suitable regularization procedure, reads as
\begin{equation}\label{finisoSTAR}
\rra\approx2\pi\,e^{-\beta E(\bm{q}^{\alpha})}\sqrt{\frac{[\bm{\mathsf{\mu}}_{{\bm{\mathsf{\psi}}}^{\alpha}}(0)]_i}{]{\hha(0)}[_{i,i}}}\;,
\end{equation}
and we are interested in the cyclization values $\rrf(\bm{q}_0,L|\bm{q}_0,0)$, $\rrm(\bm{0},L|\bm{q}_0,0)$. In particular, $\bm{\mathsf{\mu}}_{{\bm{\mathsf{\psi}}}^{\alpha}}\in\mathbb{R}^6$ and $\hha\in\mathbb{R}^{6\times 6}$ are respectively the conjugate momentum of the zero mode and the Jacobi fields associated to ${\bm{\mathsf{S}}}^{\alpha}$, both computed by means of Eq.~(\ref{JacHam}) but recalling an extra contribution of $\frac{\beta}{2\pi}$ (see next section). Moreover, here we denote with $[\cdot]_i$ the $i$-th component of a vector, with $]\cdot[_{i,i}$ the principal minor of a square matrix removing the $i$-th row and the $i$-th column, and the index $i$ depends on the choice of the boundary regularization, based on the non-zero components of $\bm{\mathsf{\mu}}_{{\bm{\mathsf{\psi}}}^{\alpha}}$. The appropriate ICs for $\hha$ are given at $s=L$ as: 
\begin{equation}\label{inFMisoSTAR}
\begin{split}
\hhf(L)=\bm{\mathbb{0}}\;,\,\,\mmf(L)=\bm{\mathsf{\chi}}\;;\qquad\qquad\quad\\
\hhm(L)=\begin{pmatrix}
      \bm{\mathsf{X}}_{1,1} &   \bm{\mathsf{X}}_{1,2} \\
     \bm{\mathbb{0}} & \bm{\mathbb{0}}
     \end{pmatrix}\;,\,\,\mmm(L)=\begin{pmatrix}
      \bm{\mathbb{0}} &  \bm{\mathbb{0}} \\
      \bm{\mathsf{X}}_{2,1} &  \bm{\mathsf{X}}_{2,2}
     \end{pmatrix}\;,
\end{split}
\end{equation}
where $\bm{\mathsf{\chi}}$ is an arbitrary matrix with unit determinant such that the $i$-th column corresponds to $\bm{\mathsf{\mu}}_{{\bm{\mathsf{\psi}}}^f}(L)$ and $\bm{\mathsf{X}}=\begin{small}\begin{pmatrix}
      \bm{\mathsf{X}}_{1,1} &   \bm{\mathsf{X}}_{1,2} \\
     \bm{\mathsf{X}}_{2,1} & \bm{\mathsf{X}}_{2,2}
     \end{pmatrix}\end{small}\in\mathbb{R}^{6\times 6}$, partitioned in $3$ by $3$ blocks, is an arbitrary matrix with determinant equal to $-1$ such that the $i$-th column corresponds to $([{\bm{\mathsf{\psi}}}^m]_{1:3},[\bm{\mathsf{\mu}}_{{\bm{\mathsf{\psi}}}^m}]_{4:6})^T(L)$.

\section{Fluctuating elastic rods and the path integral formulation}\label{s5}
If a polymer interacts with a solvent heat bath, the induced thermal motion gives rise to a stochastic equilibrium that we model making use of a Boltzmann distribution on rod configurations satisfying $\bm{q}(0)=\bm{q}_0$ \cite{LUDT, LUD}, of the form ${\mathcal{Z}^{-1}}{e^{-\beta E(\bm{q}(s))}}$, with $\beta$ the inverse temperature and $\mathcal{Z}$ the partition function of the system. A precise treatment of the previous expression requires the introduction of the path integral formalism \cite{BookFeynman, BookChaichian, BookSchulman, BookWiegel}. Namely, the $SE(3)$ and $\mathbb{R}^3$ densities $\rrf$ and $\rrm$ are respectively given as the ratios of infinite dimensional Wiener integrals \cite{2020}: 
\begin{equation}\label{dens}
\rrf(\bm{q}_L,L|\bm{q}_0,0)=\frac{\mathcal{K}_f}{\mathcal{Z}}\;,\quad\rrm(\bm{r}_L,L|\bm{q}_0,0)=\frac{\mathcal{K}_m}{\mathcal{Z}}\;,
\end{equation}
\begin{equation}\label{pathint}
{\mathcal{K}_f}=\int\limits_{\bm{q}(0)=\bm{q}_0}^{\bm{q}(L)=\bm{q}_L}{e^{-\beta E(\bm{q})}\,\mathcal{D}\bm{q}}\;,\,\,\,{\mathcal{K}_m}=\int\limits_{\bm{q}(0)=\bm{q}_0}^{\bm{r}(L)=\bm{r}_L}{e^{-\beta E(\bm{q})}\,\mathcal{D}\bm{q}}\;.
\end{equation}%
The limits of integration are dictated by the BCs Eq.~(\ref{f}) and Eq.~(\ref{m}) respectively  and $\mathcal{Z}$ is a path integral over all paths with BCs given in Eq.~(\ref{hat}) that guarantees the normalisation condition:
\begin{equation}\label{normex}
\mathcal{Z} = \int\limits_{\bm{q}(0)=\bm{q}_0}{e^{-\beta E(\bm{q})}\,\mathcal{D}\bm{q}}\;,
\end{equation}
$\int_{SE(3)}{\rrf(\bm{q}_L,L|\bm{q}_0,0)}\,\dd \bm{q}_L=\int_{\mathbb{R}^3}{\rrm(\bm{r}_L,L|\bm{q}_0,0)}\,\dd \bm{r}_L=1$.
The prescriptions $\bm{m}(L)=\bm{0}$ for $\mathcal{K}_m$ and $\bm{m}(L)=\bm{n}(L)=\bm{0}$ for $\mathcal{Z}$ account for Neumann natural BCs at $s=L$ and concern the minimizers. We stress that it is key that at this stage the model is an extensible, shearable rod, namely with $(\mathtt{C})$ energy Eq.~(\ref{energy}), otherwise the problem could not be expressed as simple BCs at $s=0$ and $s=L$. Moreover, to apply all the path integral machinery, we first have to deal with the rotation group $SO(3)$, being part of the configuration variable $\bm{q}(s)=(\bm{R}(s),\bm{r}(s))$, which gives rise to a manifold structure that should be treated carefully in order to recover eventually a ``flat space" formulation.

Following \cite{LUDT}, we show in appendix \ref{CoorSO3} how to build an $\mathbb{R}^6$ parametrization of $SE(3)$ ($SO(3)$ is not simply connected and a zero measure set of rotations is neglected) adapted to a given unit quaternion $\bar{\bm{\gamma}}\in\mathbb{R}^4$. In particular we make use of the Haar measure on $SO(3)$ and derive the metric tensor associated to the parametrization. Namely, the $\bar{\bm{\gamma}}$-adapted parametrization of $SE(3)$ denoted by $\bm{\mathsf{q}}(s)=(\bm{\mathsf{c}}(s),\bm{\mathsf{t}}(s))\in\mathbb{R}^6$ exploits the relation between unit quaternions (or Euler parameters) $\bm{\gamma}$ and elements of $SO(3)$ and is given by
\begin{equation}\label{par}
\bm{\gamma}(\bm{\mathsf{c}})=\frac{1}{\sqrt{1+\Vert \bm{\mathsf{c}}\Vert^2}}\left(\sum\limits_{i=1}^3{\mathsf{c}_i\bm{B}_i\bar{\bm{\gamma}}}+\bar{\bm{\gamma}}\right)\;,\,\,\,\,\bm{\mathsf{t}}=\bm{R}(\bar{\bm{\gamma}})^T\bm{r}\;,
\end{equation}
with $\bm{\mathsf{c}}=(\mathsf{c}_1,\mathsf{c}_2,\mathsf{c}_3)\in\mathbb{\bm{R}}^3$, $\bm{R}(\bar{\bm{\gamma}})$ the rotation matrix expressed by $\bar{\bm{\gamma}}$, and $\bm{B}_1$, $\bm{B}_2$, $\bm{B}_3$ in $\mathbb{R}^{4\times 4}$ reported in Eq.~(\ref{BMat}). Moreover, by means of the Feynman discrete interpretation of the path integral measure \cite{BookFeynman}, the metric tensor and the infinitesimal volume measure read respectively
\begin{equation}\label{par2}
\begin{split}
\bm{\mathsf{g}}(\bm{\mathsf{c}})=\frac{\bm{\mathbb{1}}}{1+\Vert\bm{\mathsf{c}}\Vert^2}-\frac{\bm{\mathsf{c}}\otimes\bm{\mathsf{c}}}{\left(1+\Vert\bm{\mathsf{c}}\Vert^2\right)^2}\;,\quad\qquad\\
\dd \bm{q}_j=\sqrt{\det{[\bm{\mathsf{g}}(\bm{\mathsf{c}}_j)]}}\,\dd \bm{\mathsf{c}}_j\,\dd \bm{\mathsf{t}}_j=\frac{1}{\left(1+\Vert\bm{\mathsf{c}}_j\Vert^2\right)^2}\,\dd \bm{\mathsf{c}}_j\,\dd \bm{\mathsf{t}}_j\;.
\end{split}
\end{equation}

The latter results are implemented by choosing three different curves of unit quaternions $\bar{\bm{\gamma}}(s)$ to be the curves defined by the rotation component ${\bm{R}}(\bar{\bm{\gamma}})$ of the minimizers $\bm{q}^f$, $\bm{q}^m$ and $\hat{\bm{{q}}}$ respectively, which characterise the three different parametrisations involved in the computation of $\mathcal{K}_f$, $\mathcal{K}_m$ and $\mathcal{Z}$ in view of the semi-classical approximation. Then, replacing the configuration variable $\bm{q}(s)\in SE(3)$ with the sans-serif fonts $\bm{\mathsf{q}}(s)\in \mathbb{R}^6$, we can formally write the integrand and measure in Eq.~(\ref{pathint}) and Eq.~(\ref{normex}) as $e^{-\beta E(\bm{\mathsf{q}})}\sqrt{\det{[\bm{\mathsf{g}}(\bm{\mathsf{c}})]}}\,\mathcal{D}\bm{\mathsf{q}}$. The treatment of the metric factor relies on the introduction of real-valued ghost fields for exponentiating the measure, as can be found in \cite{GHO}. This means rewriting the factor as a Gaussian path integral in the ghost field $\bm{\mathsf{z}}(s)\in\mathbb{R}^3$ satisfying $\bm{\mathsf{z}}(0)=\bm{0}$ with energy $\frac{1}{2}\int_0^L{\bm{\mathsf{z}}^T\bm{\mathsf{g}}^{-1}(\bm{\mathsf{c}})\bm{\mathsf{z}}\,\dd s}$. After that, we consider the path integral expressions in the joint variable $\bm{\mathsf{w}}=(\bm{\mathsf{q}},\bm{\mathsf{z}})$, \eg
\begin{equation}\label{curved}
\mathcal{K}_f=\int\limits_{\bm{\mathsf{w}}(0)=(\bm{\mathsf{q}}_0,\bm{0})}^{\bm{\mathsf{q}}(L)=\bm{\mathsf{q}}_L}{e^{-\beta\big[E(\bm{\mathsf{q}}(s))+\frac{1}{2}\int_0^L{\bm{\mathsf{z}}(s)^T\bm{\mathsf{g}}^{-1}(\bm{\mathsf{c}}(s))\bm{\mathsf{z}}(s)\,\dd s}\big]}\,\mathcal{D}\bm{\mathsf{w}}}\;.
\end{equation}
In the following, even if the theory could be given in principle for a general strain energy density $W$, in order to perform concrete computations we refer to the case of linear elasticity, where $W$ is a quadratic function, driven by the stiffness matrix $\bm{\mathcal{P}}(s)$:
\begin{equation}\label{energylin}
E(\bm{\mathsf{q}})=\frac{1}{2}\int_0^L{     
\begin{pmatrix}
       \bm{\mathsf{u}}-\hat{\bm{\mathsf{u}}} \\
        {\bm{\mathsf{v}}-\hat{\bm{\mathsf{v}}}}
\end{pmatrix}^T 
\begin{pmatrix}
       \bm{\mathcal{K}} & \bm{\mathcal{B}}  \\
       \bm{\mathcal{B}}^T & \bm{\mathcal{A}}
\end{pmatrix}
\begin{pmatrix}
       \bm{\mathsf{u}}-\hat{\bm{\mathsf{u}}} \\
        {\bm{\mathsf{v}}-\hat{\bm{\mathsf{v}}}}
\end{pmatrix}}\dd s\;.
\end{equation}
Moreover, we also refer to the particular looping case of ring-closure or cyclization, evaluating $\rrf$ at $\bm{q}_L=\bm{q}_0$ and the marginal $\rrm$ at $\bm{r}_L=\bm{0}$; the same conditions apply to the minimizers.

\subsection{Looping probabilities in the case of isolated minimizers}
Since the elastic energy functional Eq.~(\ref{energylin}) is non quadratic in $\bm{\mathsf{q}}$, after the parametrisation we approximate $\mathcal{K}_f$, $\mathcal{K}_m$ and $\mathcal{Z}$ by means of a second-order expansion about a minimal energy configuration \cite{PAP1, MOR, BookWiegel, BookChaichian, BookSchulman, 2020}, known as the semi-classical method, or, in our real-valued context, Laplace expansion \cite{PIT}. The present work follows the set-up of \citep{2020}. We further recall that such an approximation holds when the energy required to deform the system is large with respect to the temperature of the heat bath, \ie in the short-length scale, or stiff, regimes.

First, note that there is no contribution to the result coming from the ghost energy when approximating path integrals of the kind of Eq.~(\ref{curved}) to second order in the joint variable $\bm{\mathsf{w}}$. This is a consequence of the structure of the metric tensor Eq.~(\ref{par2}), \ie $\bm{\mathsf{g}}^{-1}(\bm{\mathsf{c}})=(1+\bm{\mathsf{c}}\cdot\bm{\mathsf{c}})(\bm{\mathbb{1}}+\bm{\mathsf{c}}\otimes\bm{\mathsf{c}})$, and therefore we can consider only the elastic energy Eq.~(\ref{energylin}) in the variable $\bm{\mathsf{q}}$. In fact, the minima $\bm{q}^f$ and $\bm{q}^m$ (here assumed to be isolated) encoded within the associated adapted parametrisations lead to the minimizers $\qqf$ and $\qqm$ (denoted generically by $\qqa$, $\alpha$ standing both for $f$ and $m$) characterised by $\bm{\mathsf{c}}^{\alpha}=\bm{0}$. In particular, the Neumann natural BC $\bm{m}(L)=\bm{0}$ for $\bm{q}^m$  translates into $\frac{\partial W}{\partial\bm{\mathsf{c}}'}(L)=2[(\bm{\mathbb{1}}+\bm{\mathsf{c}}^{\times})/(1+\Vert\bm{\mathsf{c}}\Vert^2)\bm{\mathsf{m}}](L)=\bm{0}$ for $\qqm$. In the semi-classical approximation for $\mathcal{K}_f$ and $\mathcal{K}_m$ the energy is expanded about the associated $\qqa$ as $E(\bm{\mathsf{q}})\sim E(\qqa)+\frac{1}{2}\delta^2E(\bm{\mathsf{h}};\qqa)$, $\bm{\mathsf{q}}=\qqa+\bm{\mathsf{h}}$, being the first variation zero. The second variation $\delta^2E$ is reported in Eq.~(\ref{sec}), with $\bm{\mathsf{h}}=(\delta\bm{\mathsf{c}},\delta\bm{\mathsf{t}})$ the perturbation field describing fluctuations about the minimizer $\qqa$ and satisfying the linearised version of the parametrised BCs, \ie $\bm{\mathsf{h}}(0)=\bm{\mathsf{h}}(L)=\bm{0}$ for $(\mathtt{f})$, or $\bm{\mathsf{h}}(0)=\bm{0}$, $\delta\bm{\mathsf{t}}(L)=\bm{0}$, $\delta\frac{\partial W}{\partial\bm{\mathsf{c}}'}(L)=2[\delta\bm{\mathsf{m}}-\bm{\mathsf{m}}^m\times\delta\bm{\mathsf{c}}](L)=\bm{0}$ for $(\mathtt{m})$. Analogously, for $\mathcal{Z}$ the energy is expanded about $\qqh$, being $\frac{\partial W}{\partial\bm{\mathsf{t}}'}(L)=[\hat{\bm{R}}^T\bm{R}\,\bm{\mathsf{n}}](L)=\bm{0}$ the associated Neumann natural BC arising from $\bm{n}(L)=\bm{0}$ in $\hat{\bm{q}}$ (in addition to the BC for the moment as described for $\qqm$). In this case, the linearised parametrised BCs are given by $\bm{\mathsf{h}}(0)=\bm{0}$, $\bm{\mathsf{\mu}}(L)=\bm{0}$, with $\bm{\mathsf{\mu}}(L)=(\delta\frac{\partial W}{\partial\bm{\mathsf{c}}'},\delta\frac{\partial W}{\partial\bm{\mathsf{t}}'})(L)=(2(\delta\bm{\mathsf{m}}-\hat{\bm{\mathsf{m}}}\times\delta\bm{\mathsf{c}}),\delta\bm{\mathsf{n}}-2\hat{\bm{\mathsf{n}}}\times\delta\bm{\mathsf{c}})(L)$.

In the present case of linear elasticity, the second variation Eq.~(\ref{sec}) is characterised by $\bm{\mathsf{P}}$, related to the stiffness matrix $\bm{\mathcal{P}}$, and ${\bm{\mathsf{C}}}$, ${\bm{\mathsf{Q}}}$ which can be computed as follows in terms of strains, forces and moments of the minimizer involved, generically denoted by $\bar{\bm{q}}=({\bm{R}}(\bar{\bm{\gamma}}),\bar{\bm{r}})$. In elastic rod theory, the natural parametrisation for the variation field around $\bar{\bm{q}}$ is directly provided by the Lie algebra $so(3)$ of the rotation group in the director frame, namely $\delta\bm{R}={\bm{R}}(\bar{\bm{\gamma}})\delta\bm{\mathsf{\eta}}^{\times}$, where $\delta\bm{\mathsf{\eta}}^{\times}$ denotes the skew-symmetric matrix or cross product matrix of $\delta\bm{\mathsf{\eta}}\in\mathbb{R}^3$. In order to show the relation between $\delta\bm{\mathsf{\eta}}$ and the variation field $\delta\bm{\mathsf{c}}$, we use the formula $\delta\bm{\mathsf{\eta}}=2\left(\sum\limits_{i=1}^3{\bm{e}_i\otimes\bm{B}_i\bar{\bm{\gamma}}}\right)\delta\bm{\gamma}$ (which is substantially the relation between the Darboux vector and Euler parameters, see \eg \cite{HAM}) with $\delta\bm{\gamma}=\frac{\partial\bm{\gamma}}{\partial\bm{\mathsf{c}}}\big|_{\bm{\mathsf{c}}=\bm{0}}\delta\bm{\mathsf{c}}=\sum\limits_{j=1}^3\bm{B}_j\bar{\bm{\gamma}}\delta\mathsf{c}_j$ referring to Eq.~(\ref{BMat}), Eq.~(\ref{par}), and we conclude that $\delta\bm{\mathsf{\eta}}(s)=2\,\delta\bm{\mathsf{c}}(s)$. 

With reference to \cite{NAD}, the second variation of the linear hyper-elastic energy Eq.~(\ref{energylin}) in the director variable $\bm{\mathsf{\omega}}=(\delta\bm{\mathsf{\eta}},\delta\bm{\mathsf{t}})$ is $\delta ^2 E=\int_0^L{[{\bm{\mathsf{\omega}}'}^T\bm{\mathcal{P}}\bm{\mathsf{\omega}}'+2{\bm{\mathsf{\omega}}'}^T{\bm{\mathcal{C}}}\bm{\mathsf{\omega}}+\bm{\mathsf{\omega}}^T{\bm{\mathcal{Q}}}\bm{\mathsf{\omega}}]}\,\dd s,$
where $\bm{\mathcal{P}}$ is the stiffness matrix and $\bm{\mathcal{C}}$, $\bm{\mathcal{Q}}$ are respectively given in terms of strains, forces and moments by Eq.~(\ref{Cmat}) and Eq.~(\ref{Qmat}). Finally, introducing the matrix $\bm{\mathcal{D}}=\begin{small}\begin{pmatrix}
       2\mathbb{1} & \mathbb{0}  \\
       \mathbb{0} & \mathbb{1}
     \end{pmatrix}\end{small}$,
we have that the second variation in the variable $\bm{\mathsf{h}}=(\delta\bm{\mathsf{c}},\delta\bm{\mathsf{t}})$ Eq.~(\ref{sec}) is given by $\delta ^2 E=\int_0^L{[{\bm{\mathsf{h}}'}^T\bm{\mathsf{P}}\bm{\mathsf{h}}'+2{\bm{\mathsf{h}}'}^T{\bm{\mathsf{C}}}\bm{\mathsf{h}}+\bm{\mathsf{h}}^T{\bm{\mathsf{Q}}}\bm{\mathsf{h}}]}\,\dd s$, with $\bm{\mathsf{P}}=\bm{\mathcal{D}}\bm{\mathcal{P}}\bm{\mathcal{D}}$, $\bm{\mathsf{C}}=\bm{\mathcal{D}}\bm{\mathcal{C}}\bm{\mathcal{D}}$ and $\bm{\mathsf{Q}}=\bm{\mathcal{D}}\bm{\mathcal{Q}}\bm{\mathcal{D}}$.

The Jacobi equations in first order Hamiltonian form associated to the latter second variation functional are given in Eq.~(\ref{JacHam}) and are driven by the symmetric matrix $\bm{\mathsf{E}}(s)\in\mathbb{R}^{12\times 12}$ detailed in Eq.~(\ref{E1}). The Jacobi fields $\bm{\mathsf{H}}(s)\in\mathbb{R}^{6\times 6}$, together with the conjugate variable under the Legendre transform $\bm{\mathsf{M}}(s)\in\mathbb{R}^{6\times 6}$ represent the solutions of the Jacobi equations once prescribed appropriate ICs. The columns $\bm{\mathsf{h}}$ of $\bm{\mathsf{H}}$ and the ones $\bm{\mathsf{\mu}}$ of $\bm{\mathsf{M}}$ are related by $\bm{\mathsf{\mu}}=\bm{\mathsf{P}}\bm{\mathsf{h}}'+\bm{\mathsf{C}}\bm{\mathsf{h}}$.

Note that until now the formulation adopted is for the general $(\mathtt{C})$ rod with extension, shear and hence an invertible stiffness matrix $\bm{\mathcal{P}}$. The constrained inextensible and unshearable case $(\mathtt{K})$ requires the stiffness components $\bm{\mathcal{B}}$ and $\bm{\mathcal{A}}$ to diverge (as discussed in \cite{HAM, LUDT, LUD}), specifically as $\bm{\mathcal{B}}/{\epsilon}$ and $\bm{\mathcal{A}}/{\epsilon^2}$, for $\epsilon\rightarrow 0$. Switching to the Hamiltonian formulation, given a $(\mathtt{C})$ rod the compliance matrix $\bm{\mathcal{R}}$ (which is the inverse of $\bm{\mathcal{P}}$) has a smooth limit for $\epsilon\rightarrow 0$. Namely, for a $(\mathtt{K})$ rod we recover $\bm{\mathcal{R}}(s)=\begin{small}\begin{pmatrix}
       \bm{\mathcal{R}}_{1,1} & \bm{\mathcal{R}}_{1,2}=\bm{\mathbb{0}}\\
        \bm{\mathcal{R}}_{2,1}=\bm{\mathbb{0}} &  \bm{\mathcal{R}}_{2,2}=\bm{\mathbb{0}}
     \end{pmatrix}\end{small}$, with  $\bm{\mathcal{R}}_{1,1}(s)=(\bm{\mathcal{K}}-\bm{\mathcal{B}}\bm{\mathcal{A}}^{-1}\bm{\mathcal{B}}^T)^{-1}$. In conclusion, once prescribed a symmetric and positive definite matrix $\bm{\mathcal{K}}^{(\mathtt{K})}=\bm{\mathcal{R}}_{1,1}^{-1}$, there exists a sequence of positive definite and symmetric compliance matrices for the $(\mathtt{C})$ case converging smoothly to the $(\mathtt{K})$ case, implying that the expressions Eq.~(\ref{E2}) and Eq.~(\ref{E3}) for the blocks of the matrix $\bm{\mathsf{E}}(s)$ Eq.~(\ref{E1}) hold for both $(\mathtt{C})$ and $(\mathtt{K})$ rods. We emphasise that for the $(\mathtt{K})$ case $\delta\frac{\partial W}{\partial\bm{\mathsf{t}}'}$ is a basic unknown of the Jacobi equations and cannot be found using the relation $\bm{\mathsf{\mu}}=\bm{\mathsf{P}}\bm{\mathsf{h}}'+\bm{\mathsf{C}}\bm{\mathsf{h}}$, since the latter is not defined.

The resulting path integrals arising from the semi-classical method are of the form, \eg
\begin{equation}\label{Gauss}
\mathcal{K}_f\approx e^{-\beta E(\qqf)}\int\limits_{\bm{\mathsf{h}}(0)=\bm{0}}^{\bm{\mathsf{h}}(L)=\bm{0}}{e^{-\frac{\beta}{2}\delta^2 E(\bm{\mathsf{h}};\qqf)}\,\mathcal{D}\bm{\mathsf{h}}}\;,
\end{equation}
and similarly for $\mathcal{K}_m$ and $\mathcal{Z}$ but considering the different minimizers and linearised BCs. Then, applying the results derived in \cite{2020} for Gaussian path integrals, which are in turn extensions of the work of Papadopoulos \cite{PAP1}, we recover the approximate form of the conditional probability density Eq.~(\ref{fin}). In principle, denoting by $\hhh(s)$ $\in\mathbb{R}^{6\times 6}$ the Jacobi fields computed at $\hat{\bm{\mathsf{q}}}$ subject to the ICs $\hhh(L)=\mathbb{1}$, $\mmh(L)=\mathbb{0}$ \cite{2020}, the numerator and denominator in Eq.~(\ref{fin}) should be respectively $e^{-\beta\left(E\left(\qqa\right)-E\left(\qqh\right)\right)}$ and $\sqrt{\det{[\hha\hhh^{-1}(0)]}}$, in order to include the contribution coming from the evaluation of the partition function $\mathcal{Z}$. However, the result simplifies since $E(\hat{\bm{\mathsf{q}}})=0$, being $\hat{\bm{\mathsf{q}}}$ the intrinsic configuration of the rod. At the same time $\bm{\mathsf{E}}_{1,1}$ is the zero matrix for this case, which implies $\mmh(s)=\bm{\mathbb{0}}\,\,\forall s$ (according to the IC $\mmh(L)=\bm{\mathbb{0}}$) and consequently $\hhh(s)$ must satisfy a linear system whose matrix has zero trace. Thus, by application of the generalized Abel's identity or Liouville's formula, $\forall s$ we have that $\det[{{\hhh}}(s)]=\det[{{\hhh}}(L)]=\det[\bm{\mathbb{1}}]=1$. Furthermore, it is worth to mention that here the partition function computation is not affected by approximations, even if it apparently undergoes the semi-classical expansion. In fact, there exists a change of variables presented in \cite{LUDT, LUD} which allows an equivalent exact computation exploiting the specific BCs involved in $\mathcal{Z}$. In general, the latter change of variables is not applicable and the present method must be used, \eg for non-linear elasticity or in the case of a linearly elastic polymer subject to external end-loadings, for which the shape of the energy leads to a non-trivial contribution of the partition function that must be approximated.

\subsection{Looping probabilities in the case of non-isolated minimizers}
In this section we consider non-isolated minimizers arising as a consequence of continuous symmetries of the problem. In particular, we provide a theory for one symmetry parameter, namely $\theta\in[0,2\pi)$ (as we want to deal with isotropic rods), but the same scheme can be suitably generalised to more symmetry parameters. The presence of a family of minimizers denoted by $\qqa(s;\theta)$ translates into a zero mode $\bm{\mathsf{\psi}}^{\alpha}(s;\theta)=\frac{\partial}{\partial\theta}\qqa(s;\theta)$ \cite{MOR} of the self-adjoint operator $\bm{\mathsf{S}}=-\bm{\mathsf{P}}\frac{d^2}{d s^2}+(\bm{\mathsf{C}}^T-\bm{\mathsf{C}}-\bm{\mathsf{P}}')\frac{\dd}{\dd s}+\bm{\mathsf{Q}}-\bm{\mathsf{C}}'$ associated to the second variation Eq.~(\ref{sec}), namely $\delta ^2 E=(\bm{\mathsf{h}},\bm{\mathsf{S}}\bm{\mathsf{h}})$, where $(\cdot,\cdot)$ is the scalar product in the space of square-integrable functions $L^2([0,L];\mathbb{R}^6)$. Consequently, we cannot proceed as before, for otherwise expression Eq.~(\ref{fin}) will diverge for the existence of a conjugate point at $s=0$.

Thus, in evaluating expression Eq.~(\ref{pathint}) for $\mathcal{K}_f$ and $\mathcal{K}_m$, we adapt the parametrization to the minimizer corresponding to $\theta=0$, our choice of the gauge in applying the collective coordinates method, which amounts to a Faddeev-Popov-type procedure \cite{FADPOP}, widely used in the context of quantum mechanics for solitons or instantons \cite{POLYA, COL1, BER, JAR}, of inserting the Dirac delta transformation identity
\begin{equation}\label{Fad-Pop}
1=\left\lvert\frac{\partial}{\partial\theta}F\right\rvert_{\theta=0}\int{\delta(F(\theta))}\,\dd \theta,\,\,\,F(\theta)=\left(\bm{\mathsf{q}}-\qqa,\frac{\bm{\mathsf{\psi}}^{\alpha}}{\Vert\bm{\mathsf{\psi}}^{\alpha}\Vert}\right)
\end{equation}
within the path integral, in order to integrate over variations which are orthogonal to the zero mode. Once performed the semi-classical expansion as before about $\qqa$, exchanged the order of integration $\mathcal{D}\bm{\mathsf{h}}\leftrightarrow\dd \theta$ to get a contribution of $2\pi$, and having approximated to leading order both the metric tensor and the factor $\vert\partial/\partial\theta F\vert_{\theta=0}\approx \Vert\bm{\mathsf{\psi}}^{\alpha}(s;0)\Vert$, we are left with the computation of a ratio of Gaussian path integrals
\begin{equation}\label{Isopathint}
\frac{1}{\mathcal{Z}_g}\int{e^{-\pi(\bm{\mathsf{h}},\frac{\beta}{2\pi}\bm{\mathsf{S}}\bm{\mathsf{h}})}\delta\left[\left(\bm{\mathsf{h}},\frac{\bm{\mathsf{\psi}}^{\alpha}}{\Vert\bm{\mathsf{\psi}}^{\alpha}\Vert}\right)\right]\,\mathcal{D}\bm{\mathsf{h}}}\;,
\end{equation}
for the linearised parametrised BCs associated to Eq.~(\ref{f}) and Eq.~(\ref{m}) respectively. For notation simplicity, throughout this section $\bm{\mathsf{S}}$ stands for $\frac{\beta}{2\pi}\bm{\mathsf{S}}$ and $\hat{\bm{\mathsf{S}}}$ for $\frac{\beta}{2\pi}\hat{\bm{\mathsf{S}}}$, the latter operator driving the Gaussian path integral $\mathcal{Z}_g$ arising from the partition function $\mathcal{Z}$, in which the minimizer $\qqh$ is isolated. Note that, since the argument of the delta distribution must vanish for $\theta=0$ according to Eq.~(\ref{Fad-Pop}), then the integration for the numerator is performed on the minimizer $\qqa(s;0)$ with associated zero mode $\bm{\mathsf{\psi}}^{\alpha}(s;0)$; in the following they will both be denoted simply by $\qqa$ and $\bm{\mathsf{\psi}}^{\alpha}$.

Interpreting Eq.~(\ref{Isopathint}) as $\sqrt{\text{Det}(\hat{\bm{\mathsf{S}}})/{\text{Det}^{\star}({\bm{\mathsf{S}}})}}$, \textit{i.e.}, the square root of the ratio of the functional determinants for the operators $\hat{\bm{\mathsf{S}}}$ and ${\bm{\mathsf{S}}}$, the latter with removed zero eigenvalue (thus the $^{\star}$ sign) \cite{MCK, FAL}, we consider the following general strategy for its evaluation. Given the second variation operator ${\bm{\mathsf{S}}}$ acting on $\bm{\mathsf{h}}(s)\in\mathbb{R}^6$, with $s\in[0,L]$ and BCs determined by the square matrices $\bm{\mathsf{T}}_0$ and $\bm{\mathsf{T}}_L$ as $\bm{\mathsf{T}}_0\begin{small}\begin{pmatrix}
      \bm{\mathsf{h}}(0) \\
       \bm{\mathsf{\mu}}(0)
     \end{pmatrix}\end{small}+\bm{\mathsf{T}}_L\begin{small}\begin{pmatrix}
      \bm{\mathsf{h}}(L) \\
       \bm{\mathsf{\mu}}(L)
     \end{pmatrix}\end{small}=\bm{0}$, we state Forman's theorem \cite{FORM} in Hamiltonian form as 
\begin{equation}\label{For}
\frac{\text{Det}({\bm{\mathsf{S}}})}{\text{Det}(\hat{\bm{\mathsf{S}}})}=\frac{\det{[\bm{\mathsf{T}}_0\bm{\mathsf{W}}(0)+\bm{\mathsf{T}}_L\bm{\mathsf{W}}(L)]}}{\det[{\bm{\mathsf{W}}(L)}]}\;,
\end{equation}
for $\bm{\mathsf{W}}(s)\in\mathbb{R}^{12\times 12}$ whose columns $
(\bm{\mathsf{h}},\bm{\mathsf{\mu}})^T$ solve the homogeneous problem ${\bm{\mathsf{S}}}\bm{\mathsf{h}}=\bm{0}$ (\textit{i.e.}, the Jacobi equations Eq.~(\ref{JacHam}) with the extra $\frac{\beta}{2\pi}$ factor, completed as $\bm{\mathsf{W}}'=\bm{J}\bm{\mathsf{E}}\bm{\mathsf{W}}$), and the trivial partition function contribution has already been evaluated. It is important to note the freedom of choosing $\bm{\mathsf{W}}(0)$, $\bm{\mathsf{W}}(L)$ consistently; the latter statements are justified by the following considerations.

Given two matrix differential operators $\bm{\mathsf{\Omega}}=\bm{\mathsf{G}}_0(s)\frac{\dd^2}{\dd s^2}+\bm{\mathsf{G}}_1(s)\frac{\dd}{\dd s}+\bm{\mathsf{G}}_2(s)$ and $\hat{\bm{\mathsf{\Omega}}}=\bm{\mathsf{G}}_0(s)\frac{\dd^2}{\dd s^2}+\hat{\bm{\mathsf{G}}}_1(s)\frac{\dd}{\dd s}+\hat{\bm{\mathsf{G}}}_2(s)$ with non-zero eigenvalues (with respect to the BCs), acting on $\bm{\mathsf{h}}(s)\in\mathbb{R}^d$, where $\bm{\mathsf{G}}_0$, $\bm{\mathsf{G}}_1$, $\hat{\bm{\mathsf{G}}}_1$, $\bm{\mathsf{G}}_2$, $\hat{\bm{\mathsf{G}}_2}\in\mathbb{R}^{d\times d}$, $\bm{\mathsf{G}}_0$ is invertible and $s\in[a,b]$, the results of Forman \cite{FORM} provide a simple way of computing the ratio of functional determinants $\text{Det}(\bm{\mathsf{\Omega}})/\text{Det}(\hat{\bm{\mathsf{\Omega}}})$, once prescribed the BCs $\bm{\mathsf{I}}_a\begin{small}\begin{pmatrix}
      \bm{\mathsf{h}}(a) \\
      \bm{\mathsf{h}}'(a)
     \end{pmatrix}\end{small}+\bm{\mathsf{I}}_b\begin{small}\begin{pmatrix}
      \bm{\mathsf{h}}(b) \\
       \bm{\mathsf{h}}'(b)
     \end{pmatrix}\end{small}=\bm{0}$
for $\bm{\mathsf{\Omega}}$ and $\hat{\bm{\mathsf{I}}}_a\begin{small}\begin{pmatrix}
      \bm{\mathsf{h}}(a) \\
      \bm{\mathsf{h}}'(a)
     \end{pmatrix}\end{small}+\hat{\bm{\mathsf{I}}}_b\begin{small}\begin{pmatrix}
      \bm{\mathsf{h}}(b) \\
       \bm{\mathsf{h}}'(b)
     \end{pmatrix}\end{small}=\bm{0}$
for $\hat{\bm{\mathsf{\Omega}}}$, being $\bm{\mathsf{I}}_a$, $\bm{\mathsf{I}}_b$, $\hat{\bm{\mathsf{I}}}_a$, $\hat{\bm{\mathsf{I}}}_b\in\mathbb{R}^{2d\times 2d}$. Namely
\begin{equation}\label{For1}
\frac{\text{Det}(\bm{\mathsf{\Omega}})}{\text{Det}(\hat{\bm{\mathsf{\Omega}}})}=\frac{\det{[\bm{\mathsf{I}}_a+\bm{\mathsf{I}}_b\bm{\mathsf{F}}(b)]}}{\sqrt{\det[{\bm{\mathsf{F}}(b)}]}}\frac{\sqrt{\det[{\hat{\bm{\mathsf{F}}}(b)}]}}{\det{[\hat{\bm{\mathsf{I}}}_a+\hat{\bm{\mathsf{I}}}_b\hat{\bm{\mathsf{F}}}(b)]}}\;,
\end{equation}
with $\bm{\mathsf{F}}(s)$ ($\hat{\bm{\mathsf{F}}}(s)$) in $\mathbb{R}^{2d\times 2d}$ the fundamental solution of the linear differential system $\bm{\mathsf{F}}'=\bm{\mathsf{\Gamma}}\bm{\mathsf{F}}$, $\bm{\mathsf{F}}(a)=\bm{\mathbb{1}}$ ($\hat{\bm{\mathsf{F}}}'=\hat{\bm{\mathsf{\Gamma}}}\hat{\bm{\mathsf{F}}}$, $\hat{\bm{\mathsf{F}}}(a)=\bm{\mathbb{1}}$) associated to the homogeneous problem $\bm{\mathsf{\Omega}}\bm{\mathsf{h}}=\bm{0}$ ($\hat{\bm{\mathsf{\Omega}}}\bm{\mathsf{h}}=\bm{0}$) and $\bm{\mathsf{\Gamma}}$ ($\hat{\bm{\mathsf{\Gamma}}}$) the matrix of first order reduction interpreting $\bm{\mathsf{h}}'$ as an independent variable \cite{MCK, FAL}. 

In particular, we specialize to general second variation operators for $s\in[0,L]$, $d=6$, and we make the choice $\bm{\mathsf{\Omega}}={\bm{\mathsf{S}}}=-{\bm{\mathsf{P}}}(s)\frac{\dd^2}{\dd s^2}+({\bm{\mathsf{C}}}^T(s)-{\bm{\mathsf{C}}}(s)-{\bm{\mathsf{P}}}'(s))\frac{\dd}{\dd s}+{\bm{\mathsf{Q}}}(s)-{\bm{\mathsf{C}}}'(s)$ computed in either $\bm{\mathsf{q}}^f$ or $\bm{\mathsf{q}}^m$ and $\hat{\bm{\mathsf{\Omega}}}={\hat{\bm{\mathsf{S}}}}$ computed in $\hat{\bm{\mathsf{q}}}$. Note that, for notation convenience, throughout this section $\bm{\mathsf{P}}$, $\bm{\mathsf{C}}$ and $\bm{\mathsf{Q}}$ stand for $\frac{\beta}{2\pi}\bm{\mathsf{P}}$, $\frac{\beta}{2\pi}\bm{\mathsf{C}}$ and $\frac{\beta}{2\pi}\bm{\mathsf{Q}}$. Moreover, defining $\bm{\mathsf{Y}}(s)=\bm{\mathsf{F}}(s)\bm{\mathsf{Y}}(0)$ for a given non-singular matrix $\bm{\mathsf{Y}}(0)$, changing variables in Hamiltonian form by means of $\bm{\mathsf{Y}}=\bm{\mathsf{O}}\bm{\mathsf{W}}$, $\bm{\mathsf{O}}(s)=\begin{small}\begin{pmatrix}
      \bm{\mathbb{1}} &  \bm{\mathbb{0}} \\
     -{\bm{\mathsf{P}}}^{-1}{\bm{\mathsf{C}}} & {\bm{\mathsf{P}}}^{-1}
     \end{pmatrix}\end{small}$ being $\bm{\mathsf{W}}$ partitioned in $6$ by $6$ blocks as $\bm{\mathsf{W}}(s)=\begin{small}\begin{pmatrix}
      \bm{\mathsf{H}} &  \bm{\mathsf{H}}^* \\
    \bm{\mathsf{M}} & \bm{\mathsf{M}}^*
     \end{pmatrix}\end{small}$,
and doing the same in terms of $\hat{\bm{\mathsf{F}}}$, it is easily shown that Forman's theorem Eq.~(\ref{For1}) for $\bm{\mathsf{S}}$, $\hat{\bm{\mathsf{S}}}$ becomes Eq.~(\ref{For}) multiplied by ${\det[{\hat{\bm{\mathsf{W}}}(L)}]}/{\det{[\hat{\bm{\mathsf{T}}}_0\hat{\bm{\mathsf{W}}}(0)+\hat{\bm{\mathsf{T}}}_L\hat{\bm{\mathsf{W}}}(L)]}}$,
with the Hamiltonian version of the BCs being equal to $\bm{\mathsf{T}}_0=\bm{\mathsf{I}}_0\bm{\mathsf{O}}(0)$, $\bm{\mathsf{T}}_L=\bm{\mathsf{I}}_L\bm{\mathsf{O}}(L)$ and $\bm{\mathsf{W}}'=\bm{J}\bm{\mathsf{E}}\bm{\mathsf{W}}$ (the same is done for the ``hat'' term). Since the trace of $\bm{J}\bm{\mathsf{E}}$ is always zero, the so-called generalized Abel's identity or Liouville’s formula implies that $\det[{\bm{\mathsf{W}}}]$ ($\det[{\hat{\bm{\mathsf{W}}}}]$) is constant. We further observe that for ${\hat{\bm{\mathsf{S}}}}$ the BCs on the paths (being the ones entering the path integral for the partition function) must be given by the matrices $\hat{\bm{\mathsf{T}}}_0=\begin{small}\begin{pmatrix}
      \bm{\mathbb{1}} &  \bm{\mathbb{0}} \\
     \bm{\mathbb{0}} & \bm{\mathbb{0}}
     \end{pmatrix}\end{small}$, $\hat{\bm{\mathsf{T}}}_L=\begin{small}\begin{pmatrix}
      \bm{\mathbb{0}} &  \bm{\mathbb{0}} \\
     \bm{\mathbb{0}} & \bm{\mathbb{1}}
     \end{pmatrix}\end{small}$,
and choosing ${\hat{\bm{\mathsf{H}}}}(L)=\bm{\mathbb{1}}$, ${\hat{\bm{\mathsf{M}}}}(L)= \bm{\mathbb{0}}$ within $\hat{\bm{\mathsf{W}}}(L)$, the ``hat'' contribution reduces to $\text{det}[\hat{\bm{\mathsf{H}}}(0)]$, which is equal to $1$ by direct inspection (see previous section).

The idea is now to compute expression Eq.~(\ref{For}) for the operator ${\bm{\mathsf{S}}}$ subject to carefully chosen perturbed BCs $\bm{\mathsf{T}}^{(\varepsilon)}_0$, in order to avoid the zero mode. This gives rise to a quasi-zero eigenvalue that can be found analytically using our extension to general second variation operators (including cross-terms) of the trick introduced in \cite{MCK}. Finally, by taking the limit for $\varepsilon\rightarrow 0$ in the ratio of the regularized expression Eq.~(\ref{For}) to the regularized quasi-zero eigenvalue, we recover the desired quantity ${\text{Det}^{\star}({\bm{\mathsf{S}}})}/{\text{Det}(\hat{\bm{\mathsf{S}}})}$. 

We anticipate here the results for the approximation formulas of the probability densities in the case of non-isolated minimizers (already stated in Eq.~(\ref{finisoSTAR}), Eq.~(\ref{inFMisoSTAR}) when presenting the final formulas), valid also for $(\mathtt{K})$ rods as detailed in the previous section (note that the factor $\Vert\bm{\mathsf{\psi}}^{\alpha}\Vert$ simplifies out within the regularization procedure)
\begin{equation}\label{finiso}
\rra\approx2\pi\,e^{-\beta E(\qqa)}\sqrt{\frac{[\bm{\mathsf{\mu}}_{{\bm{\mathsf{\psi}}}^{\alpha}}(0)]_i}{]{\hha(0)}[_{i,i}}}\;,
\end{equation}
and we are interested in the cyclization values $\rrf(\bm{q}_0,L|\bm{q}_0,0)$, $\rrm(\bm{0},L|\bm{q}_0,0)$. In particular, $\bm{\mathsf{\mu}}_{{\bm{\mathsf{\psi}}}^{\alpha}}\in\mathbb{R}^6$ and $\hha\in\mathbb{R}^{6\times 6}$ are respectively the conjugate momentum of the zero mode and the Jacobi fields associated to ${\bm{\mathsf{S}}}^{\alpha}$, both computed by means of Eq.~(\ref{JacHam}) but recalling the contribution of $\frac{\beta}{2\pi}$. Moreover, here we denote with $[\cdot]_i$ the $i$-th component of a vector, with $]\cdot[_{i,i}$ the principal minor of a square matrix removing the $i$-th row and the $i$-th column, and the index $i$ depends on the choice of the boundary regularization, based on the non-zero components of $\bm{\mathsf{\mu}}_{{\bm{\mathsf{\psi}}}^{\alpha}}$. The appropriate ICs for $\hha$ are given at $s=L$ as: 
\begin{equation}\label{inFMiso}
\begin{split}
\hhf(L)=\bm{\mathbb{0}}\;,\,\,\mmf(L)=\bm{\mathsf{\chi}}\;;\qquad\qquad\quad\\
\hhm(L)=\begin{pmatrix}
      \bm{\mathsf{X}}_{1,1} &   \bm{\mathsf{X}}_{1,2} \\
     \bm{\mathbb{0}} & \bm{\mathbb{0}}
     \end{pmatrix}\;,\,\,\mmm(L)=\begin{pmatrix}
      \bm{\mathbb{0}} &  \bm{\mathbb{0}} \\
      \bm{\mathsf{X}}_{2,1} &  \bm{\mathsf{X}}_{2,2}
     \end{pmatrix}\;,
\end{split}
\end{equation}
where $\bm{\mathsf{\chi}}$ is an arbitrary matrix with unit determinant such that the $i$-th column corresponds to $\bm{\mathsf{\mu}}_{{\bm{\mathsf{\psi}}}^f}(L)$ and $\bm{\mathsf{X}}=\begin{small}\begin{pmatrix}
      \bm{\mathsf{X}}_{1,1} &   \bm{\mathsf{X}}_{1,2} \\
     \bm{\mathsf{X}}_{2,1} & \bm{\mathsf{X}}_{2,2}
     \end{pmatrix}\end{small}\in\mathbb{R}^{6\times 6}$, partitioned in $3$ by $3$ blocks, is an arbitrary matrix with determinant equal to $-1$ such that the $i$-th column corresponds to $([{\bm{\mathsf{\psi}}}^m]_{1:3},[\bm{\mathsf{\mu}}_{{\bm{\mathsf{\psi}}}^m}]_{4:6})^T(L)$.

We are now ready to explain how to regularize the functional determinants for ${\bm{\mathsf{S}}}^f$ and ${\bm{\mathsf{S}}}^m$ respectively, in order to get rid of the zero eigenvalue. Starting from the pure Dirichlet case, the BCs are given as ${\bm{\mathsf{T}}}_0^{(\varepsilon)}=\begin{small}\begin{pmatrix}
      \bm{\mathbb{1}} &  \bm{\mathcal{E}} \\
     \bm{\mathbb{0}} & \bm{\mathbb{0}}
     \end{pmatrix}\end{small}$, ${\bm{\mathsf{T}}}_L=\begin{small}\begin{pmatrix}
      \bm{\mathbb{0}} &  \bm{\mathbb{0}} \\
     \bm{\mathbb{1}} & \bm{\mathbb{0}}
     \end{pmatrix}\end{small}$,
with $\bm{\mathcal{E}}$ the zero matrix with a non-zero diagonal entry $\varepsilon$ in position $i,i$ serving as a perturbation to avoid the zero mode. Then, choosing ${\bm{\mathsf{H}}}^f(L)$, ${\bm{\mathsf{M}}}^f(L)$ as given in Eq.~(\ref{inFMiso}), and applying the formulas for the determinant of a block matrix, from Eq.~(\ref{For}) we get ${\text{Det}^{(\varepsilon)}({\bm{\mathsf{S}}^f})}/{\text{Det}(\hat{\bm{\mathsf{S}}})}={\det{[\bm{\mathsf{H}}^f(0)+\bm{\mathcal{E}}\bm{\mathsf{M}}^f(0)]}}/{\det[{\bm{\mathsf{M}}^f(L)}]}$ $=\varepsilon [\bm{\mathsf{\mu}}_{{\bm{\mathsf{\psi}}}^f}(0)]_i\,]{{\bm{\mathsf{H}}}^f(0)}[_{i,i}$. By construction the zero mode represents the $i$-th column of $\bm{\mathsf{H}}^f(s)$ and satisfies the linearised BC ${{\bm{\mathsf{\psi}}}^f}(0)=\bm{0}$, hence the last equality.

On the other hand, for the marginalized case, the BCs are given by ${\bm{\mathsf{T}}}_0^{(\varepsilon)}$ as before and ${\bm{\mathsf{T}}}_L=\begin{small}\begin{pmatrix}
      \bm{\mathbb{0}} &  \bm{\mathbb{0}} \\
     \bm{\mathbb{1}}^{\bm{\mathbb{0}}} & \bm{\mathbb{1}}_{\bm{\mathbb{0}}}
     \end{pmatrix}\end{small}$, being $ \bm{\mathbb{1}}^{\bm{\mathbb{0}}}=\begin{small}\begin{pmatrix}
      \bm{\mathbb{0}} &  \bm{\mathbb{0}} \\
     \bm{\mathbb{0}} & \bm{\mathbb{1}}
     \end{pmatrix}\end{small}$, $\bm{\mathbb{1}}_{\bm{\mathbb{0}}}=\begin{small}\begin{pmatrix}
      \bm{\mathbb{1}} &  \bm{\mathbb{0}} \\
     {\bm{\mathbb{0}}} & {\bm{\mathbb{0}}}
     \end{pmatrix}\end{small}$ partitioned in $3$ by $3$ blocks. Then, choosing ${\bm{\mathsf{H}}}^m(L)$, ${\bm{\mathsf{M}}}^m(L)$ as given in Eq.~(\ref{inFMiso}), and applying the formulas for the determinant of a block matrix, from Eq.~(\ref{For}) we get ${\text{Det}^{(\varepsilon)}({\bm{\mathsf{S}}^m})}/{\text{Det}(\hat{\bm{\mathsf{S}}})}=-{\det{[\bm{\mathsf{H}}^m(0)+\bm{\mathcal{E}}\bm{\mathsf{M}}^m(0)]}}/{\det[{\bm{\mathsf{X}}}]}=\varepsilon [\bm{\mathsf{\mu}}_{{\bm{\mathsf{\psi}}}^m}(0)]_i\,]{{\bm{\mathsf{H}}}^m(0)}[_{i,i}$. By construction the zero mode represents the $i$-th column of $\bm{\mathsf{H}}^m(s)$ and satisfies the linearised BC ${{\bm{\mathsf{\psi}}}^m}(0)=\bm{0}$, hence the last equality. In addition, when computing $\text{det}[\bm{\mathsf{W}}^m(L)]$, we have used the determinant identity Eq.~(\ref{Id}) for $n=3$.

The last step consists of finding the non-zero eigenvalue $\lambda^{\alpha}_{(\varepsilon)}$ associated to the eigenfunction $\bm{\mathsf{\psi}}^{\alpha}_{(\varepsilon)}$ (arising from the zero mode $\bm{\mathsf{\psi}}^{\alpha}$) of the operator $\bm{\mathsf{S}}^{\alpha}$ with perturbed BCs. First we have that $(\bm{\mathsf{\psi}}^{\alpha},\bm{\mathsf{S}}^{\alpha}\bm{\mathsf{\psi}}^{\alpha}_{(\varepsilon)})=\lambda^{\alpha}_{(\varepsilon)}(\bm{\mathsf{\psi}}^{\alpha},\bm{\mathsf{\psi}}^{\alpha}_{(\varepsilon)})$, and the left hand side can be rewritten as $(\bm{\mathsf{\psi}}^{\alpha},\bm{\mathsf{S}}^{\alpha}\bm{\mathsf{\psi}}^{\alpha}_{(\varepsilon)})=(\bm{\mathsf{\psi}}^{\alpha},\bm{\mathsf{S}}^{\alpha}\bm{\mathsf{\psi}}^{\alpha}_{(\varepsilon)})-(\bm{\mathsf{S}}^{\alpha}\bm{\mathsf{\psi}}^{\alpha},\bm{\mathsf{\psi}}^{\alpha}_{(\varepsilon)})=[\bm{\mathsf{\mu}}_{\bm{\mathsf{\psi}}^{\alpha}}\cdot\bm{\mathsf{\psi}}^{\alpha}_{(\varepsilon)}]_0^L-[\bm{\mathsf{\mu}}_{\bm{\mathsf{\psi}}^{\alpha}_{(\varepsilon)}}\cdot\bm{\mathsf{\psi}}^{\alpha}]_0^L=-(\bm{\mathsf{\mu}}_{\bm{\mathsf{\psi}}^{\alpha}}\cdot\bm{\mathsf{\psi}}^{\alpha}_{(\varepsilon)})(0)=\varepsilon [\bm{\mathsf{\mu}}_{\bm{\mathsf{\psi}}^{\alpha}}(0)]_i [\bm{\mathsf{\mu}}_{\bm{\mathsf{\psi}}^{\alpha}_{(\varepsilon)}}(0)]_i$, where the second equality comes after integration by parts and the third and fourth ones are a consequence of the BCs. Finally, being $\lambda^{\alpha}_{(\varepsilon)}=\varepsilon [\bm{\mathsf{\mu}}_{\bm{\mathsf{\psi}}^{\alpha}}(0)]_i [\bm{\mathsf{\mu}}_{\bm{\mathsf{\psi}}^{\alpha}_{(\varepsilon)}}(0)]_i/(\bm{\mathsf{\psi}}^{\alpha},\bm{\mathsf{\psi}}^{\alpha}_{(\varepsilon)})$, then $\text{Det}^{\star}(\bm{\mathsf{S}}^{\alpha})/\text{Det}({\hat{\bm{\mathsf{S}}}})=\lim_{\varepsilon\rightarrow 0}(\lambda^{\alpha}_{(\varepsilon)})^{-1}\text{Det}^{(\varepsilon)}(\bm{\mathsf{S}}^{\alpha})/\text{Det}({\hat{\bm{\mathsf{S}}}})=\Vert\bm{\mathsf{\psi}}^{\alpha}\Vert^2\,]{{{\bm{\mathsf{H}}^{\alpha}}}(0)}[_{i,i}/[\bm{\mathsf{\mu}}_{\bm{\mathsf{\psi}}^{\alpha}}(0)]_i$.

We conclude with a technical remark. That is, we observe that a priori the solution formulas for isolated minimizers could be recovered by applying Forman's theorem in the framework of functional determinants (as done here for the non-isolated case); however, there we exploit the insightful connection with the more standard theory of path integrals via ``time slicing''. Exploring both possibilities not only allows us to gain a deeper understanding of the subject, but is crucial to developing the right ideas for solving the problems.

\section{A Monte Carlo algorithm for stochastic elastic rods}\label{s6}
In this section we refer to the approach of \cite{MCDNA, ALEX2, MCD} for DNA MC simulations of J-factors, using the ``half-molecule'' technique \cite{ALEX1} for enhancing the efficiency. Namely we give a Monte Carlo sampling algorithm for fluctuating linearly elastic rods according to the Boltzmann distribution having partition function Eq.~(\ref{normex}), \ie $\mathcal{Z}=\int_{\bm{q}(0)=\bm{q}_0}{e^{-\beta E(\bm{q})}\,\mathcal{D}\bm{q}}$ with energy Eq.~(\ref{energylin}), and we use the compact notation $\bm{\mathsf{u}}_{\Delta}=\bm{\mathsf{u}}-\hat{\bm{\mathsf{u}}}$, $\bm{\mathsf{v}}_{\Delta}=\bm{\mathsf{v}}-\hat{\bm{\mathsf{v}}}$ for the shifted strains. First of all, we need to rewrite the infinite-dimensional problem as a finite-dimensional one by means of a ``parameter slicing method". This is achieved, after parametrizing the configuration variable as $\bm{\mathsf{q}}(s)=(\bm{\mathsf{c}}(s),\bm{\mathsf{t}}(s))\in\mathbb{R}^6$, setting $\epsilon=\frac{L}{n}$ with $n$ a large positive integer and $s_j=j\epsilon$ for $j=0,...,n$. Moreover, by exploiting the change of variables $(\bm{\mathsf{c}}_j,\bm{\mathsf{t}}_j)\rightarrow(\bm{\mathsf{u}}_j,\bm{\mathsf{v}}_j)$ as presented in \cite{LUDT}, we get the following equality up to a constant factor for the discrete version of the partition function $\mathcal{Z}$
\begin{equation}\label{partfin}
\begin{split}
\int{e^{-\beta\epsilon\sum\limits_{j=0}^{n}{W(\bm{\mathsf{c}},\bm{\mathsf{t}})_j}}}\prod\limits_{j=1}^n\left(1+\Vert\bm{\mathsf{c}}_j\Vert^2\right)^{-2}\dd\bm{\mathsf{c}}_j\dd\bm{\mathsf{t}}_j\\
\sim\int{e^{-{\beta\epsilon}\sum\limits_{j=0}^{n-1}{W({\bm{\mathsf{u}}_{\Delta}},{\bm{\mathsf{v}}_{\Delta}})_j}}}\prod\limits_{j=0}^{n-1}\mathcal{J}(\bm{\mathsf{u}}_j)\dd\bm{\mathsf{u}}_j\dd\bm{\mathsf{v}}_j\;,
\end{split}
\end{equation}
with $W_j=\frac{1}{2}\left[{\bm{\mathsf{u}}_{\Delta}}_j^T \bm{\mathcal{K}}_j{\bm{\mathsf{u}}_{\Delta}}_j+2{\bm{\mathsf{u}}_{\Delta}}_j^T \bm{\mathcal{B}}_j{\bm{\mathsf{v}}_{\Delta}}_j+{\bm{\mathsf{v}}_{\Delta}}_j^T \bm{\mathcal{A}}_j{\bm{\mathsf{v}}_{\Delta}}_j\right]$, $\mathcal{J}(\bm{\mathsf{u}}_j)=\left(1- {\epsilon^2}\Vert\bm{\mathsf{u}}_j\Vert^2/{4}\right)^{-\frac{1}{2}}$ and the subscript $j$ indicates that the associated term is evaluated in $s_j$. We observe that the Jacobian factor $\mathcal{J}$ can be neglected, as discussed in \cite{PL}, leading to the Gaussian distribution $\rho_{\mathcal{Z}}=e^{-{\beta\epsilon}\sum\limits_{j=0}^{n-1}{W({\bm{\mathsf{u}}_{\Delta}},{\bm{\mathsf{v}}_{\Delta}})_j}}/\int e^{-{\beta\epsilon}\sum\limits_{j=0}^{n-1}{W({\bm{\mathsf{u}}_{\Delta}},{\bm{\mathsf{v}}_{\Delta}})_j}}$ $\prod\limits_{j=0}^{n-1}\dd\bm{\mathsf{u}}_j\dd\bm{\mathsf{v}}_j$ which can be easily sampled by a direct MC method in order to get random instances of $\bm{\mathsf{u}}_j$, $\bm{\mathsf{v}}_j$, $j=0,...,n-1$, associated to a random framed curve with initial data $\bm{q}_0=(\bm{\mathbb{1}},\bm{0})$. Note that, in the proposed uniform example with diagonal stiffness matrix, the Gaussian factorises and the sampling is simply performed componentwise in terms of independent univariate Gaussians. 

Since the conditional probability density is a function of the variables $\bm{R}_L$, $\bm{r}_L$, we need to reconstruct $\bm{R}_n$, $\bm{r}_n$ from the sampled strains by discretization of the differential equations $\bm{\gamma}'(s)=\frac{1}{2}\sum\limits_{i=1}^3 [\bm{\mathsf{u}}(s)]_i\bm{B}_i\bm{\gamma}(s)$, $\bm{r}'(s)=\bm{R}(\bm{\gamma}(s))\bm{\mathsf{v}}(s)$, with $[\bm{\mathsf{u}}(s)]_i$ the $i$th component of $\bm{\mathsf{u}}$ and $\bm{R}(\bm{\gamma})$ the rotation matrix associated to the quaternion $\bm{\gamma}$. This is achieved, \eg, by application of the scalar factor method, derived in \cite{QUAT1} and discussed in \cite{QUAT2}, which is an efficient and precise one-step method for integrating the Darboux vector $\bm{\mathsf{u}}$, preserving the unit norm of the quaternion. Defining ${\delta_{\bm{\gamma}}}_j=\frac{\epsilon}{2}\sum\limits_{i=1}^3[\bm{\mathsf{u}}_j]_i\bm{B}_i\bm{\gamma}_j$, then we have that $\bm{\gamma}_{j+1}=(\bm{\gamma}_{j}+\tan{(\Vert{\delta_{\bm{\gamma}}}_j\Vert){\delta_{\bm{\gamma}}}_j/\Vert{\delta_{\bm{\gamma}}}_j\Vert})\cos{(\Vert{\delta_{\bm{\gamma}}}_j\Vert)}$ subject to the initial data $\bm{\gamma}_0=(0,0,0,1)$, and consequently $\bm{r}_{j+1}=\bm{r}_{j}+\epsilon\bm{R}(\bm{\gamma}_j)\bm{\mathsf{v}}_j$, $\bm{r}_0=\bm{0}$. 

In the spirit of \cite{MCD} for computing cyclization densities, we are now able to generate MC trajectories and assess whether or not $\bm{q}_n=(\bm{R}_n=\bm{R}(\bm{\gamma}_n),\bm{r}_n)$ is falling inside the given small region $\mathcal{R}_{\zeta,\xi}$ of $SE(3)$ centred in $(\bm{\mathbb{1}},\bm{0})$ parametrized as the Cartesian product $\mathcal{B}_{\zeta}\times\mathcal{B}_{\xi}$ of two open balls in $\mathbb{R}^3$, centred in $\bm{0}$, of radius $\zeta,\xi>0$ respectively. Namely, $(\bm{R}_n,\bm{r}_n)\in\mathcal{R}_{\zeta,\xi}$ if and only if $\Vert\bm{\mathsf{c}}(\bm{\gamma}_n)\Vert<\zeta$ and $\Vert\bm{r}_n\Vert<\xi$, with $\bm{\mathsf{c}}\in\mathbb{R}^3$ the same parametrization of $SO(3)$ presented above, adapted to $\bar{\bm{\gamma}}=(0,0,0,1)$. Note that, since $\bm{\mathsf{c}}(\bm{\gamma}_n)=[\bm{\gamma}_n]_4^{-1}([\bm{\gamma}_n]_1,[\bm{\gamma}_n]_2,[\bm{\gamma}_n]_3)$ and $\Vert\bm{\gamma}_n\Vert=1$, the condition $\Vert\bm{\mathsf{c}}(\bm{\gamma}_n)\Vert<\zeta$ is equivalent to $\sqrt{[\bm{\gamma}_n]_4^{-2}-1}<\zeta$. 

Moreover, we have the following link between the probability of the set $\mathcal{R}_{\zeta,\xi}$ ($ \mathbb{P}(\mathcal{R}_{\zeta,\xi})$) computed using MC simulations and the conditional probability density defined in the theoretical framework
\begin{equation}\label{link}
\begin{split}
&{|\lbrace\text{samples:}\,(\bm{R}_n,\bm{r}_n)\in\mathcal{R}_{\zeta,\xi}\rbrace|}\,/\,{|\lbrace\text{all samples}\rbrace|}\approx \mathbb{P}(\mathcal{R}_{\zeta,\xi})\\
&=\int_{\mathcal{R}_{\zeta,\xi}}{\rho_f(\bm{q}_L,L|\bm{q}_0,0)}\dd\bm{q}_L\approx |\mathcal{R}_{\zeta,\xi}|\,\rho_f(\bm{q}_0,L|\bm{q}_0,0)\;,
\end{split}
\end{equation}
where the notation $|\cdot |$ stands for the number of elements of a discrete set or the measure of a continuous set, and the accuracy of the approximation increases with $n\rightarrow\infty$, $|\lbrace\text{all samples}\rbrace |\rightarrow\infty$, $\zeta\rightarrow 0$, $\xi\rightarrow 0$. The set $\mathcal{R}_{\zeta,\xi}$ is measured by means of the product of the Haar measure and the Lebesgue measure for the $SO(3)$ and the $E(3)$ components. Thus, making use of the parametrisation, $|\mathcal{R}_{\zeta,\xi}|=\int_{\mathcal{R}_{\zeta,\xi}}{\dd\bm{q}_n}=\int_{\mathcal{B}_{\zeta}\times\mathcal{B}_{\xi}}{\left(1+\Vert\bm{\mathsf{c}}_n\Vert^2\right)^{-2}\,\dd \bm{\mathsf{c}}_n\,\dd\bm{r}_n}=8\pi^2\xi^3(\arctan{(\zeta)}-\zeta/(1+\zeta^2))/3$. Regarding the marginal $\rho_m(\bm{r}_0,L|\bm{q}_0,0)$, the method is applied only considering the condition on $\bm{r}_n$ for being inside the open ball $\mathcal{B}_{\xi}$ with measure $| \mathcal{B}_{\xi} |=4\pi\xi^3/3$, and neglecting all the details concerning the rotation component. 

More specifically, in order to enhance the efficiency of the algorithm, we refer to the approach adopted in \cite{MCDNA, ALEX2, MCD} for DNA MC simulations, using the ``half-molecule'' technique as developed by Alexandrowicz \cite{ALEX1}. In this technique, one computes $M$ random instances each of the first and second halves of the framed curve and then considers all first-half-second-half pairs in order to generate $M^2$ random curves, allowing a large sample size contributing for each density data point and providing the necessary accuracy to the estimation. In particular, we give here the specifications for the simulations reported in the following section. For the $(\mathtt{f})$ computations, $\sim 10^{15}$ samples were produced for each data point, choosing $n=200$ and $\zeta$, $\xi$ ranging from $2.5$ to $6.6$ \% of the parameter $L$. The estimated density value corresponds to the mean taken over $81$ ``boxes'', along with the standard deviation for these boxes defining the range of the bar for each MC data point. For the $(\mathtt{m})$ cases, $\sim 10^{13}$ samples were produced for each data point, choosing $n=200$ and $\xi$ ranging from $0.1$ to $4$ \% of the parameter $L$; $40$ different ``boxes'' were used for the final estimation.

\section{Results and discussion for the examples considered}\label{s7}
This section is dedicated to the application of formulas Eq.~(\ref{fin}) and Eq.~(\ref{finisoSTAR}) in order to predict cyclization probabilities in a concrete example of a fluctuating polymer modelled as a linearly elastic, uniform, with diagonal stiffness matrix, intrinsically straight and untwisted rod ($\bm{\mathcal{P}}(s)=\bm{\mathcal{P}}=$ diag $\lbrace k_1,k_2,k_3,a_1,a_2,a_3\rbrace$, $\hat{\bm{\mathsf{u}}}=\bm{0}$, $\hat{\bm{\mathsf{v}}}=(0,0,1)$), as presented above. The chosen example allows the physical peculiarities of the problem to be investigated in a clear and effective manner, while also providing analytical expressions for particularly simple cases and capturing the phenomena involved. We remark that the theory proposed in this article is general and can be applied to non-uniform problems, \eg to consider sequence dependent variations in stiffness in the context of DNA modelling, as well as sequence dependent intrinsic curvature.

We start with a preliminary analysis. Since in the $(\mathtt{C})$ case the compressed (isolated) solution is a minimizer for the short-length scale regimes, we evaluate analytically its contribution $\rho^c_{\alpha}$ to the cyclization probability density $(\mathtt{f})$ and $(\mathtt{m})$ for $0<L<L^f$ and $0<L<L^m$ respectively. Making use of Eq.~(\ref{fin}) with ICs Eq.~(\ref{inFM}) and setting the non-dimensional length $\tilde{L}=L/l_p$ for a given $l_p>0$, we get 
\begin{equation}\label{compr}
\rho^c_{\alpha}\approx e^{-E_p\tilde{L}}\frac{1}{{l_p}^{3}{\tilde{L}}^{\frac{1}{x}}}\sqrt{\tau\,\csc(x\,\vartheta_1\,\tilde{L})^{\frac{2}{x}}\csc(x\,\vartheta_2\,\tilde{L})^{\frac{2}{x}}}\;,
\end{equation}
where $x=x(\alpha)$ with $x(f)=1$, $x(m)=2$ and $E_p=\beta\,l_p\,a_3/2$, $\vartheta_1=(l_p\,a_3)/(2\sqrt{k_1\,a_2})$, $\vartheta_2=(l_p\,a_3)/(2\sqrt{k_2\,a_1})$, $\tau=\tau(\alpha)$ with $\tau(f)=\beta^2\,k_3\,a_3\,E_p^4/\pi^6$, $\tau(m)=l_p^2\,\sqrt{a_1\,a_2/(k_1\,k_2)}\,E_p^3/{\pi^3}$. The latter formula is valid both for isotropic (setting $k_1=k_2$, $a_1=a_2$) and non-isotropic rods. In the following we focus on the contribution $\rho_{\alpha}$ to the cyclization probability density $(\mathtt{f})$ and $(\mathtt{m})$ coming from the circular and teardrop minimizers respectively.

\subsection{Non-isotropic polymers}
\begin{figure*}
\subfigure[]{\includegraphics[width=.447\textwidth]{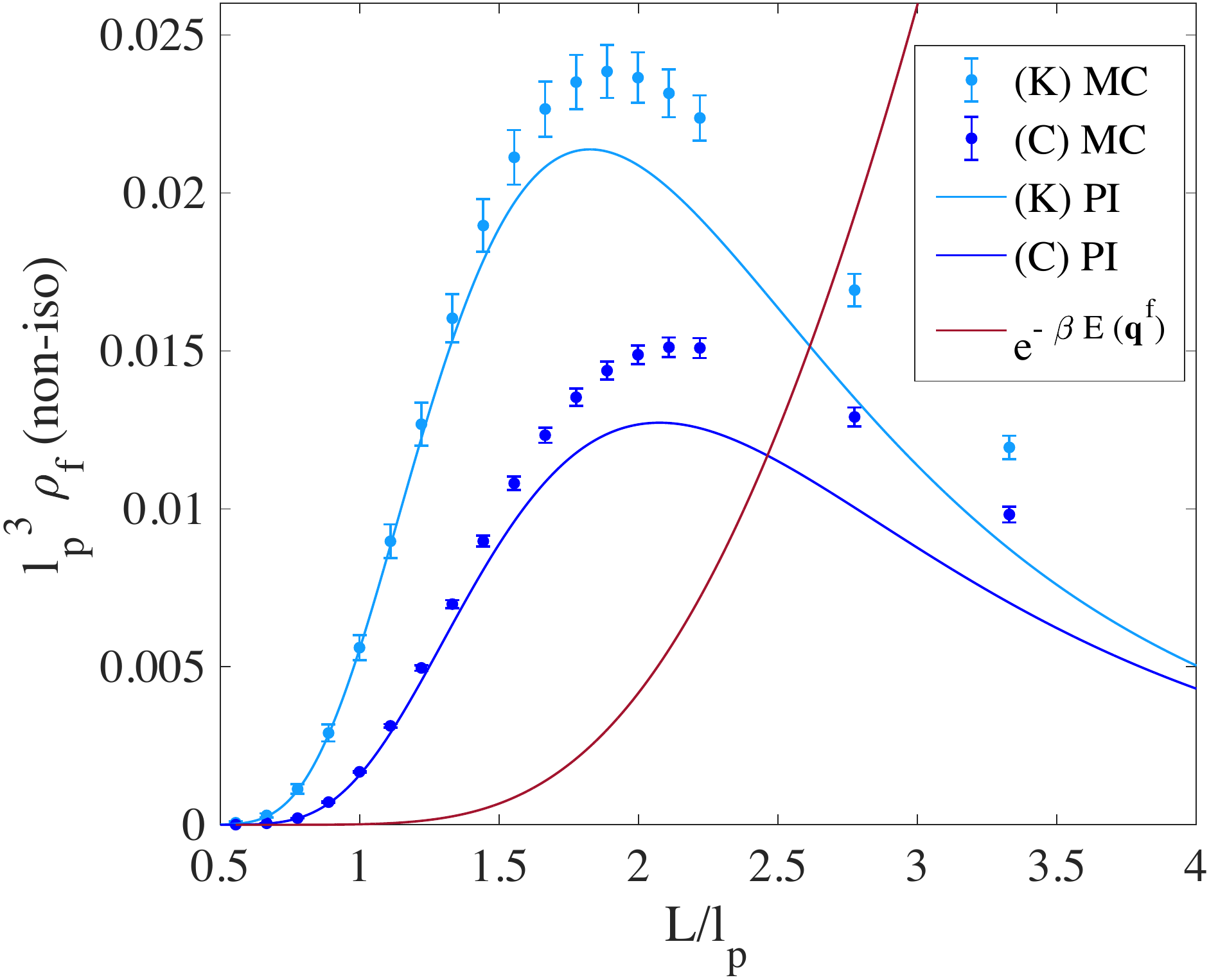}}\qquad
\subfigure[]{\includegraphics[width=.45\textwidth]{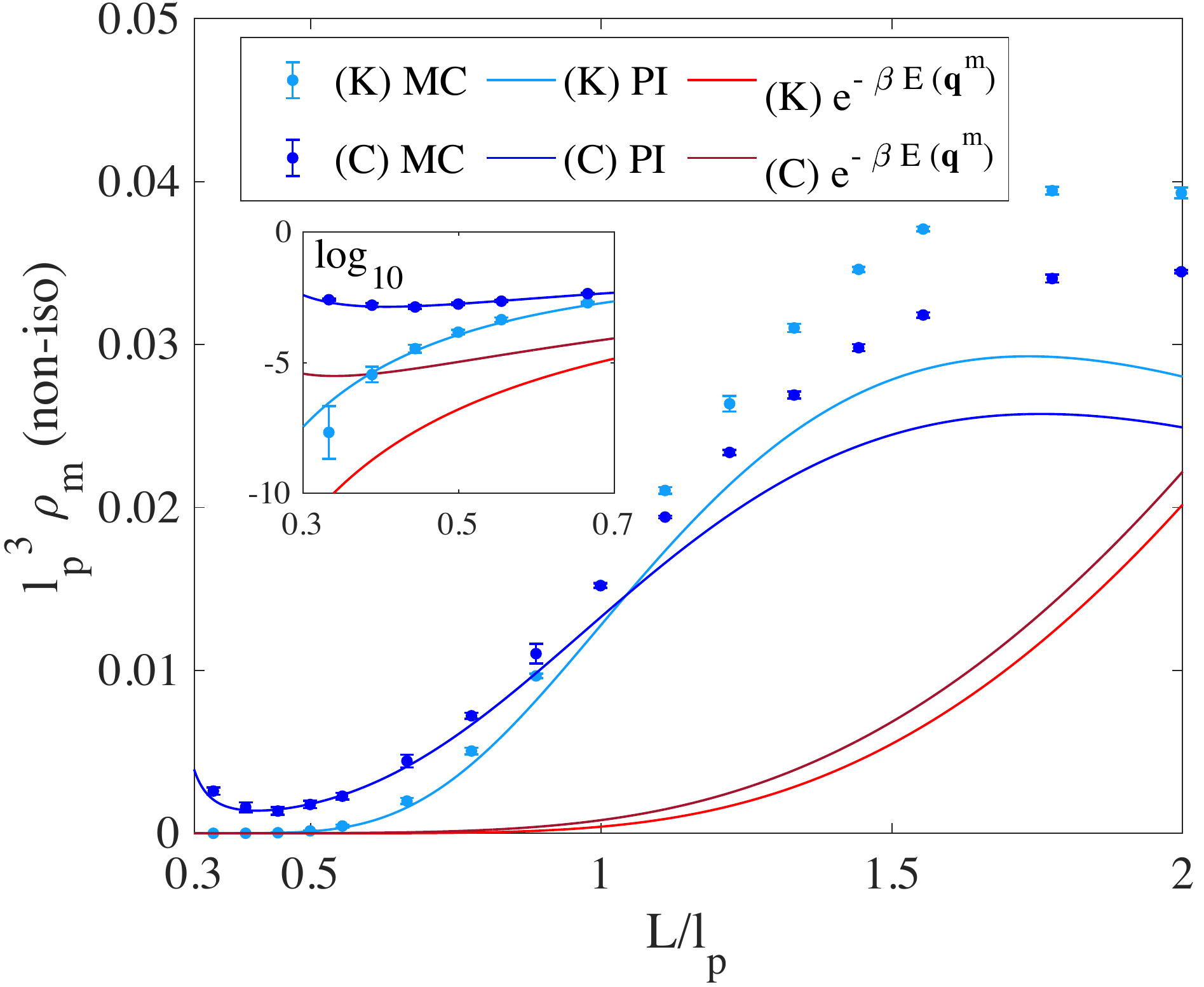}}
\caption{Comparison of cyclization densities between the path integral (PI) Laplace approximation (continuous lines) and MC (discrete points with standard deviation error bars) for a non-isotropic rod. For $(\mathtt{K})$ we set $\beta=1$, $k_1=0.5$, $k_2=5$, $k_3=10$; for the $(\mathtt{C})$ case, we also set $a_1=a_2=a_3=100$. The quantities are reported in non-dimensional form. In particular, the undeformed length of the rod is expressed in units of real persistence length $l_p\approx 0.9$, the harmonic average of $k_1$ and $k_2$. In panel (a) we address the $(\mathtt{f})$ case, reporting the values for $\rrf$ and displaying in red the zero order contribution. In panel (b) the results for the marginal density $\rrm$ are reported, with a zoom window in $\log_{10}$ scale in order to underline the peculiar small length trend. In this case, $(\mathtt{K})$ and $(\mathtt{C})$ rods differ in zero order contribution of the energy, and two different red curves are displayed.}
\label{fig67}
\end{figure*}

First, we consider a non-isotropic rod ($k_1\neq k_2$), further assuming w.o.l.o.g. that $k_1<k_2$. For the case of full looping $(\mathtt{f})$, there exist two circular, untwisted, isolated minimizers $\bm{{q}}^f$ lying on the $y-z$ plane with energy ${2\pi^2 k_1}/{L}$. The existence of a couple of reflected minima simply translates into a factor of 2 in front of Eq.~(\ref{fin}) and the semi-classical expansion is performed about one of them (\eg about the one having non-positive $y$ coordinate). For this case Eq.~(\ref{JacHam}) is a constant coefficients Jacobi system, that we solve analytically together with the first set of ICs in Eq.~(\ref{inFM}), in order to obtain the approximated formula for the cyclization probability density $\rho_f(\bm{q}_0,L|\bm{q}_0,0)$ both for $(\mathtt{C})$ and $(\mathtt{K})$ rods. Setting the length scale $l_p=2\beta k_1$, which corresponds to the planar tangent-tangent persistence length for the same rod but constrained in two dimensions \cite{PL}, and the non-dimensional length $\tilde{L}=L/l_p$, we get $ \rrf\approx 2\,e^{-\pi^2 / \tilde{L}} h_{I}\,h_{O}$, where $h_{I}$ and $h_{O}$ are the in-plane (of the minimizer) and out-of-plane contributions
\begin{small}
\begin{equation}\label{Pop}
    {h_{I}}=\frac{1}{{l_p}^{2}{\tilde{L}}^{7/2}}\sqrt{\frac{\pi}{a^2}}\;,\quad {h_{O}}_{(\mathtt{non-iso})} = \frac{1}{l_p{\tilde{L}}^{5/2}} \sqrt{\frac{8 \pi (1-\nu_3)}{b\,\nu_3(1-\cos\lambda)}}\;,
\end{equation}
\end{small}
\begin{equation}\label{An1}
{\rho_f}_{(\mathtt{non-iso})} \approx 2\,e^{-\frac{\pi^2}{\tilde{L}}}\frac{1}{{l_p}^{3}{\tilde{L}}^{6}}\sqrt{\frac{8 \pi^2 (1-\nu_3)}{a^2\,b\,\nu_3(1-\cos\lambda)}}\;,
\end{equation}
with $\lambda=2\pi\sqrt{(1-\nu_2)(1-\nu_3)}$, $a = 1 + ({2\pi}/{{\tilde{L}}})^2 (\eta_2+\eta_3)$, $b = 1 + ({2\pi}/{{\tilde{L}}})^2 (\omega_1-\eta_1)$, $\nu_2 = k_1/k_2$, $\nu_3 = k_1/k_3$, $\eta_1 = {k_1}/(a_1\,l_p^2)$, $\eta_2 = {k_1}/(a_2\,l_p^2)$, $\eta_3 = {k_1}/(a_3 \,l_p^2)$, $\omega_1 = {k_3}/(a_1\,l_p^2)$. The $(\mathtt{K})$ case is recovered setting $a=b=1$, and the density obtained disregarding the factor ${h_{O}}$ coincides with the cyclization probability density for planar rods given in \cite{LUD}. Note that the in-plane and out-of-plane contributions are computed by performing two separated Gaussian path integrals for the in-plane and out-of-plane variation fields, exploiting the decomposition of the second variation in two distinguished terms \cite{LUDT}. Moreover, the expressions in Eq.~(\ref{Pop}), Eq.~(\ref{An1}) are valid under stability assumptions for $k_1<k_2$, $k_1\neq k_3$ and equal to the limit $k_3\rightarrow k_1$, \ie $\nu_3\rightarrow 1$, if $k_3=k_1$. We further underline that Eq.~(\ref{An1}) diverges in the isotropic limit $k_2\rightarrow k_1$, \ie $\nu_2\rightarrow 1$. The results for the full looping conditional probability in the case of isolated minimizers given above were first derived in \citep{LUDT}, where the Gaussian path integrals are carried out in the variables $\bm{\mathsf{\omega}}=(\delta\bm{\mathsf{\eta}},\delta\bm{\mathsf{t}})$ instead of $\bm{\mathsf{h}}=(\delta\bm{\mathsf{c}},\delta\bm{\mathsf{t}})$ as done here. As a consequence of the latter choice, in \citep{LUDT} all the formulas have a factor of $8$ in front corresponding to the Jacobian factor of the transformation (actually in the cited work a factor of $2$ is present, but it is a typo, should be $8$). We cite this reference for the explicit evaluation of the Jacobi fields leading to expressions Eq.~(\ref{Pop}) and Eq.~(\ref{An1}).


In general, for computing the density $\rrm(\bm{0},L|\bm{q}_0,0)$ from Eq.~(\ref{fin}) together with the second set of ICs in Eq.~(\ref{inFM}), numerics must be used. In fact, for the case of marginal looping $(\mathtt{m})$, there are no simple analytical expressions for the two planar ($y-z$ plane) and untwisted teardrop shaped isolated minimizers $\bm{{q}}^m$.
However, in the $(\mathtt{K})$ case there exists a scaling argument in the variable $L$, which allows to provide a qualitative expression. Namely, given the fact that we can compute numerically a $(\mathtt{K})$ equilibrium $\bm{{q}}^m_p$ for a given rod length $l_p$, characterised by $\bm{r}_p(s_p)$, $\bm{R}_p(s_p)$, $\bm{\mathsf{v}}_p(s_p)=\hat{\bm{\mathsf{v}}}$, $\bm{\mathsf{u}}_p(s_p)$, $\bm{\mathsf{n}}_p(s_p)$, $\bm{\mathsf{m}}_p(s_p)$ for $s_p\in[0,l_p]$, then for each $L>0$ it can be easily checked that $\bm{r}(s)=\tilde{L}\,\bm{r}_p(s/\tilde{L})$, $\bm{R}(s)=\bm{R}_p(s/\tilde{L})$, $\bm{\mathsf{v}}(s)=\,\bm{\mathsf{v}}_p(s/\tilde{L})=\hat{\bm{\mathsf{v}}}$, $\bm{\mathsf{u}}(s)=1/\tilde{L}\,\bm{\mathsf{u}}_p(s/\tilde{L})$, $\bm{\mathsf{n}}(s)=1/\tilde{L}^2\,\bm{\mathsf{n}}_p(s/\tilde{L})$, $\bm{\mathsf{m}}(s)=1/\tilde{L}\,\bm{\mathsf{m}}_p(s/\tilde{L})$, $\tilde{L}=L/l_p$ define a $(\mathtt{K})$ equilibrium $\bm{{q}}^m$ for $s\in[0,L]$. This immediately implies that $E(\bm{{q}}^m)=1/\tilde{L}\,E(\bm{{q}}^m_p)$. Moreover, since the matrix $\bm{\mathsf{E}}(s)$ is given in terms of strains, forces and moments at the equilibrium by means of Eq.~(\ref{E2}), Eq.~(\ref{E3}), it is possible to obtain the scaling for the Jacobi fields as ${\text{det}[\hhm(0)]}=\tilde{L}^{9}\,{\text{det}[\hhm_p(0)]}$. Finally, defining $E_p=\beta E(\bm{{q}}^m_p)$ and $h_p=l_p^{3}({\beta}/(2\pi))^\frac{3}{2}/\sqrt{\text{det}[\hhm_p(0)]}$, we get 
\begin{equation}\label{SAn1}
{\rrm}_{(\mathtt{non-iso})}\approx 2\,e^{-\frac{E_p}{\tilde{L}}}\frac{h_p}{{l_p}^{3}{\tilde{L}}^{\frac{9}{2}}}\;,
\end{equation}
where $E_p$ and $h_p$ have to be computed numerically, and the factor $2$ accounts for the contribution of both the minimizers. By contrast, a simple scaling argument is not present for a $(\mathtt{C})$ rod, therefore allowing for more complex behaviours.

We show the results in Fig.\ref{fig67} for a specific choice of the parameters, in the range $L>L^f$ and $L>L^m$ respectively for $(\mathtt{f})$ and $(\mathtt{m})$, so that the only accounted minimizers for the computation of the cyclization probability densities are the circular and the teardrop solutions, and we can apply Eq.~(\ref{An1}) and Eq.~(\ref{SAn1}). The simulations show good agreement between the Laplace approximation and MC in the target small length domain. Even tough the second order expansion looses its quantitative power for larger lengths, the qualitative behaviour is captured and the error does not explode. We recall that looping is a rare event and MC simulations are usually expensive and unfeasible; by contrast, the method proposed in the present article is performing successfully with much higher efficiency. It is also important to underline that for the specific example considered the difference in $\rrf$ between $(\mathtt{K})$ and $(\mathtt{C})$ rods is only due to Jacobi fields, since the energy factor is the same, the circular minima having no extension and no shear deformations. The marginal case $(\mathtt{m})$ is more representative of the general behaviour where $(\mathtt{K})$ and $(\mathtt{C})$ minimizers are distinct solutions, which is true also for $(\mathtt{f})$ BCs for arbitrary (non-uniform, with non-straight intrinsic shape) elastic rods. In fact, in the short-length scale regimes, the possibility to exploit the additional degrees of freedom associated to extension and shear is crucial for minimizing the overall elastic energy, in the face of an increasingly penalizing bending contribution. This phenomenon allows the probability density to be remarkably higher than the $(\mathtt{K})$ case below the persistence length, remaining almost constant and even increasing in the range where for the $(\mathtt{K})$ rod (and therefore also for the WLC model) is exponentially vanishing. By contrast, for large lengths extension and shear become negligible. In addition, as a general statement, the Jacobi factor is fundamental to determine the peak of the density, in a domain where the energy is monotonically decreasing with length. On the other hand, the energy contribution dominates the system for smaller lengths. Finally, we clearly observe overall higher values for the marginal density compared to the full case because of the less restrictive BCs.

\subsection{Isotropic polymers}
\begin{figure*}
\centering
\subfigure[]{\includegraphics[width=.45\textwidth]{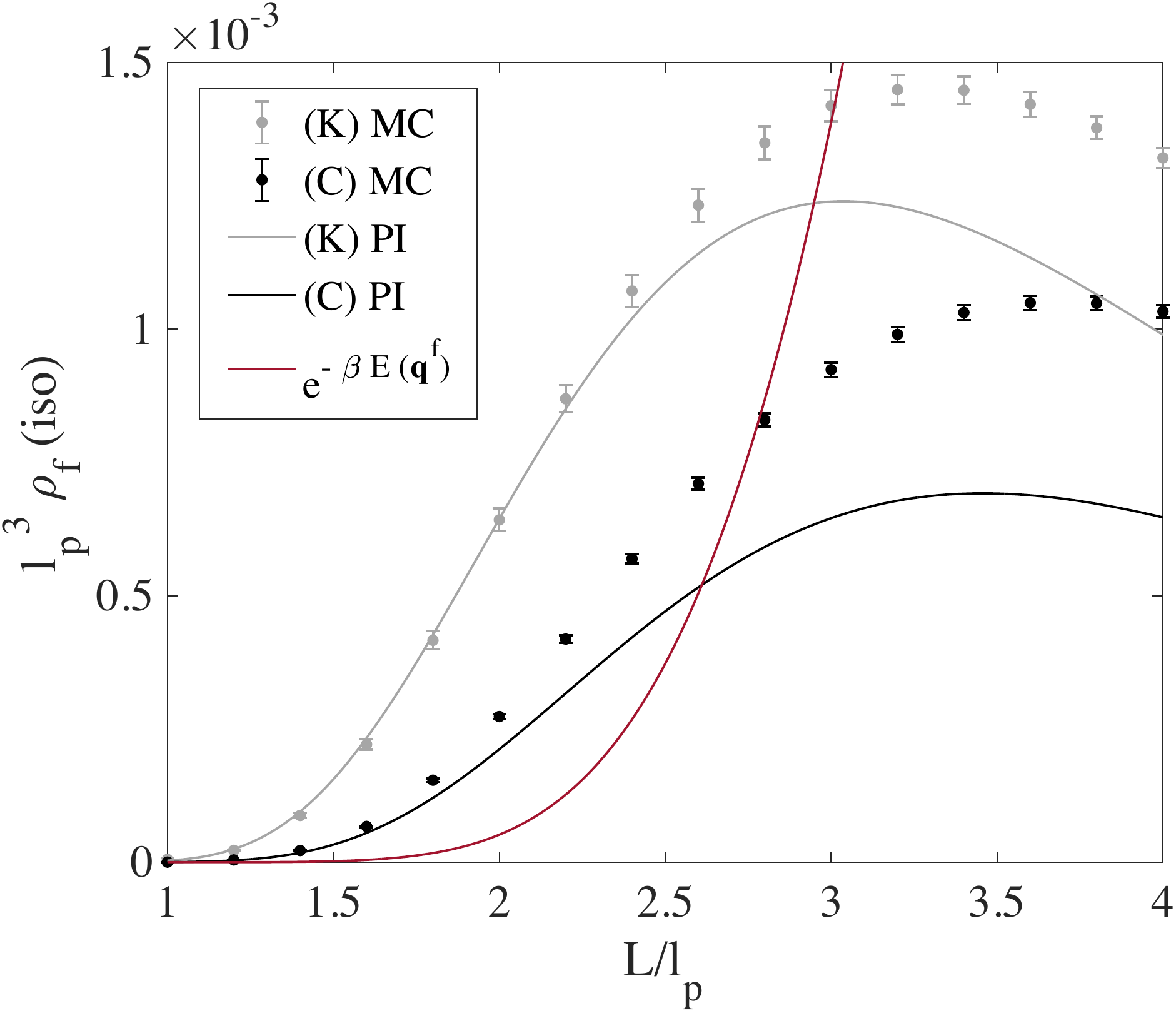}}\qquad
\subfigure[]{\includegraphics[width=.44\textwidth]{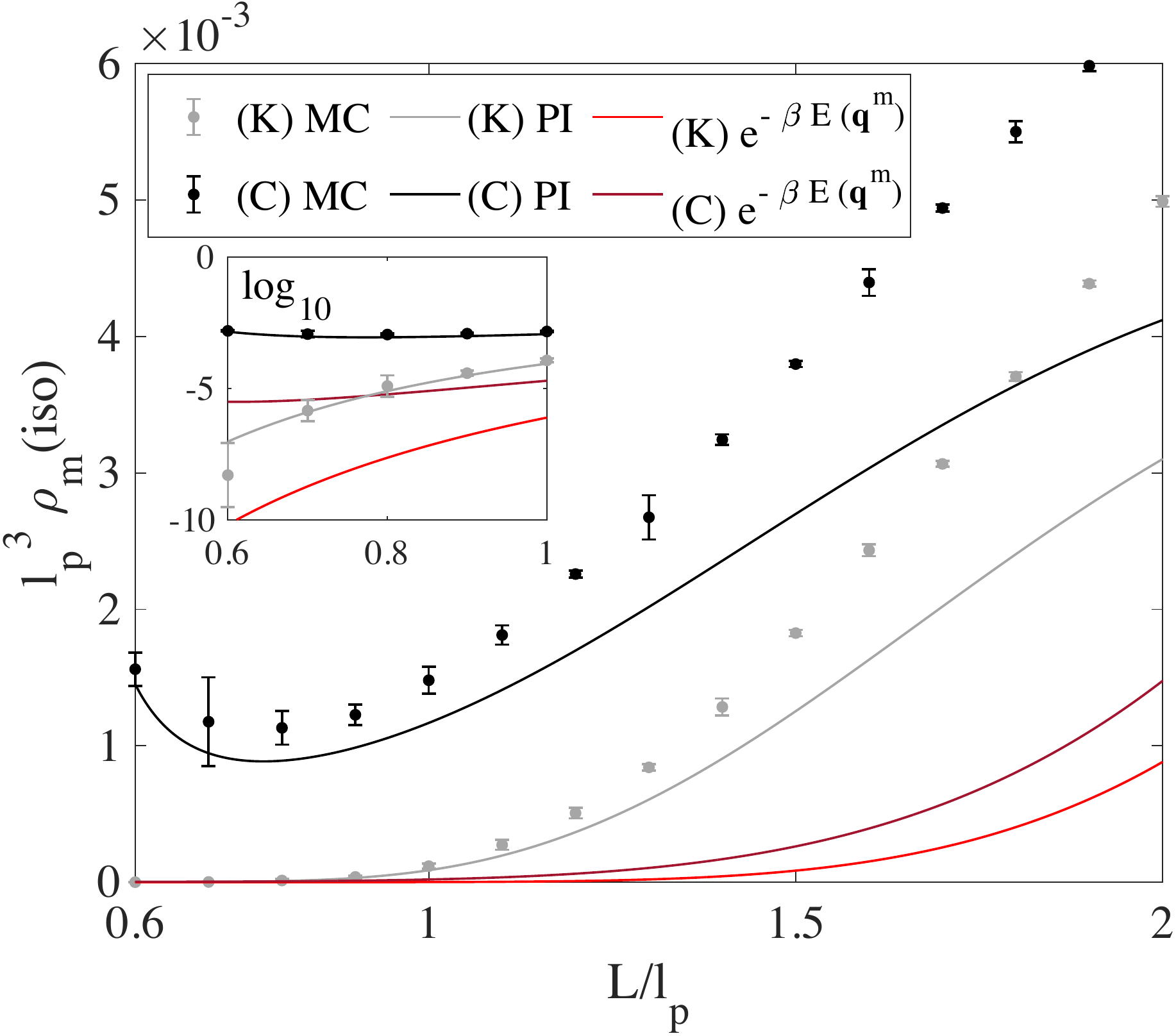}}

\subfigure[]{\includegraphics[width=.45\textwidth]{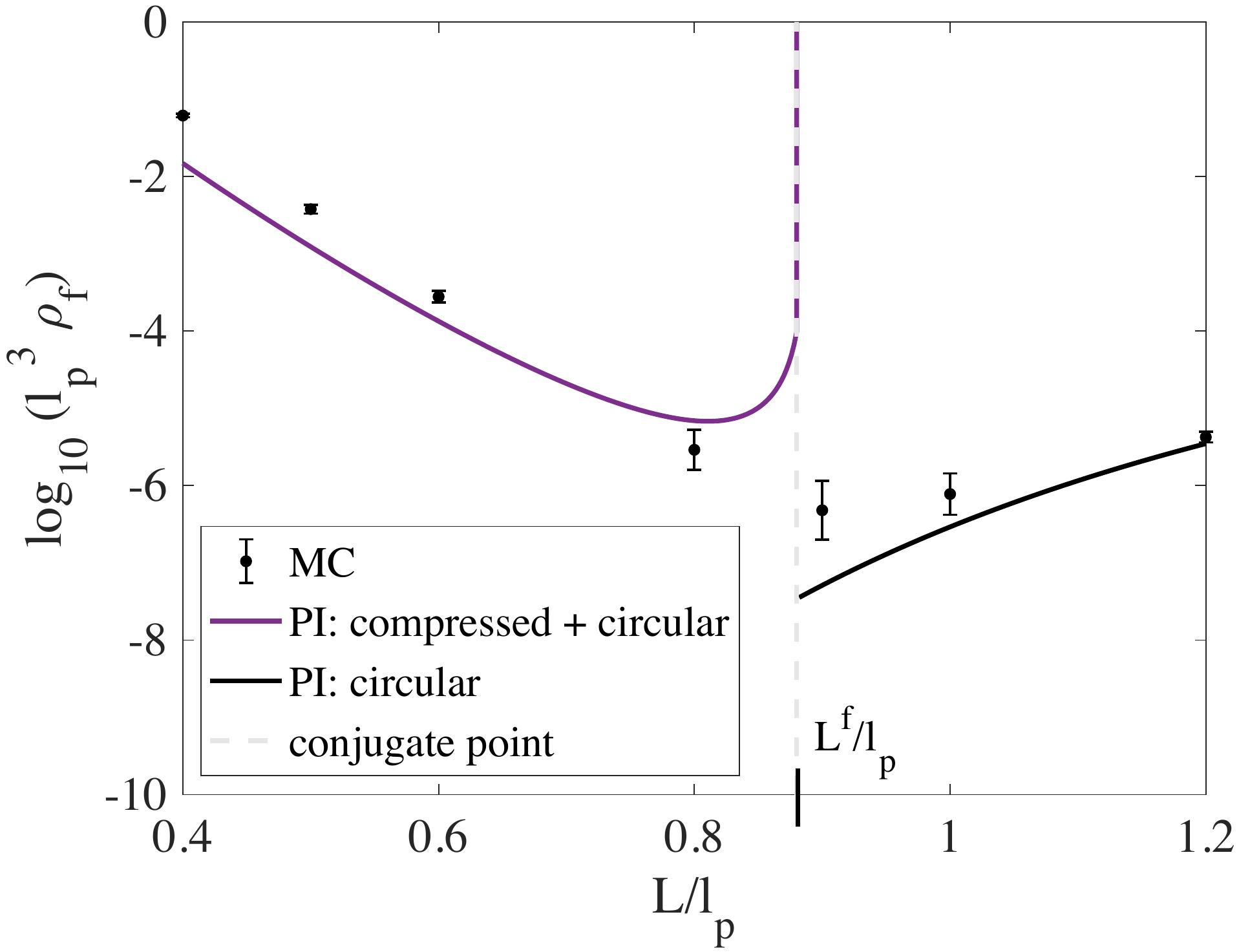}}\qquad
\subfigure[]{\includegraphics[width=.45\textwidth]{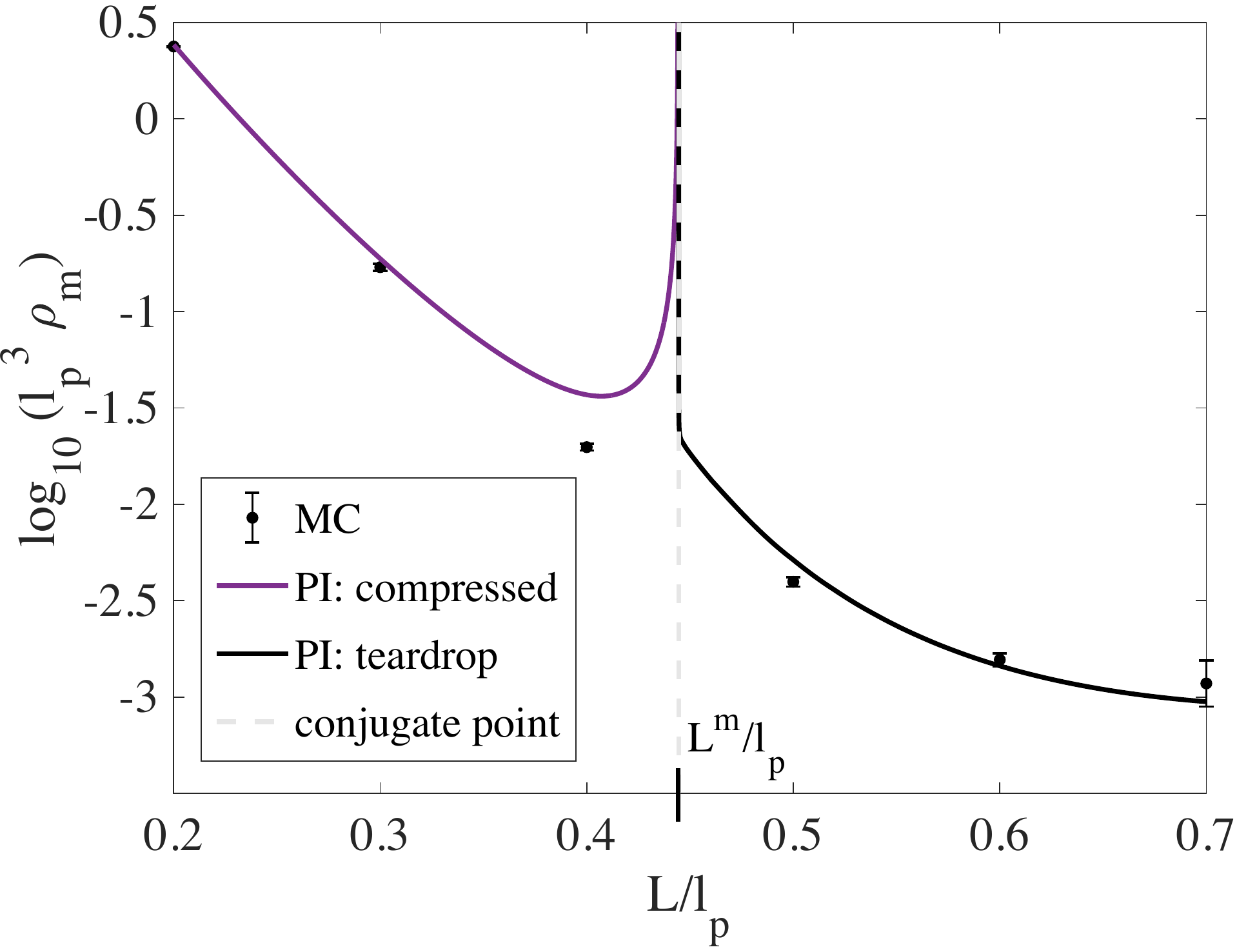}}
\caption{Comparison of cyclization densities between the path integral (PI) approximation and MC for an isotropic rod. For $(\mathtt{K})$ we set $\beta=1$, $k_1=k_2=0.5$, $k_3=10$; for the $(\mathtt{C})$ case, we also set $a_1=a_2=a_3=100$. The quantities are reported in non-dimensional form and the undeformed length of the rod is expressed in units of real persistence length $l_p\approx 0.5$. In panel (a) and (c) we address the $(\mathtt{f})$ case, reporting the values for $\rrf$ and displaying in red the zero order contribution. The behaviour for $(\mathtt{C})$ in the small length regime is shown in (c). In panel (b) and (d) the results for the marginal density $\rrm$ are reported, with a zoom window in $\log_{10}$ scale; the two different zero order contributions for $(\mathtt{K})$ and $(\mathtt{C})$ rods are displayed in red. The behaviour for $(\mathtt{C})$ in the small length regime is shown in (d).}
\label{fig891011}
\end{figure*}

\begin{figure*}[]
\centering
\subfigure[]{\includegraphics[width=.351\textwidth]{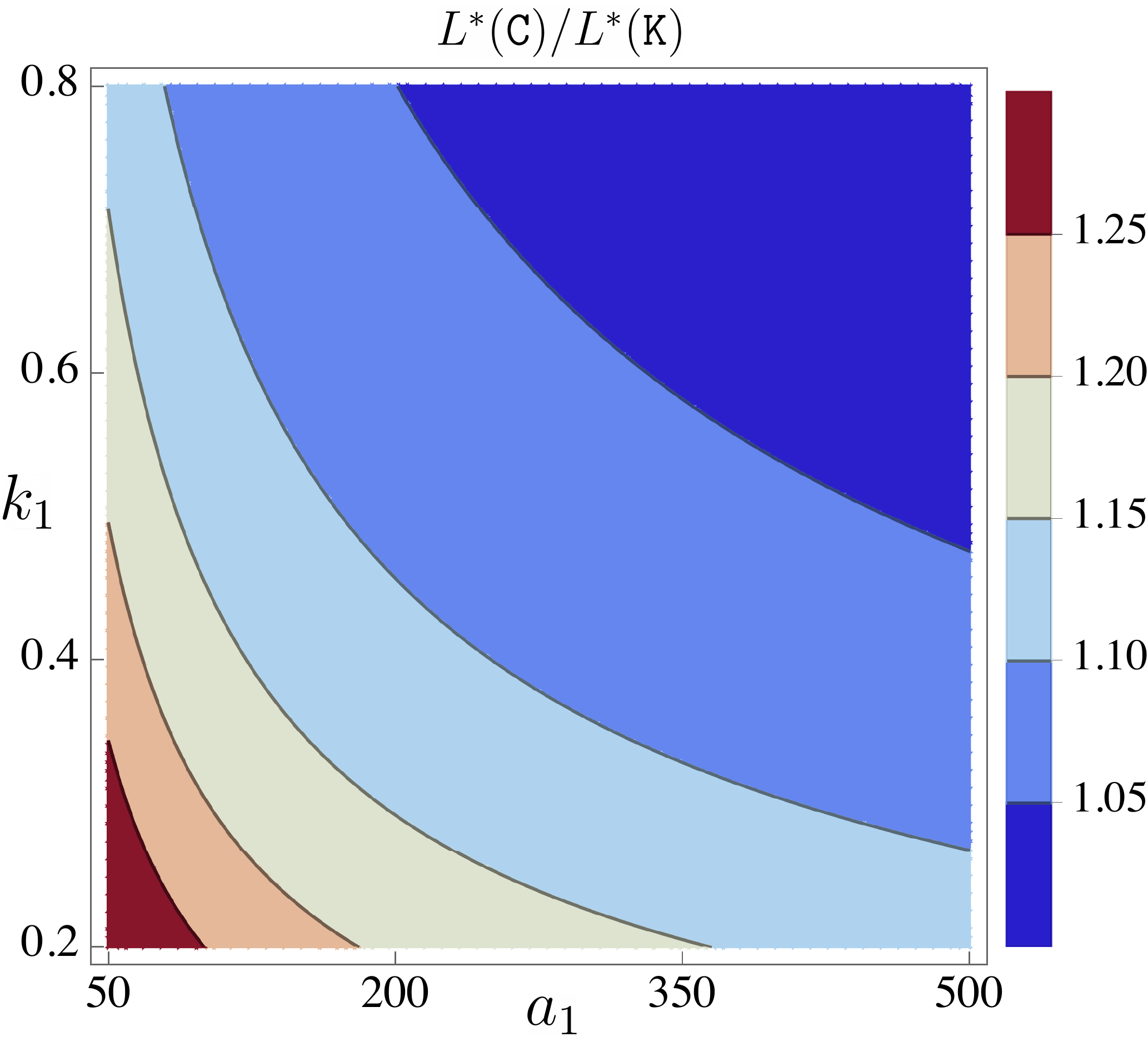}}\quad\qquad\qquad
\subfigure[]{\includegraphics[width=.35\textwidth]{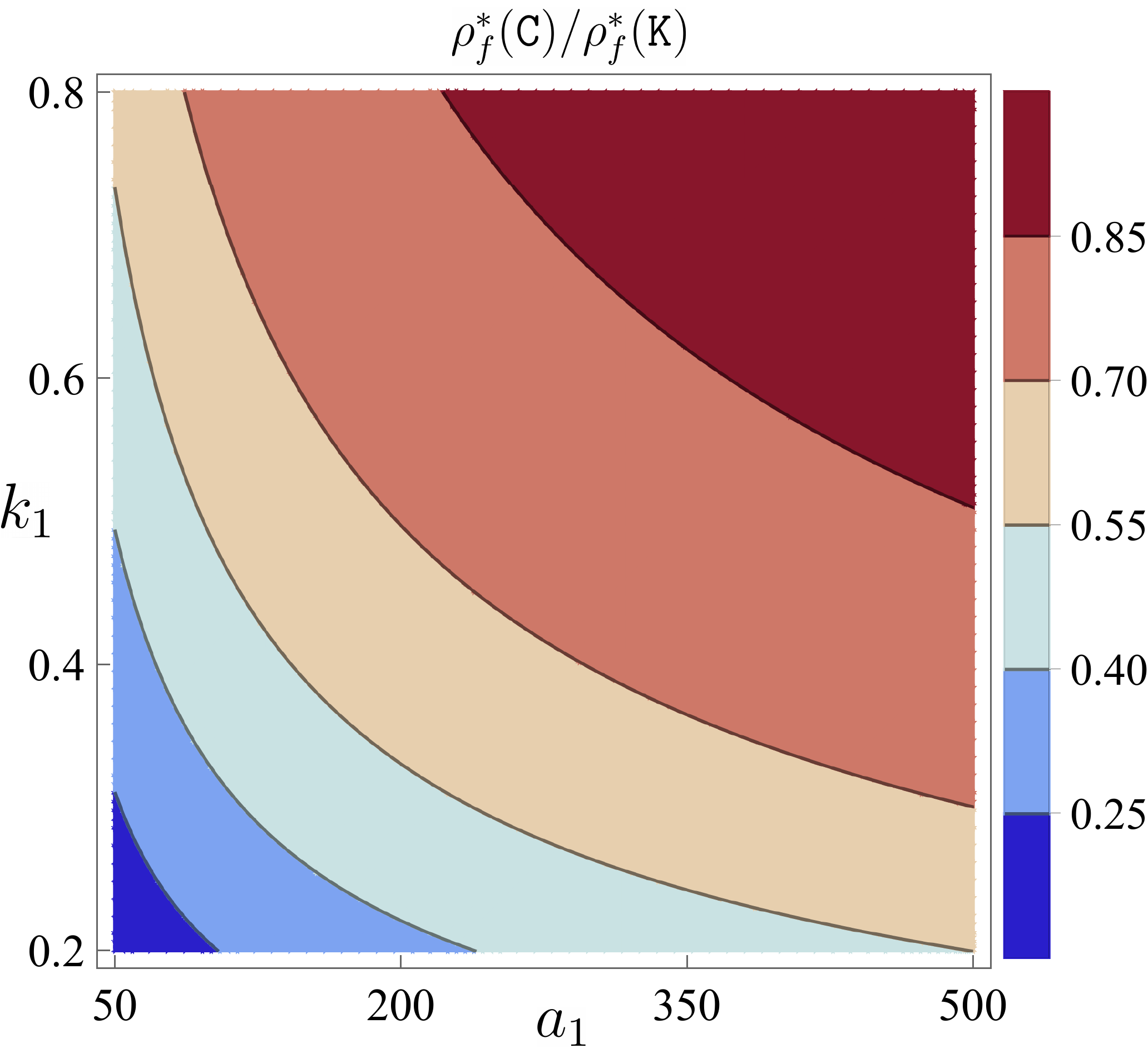}}
\caption{Contour plots for an isotropic rod (with the same parameters already used) showing the ratio of $(\mathtt{C})$ to $(\mathtt{K})$ for the length $L^*$ at which the maximum of $\rrf$ occurs ($L^*(\mathtt{C})/L^*(\mathtt{K})$ in panel (a)) and for the value $\rrf^*$ at the maximum ($\rrf^*(\mathtt{C})/\rrf^*(\mathtt{K})$ in panel (b)) as a function of $k_1=k_2$ and $a_1=a_2=a_3$.}
\label{fig1213}
\end{figure*}

Now we consider the isotropic case, \ie, $k_1=k_2$, $a_1=a_2$ and a $1$-parameter family of non-isolated circular or teardrop minima arises. Given the minimizer in the $y-z$ plane with $y\leq 0$ represented by $\bm{r}(s)=(0,r_2(s),r_3(s))$ and $\bm{R}(s)$ a counter-clockwise planar rotation about the $x$ axis of an angle $\varphi(s)$, the $1$-parameter family of minimizers can be expressed as $\bm{r}(s;\theta)=\bm{Q}_{\theta}\bm{r}(s)$ and $\bm{R}(s;\theta)=\bm{Q}_{\theta}\bm{R}(s)\bm{Q}_{\theta}^T$, where $\bm{Q}_{\theta}$ is defined as the counter-clockwise planar rotation about the $z$ axis of an angle $\theta\in [0,2\pi)$. Thus, taking the derivative of such minimizers with respect to $\theta$ and finally setting $\theta=0$, the zero mode can be easily recovered in the chosen parametrisation to be ${\bm{\mathsf{\psi}}}(s)=(0,\frac{1}{2}\sin{(\varphi)},\frac{1}{2}(\cos{(\varphi)}-1),-r_2,0,0)$. Moreover, the conjugate momentum of ${\bm{\mathsf{\psi}}}$ is derived in general (for both $(\mathtt{C})$ and $(\mathtt{K})$ rods) substituting the zero mode itself and its to-be-found moment as the unknowns of the Jacobi equations in Hamiltonian form Eq.~(\ref{JacHam}) computed on the minimizer associated to $\theta=0$ (recalling to multiply $\bm{\mathsf{E}}_{1,1}$ by $\frac{\beta}{2\pi}$ and $\bm{\mathsf{E}}_{2,2}$ by $\frac{2\pi}{\beta}$), and reads as $\bm{\mathsf{\mu}}_{{\bm{\mathsf{\psi}}}}(s)=(0,\frac{\beta k_1}{2\pi}(\cos{(\varphi)}+1)\varphi ',-\frac{\beta k_1}{2\pi}\sin{(\varphi)}\varphi ',-\frac{\beta}{2\pi} n_2,0,0)$. 

At this  point it is straightforward to apply the theory developed above for non-isolated minimizers, choosing $\bm{\mathsf{\chi}}$ to be a matrix with unit determinant such that the second column (i=2) corresponds to $\bm{\mathsf{\mu}}_{{\bm{\mathsf{\psi}}}^f}(L)$ and $\bm{\mathsf{X}}$ a matrix with determinant equal to $-1$ such that the fourth column (i=4) corresponds to $([{\bm{\mathsf{\psi}}}^m]_{1:3},[\bm{\mathsf{\mu}}_{{\bm{\mathsf{\psi}}}^m}]_{4:6})^T(L)$, according to Eq.~(\ref{inFMisoSTAR}). Consequently, the ICs for the Jacobi equations are well defined, the energy is computed, \textit{e.g.}, for the minimizer corresponding to $\theta=0$ as before, and Eq.~(\ref{finisoSTAR}) is analytical for $\rrf\approx 2\pi e^{-\pi^2 / \tilde{L}} h_{I}\,h_{O}$, where
\begin{small}
\begin{equation}\label{An2}
{h_{O}}_{(\mathtt{iso})} = \frac{1}{l_p{\tilde{L}}^{3}} \sqrt{\frac{4}{b\,\nu_3}}\;,\quad{\rho_f} _{(\mathtt{iso})}\approx2\pi e^{-\frac{\pi^2}{ \tilde{L}}}\frac{1}{{l_p}^{3}{\tilde{L}}^{\frac{13}{2}}}\sqrt{\frac{4 \pi}{a^2 b\,\nu_3}}\;,
\end{equation}
\end{small}
with $a = 1 + ({2\pi}/{{\tilde{L}}})^2 (\eta_1+\eta_3)$ and all the other quantities have been defined previously. In particular, $h_{I}$ is the same as for the non-isotropic case and therefore the zero mode arises for the out-of-plane factor for which the above regularization is applied. The $(\mathtt{K})$ limit is recovered as before setting $a=b=1$.


For the marginal density $\rrm$ numerics must be used, but in the $(\mathtt{K})$ case we can carry on the scaling argument in the variable $L$ as before, obtaining
\begin{equation}\label{SAn2}
{\rrm}_{(\mathtt{iso})}\approx 2\pi\,e^{-\frac{E_p}{\tilde{L}}}\frac{h_p}{{l_p}^{3}{\tilde{L}}^{5}}\;,
\end{equation}
for given $E_p=\beta E(\qqm_p)$, $h_p=l_p^{3}\sqrt{{[\bm{\mathsf{\mu}}_{{{\bm{\mathsf{\psi}}}_{ p}^{m}}}(0)]_i}/{]{\hhm_p(0)}[_{i,i}}}$ computed numerically.

It is interesting to note that formulas Eq.~(\ref{An1}), Eq.~(\ref{SAn1}), Eq.~(\ref{An2}) and Eq.~(\ref{SAn2}) scale differently with length as far as the second order correction term is concerned. The latter scalings naturally arise from the ones observed within simpler WLC models, see chapter $7$ in \cite{YAM00}. The comparison between Laplace and MC simulations for isotropic polymers is shown in Fig.\ref{fig891011}, (a) and (b), for the same parameters addressed in the non-isotropic case, but now sending $k_2\rightarrow k_1$. Once more time we only consider the contributions of the manifolds made of circular and teardrop minimizers, setting $L>L^f$ and $L>L^m$. The fact that now $k_2$ is ten times smaller than the same parameter adopted in Fig.\ref{fig67} implies that the overall trend of the density is shifted to the right in units of persistence length, allowing large effects of shear and extension compared to the more standard inextensible and unshearable models, as already discussed. We further observe that the approximation error is generally higher for $(\mathtt{C})$ rods and for marginal looping $(\mathtt{m})$, which is a consequence of the semi-classical expansion that depends on the stiffness values and BCs. For the simple examples considered, there clearly exist more accurate formulas for the $(\mathtt{K})$ case in the literature, \textit{e.g.}, Eq.~(\ref{SAn2}) can be related to the WLC formula (7.68), pag. 266 in \cite{YAM00}. However, the power of the method explained above lies in its generality and ability to easily provide approximation formulas for a wide range of potentially realistic and complex problems in the short length scale regimes. By contrast, since the $(\mathtt{C})$ case represents itself a novelty, we believe that basic examples are still important to understand the underlying physical behaviour.

It is natural to ask what happens for $L\leq L^f$ and $L\leq L^m$, respectively in $(\mathtt{f})$ and $(\mathtt{m})$, for $(\mathtt{C})$ rods (for $(\mathtt{K})$ the former analysis based only on circular and teardrop solutions is valid for all lengths). Due to the presence of the stable compressed solution in this range, the density diverges for vanishing length, and this is true both for isotropic and non-isotropic rods. In particular, for $(\mathtt{f})$ here we sum up the contributions coming from the compressed solution Eq.~(\ref{compr}) and the manifold of circular minimizers Eq.~(\ref{An2}); for $(\mathtt{m})$ only the compressed solution is present and we apply Eq.~(\ref{compr}). At the critical lengths $L^f$ and $L^m$ a conjugate point arises for the compressed solution (in $(\mathtt{m})$ the conjugate point arises also in the teardrop minimizer) and the Jacobi fields are singular, leading to an incorrect explosion of the probability density, which should be regularised. We do not address such regularisations, but in Fig.\ref{fig891011}, (c) and (d), we report the results for this length regime, together with MC simulations which connect our approximation formulas valid on the left and on the right of the singularities. We remark that the diverging behaviour of the conditional probability density at zero length observed for Cosserat polymers is a consequence of the linearly elastic hypothesis on the energy functional and cannot be regarded as a physical behaviour in the context of polymers made of discrete elementary units. However, we show the existence of a length-range, not affected by a compressed stable solution, where high looping probabilities occur due to an energy relaxation of the minimizers achieved by exploiting the degrees of freedom associated with extension and shear.

Finally, in order to highlight the effect of shear and extension for larger lengths, in Fig.\ref{fig1213} we compare the $(\mathtt{K})$ and the $(\mathtt{C})$ cases in terms of the length and the value of the probability density at which the maximum of $\rrf$ occurs, the first increasing and the second decreasing in presence of extension and shear.

\section{Conclusions}
In the present article we addressed the problem of computing looping probabilities from a continuum perspective, for different choices of BCs, with particular emphasis on extensible and shearable polymers, which are not generally treated in the standard literature of WLC-type models. Moreover, the proposed theoretical framework employed for deriving general looping formulas is supplemented with concrete examples, the results of which are also supported by extensive Monte Carlo simulations.

In a first approximation DNA fits the WLC hypothesis of inextensibility and unshearability. However, contradictory results have been reported for DNA below the persistence length since the studies of Cloutier and Widom \cite{CW}, actually showing enhanced cyclization of short DNA molecules not explainable by WLC-type models. In a recent study \cite{BIO5} the authors conclude that ``determining whether the high bendability of DNA at short length scales comes from transient kinks or bubbles or stems from anharmonic elasticity of DNA requires improved computational methods and further studies''. Working in this direction, and being aware  of the fact that DNA is in fact an extensible molecule \cite{DNAex}, our high cyclization predictions for small lengths in the presence of extension and shear aim to add a piece to the puzzle. Note that this is achieved even under simple linearly elastic assumptions. We believe this mechanism to be relevant and enough general to be shared by several different problems in biology. 

Furthermore, birod models \cite{BIR, PRA} with sequence-dependent parameters are more accurate in capturing DNA conformations, but the theory devised here is comprehensive and can be applied analogously to this level of complexity, allowing the computation of different ring-closure probabilities without involving expensive MC simulations. In the future, in the wide context of end-to-end probabilities, the effect of external loadings will also be investigated.

\vspace{3mm}
\section*{Acknowledgement}

We are grateful to Prof. John H. Maddocks for the fruitful discussions and insights, as well as to all the LCVMM group of Lausanne for the constant support. G.C. and R.S. acknowledge SCITAS computer facilities and grant SNF 200020-182184. 

\newpage
\appendix
\begin{widetext}

\section{Coordinates on $SE(3)$}\label{CoorSO3}
For an explicit evaluation of the path integrals in Eq.~(\ref{pathint}) and Eq.~(\ref{normex}), in the following we introduce appropriate coordinates on $SE(3)$. As done originally by Feynman \cite{BookFeynman}, a path integral can be defined via a ``time slicing'' procedure, or ``parameter slicing'' in our case, which is to replace the infinite-dimensional integral $\mathcal{D}\bm{q}$ with the limit for $n\rightarrow\infty$ of $n$ iterated finite-dimensional integrals $\prod\limits_{j=1}^{n}{\dd \bm{q}_j}$. These have to be performed on the space of framed curves, whose measure can be chosen to be the product of the Lebesgue measure on the three-dimensional euclidean space $E(3)$ and of the Haar bi-invariant measure on $SO(3)$, which may be uniquely defined up to a constant factor \cite{BookTung, BookSattinger}. 

In order to avoid difficulty that can arise from the non simple connectivity of $SO(3)$, it is often convenient to consider instead its universal (double) covering $SU(2)$. Any matrix in $SU(2)$ can be parametrized by a quadruple of real numbers $\bm{\gamma}=(\gamma_1,\gamma_2,\gamma_3,\gamma_4)$ living on the unit sphere $S^3$ in $\mathbb{R}^4$, \ie $\bm{\gamma}\cdot\bm{\gamma}=1$. The latter quadruple is know as a unit quaternion or a set of Euler parameters \cite{HAM}. Recalling that by Euler's theorem each element of SO(3) is equivalent to a rotation of an angle $\varphi$ about a unit vector $\bm{w}$, the Euler parameters are expressed as a function of $\varphi$ and $\bm{w}$ as $\gamma_4=\cos{\left(\varphi/{2}\right)}$, $\gamma_i=w_i\sin{\left({\varphi}/{2}\right)}$, $i=1,2,3$. Hence $\bm{\gamma}$ and $-\bm{\gamma}$ encode the same rotation matrix and the correspondence from $SU(2)$ to $SO(3)$ is 2 to 1. 

Referring to \cite{LUDT}, for parametrising the group of proper rotations we restrict ourself to one hemisphere of the unit sphere $S^3$ in $\mathbb{R}^4$, and we introduce the matrices $\bm{B}_1$, $\bm{B}_2$ and $\bm{B}_3$ in $\mathbb{R}^{4\times 4}$ 
\begin{equation}\label{BMat}
     \bm{B}_1=\begin{small}\begin{pmatrix}
       0 & 0 & 0 & 1 \\
       0 & 0 & 1 & 0 \\
       0 & -1 & 0 & 0 \\
       -1 & 0 & 0 & 0 \\
     \end{pmatrix}\end{small}\;, \,\,
    \bm{B}_2=\begin{small}\begin{pmatrix}
       0 & 0 & -1 & 0 \\
       0 & 0 & 0 & 1 \\
       1 & 0 & 0 & 0 \\
       0 & -1 & 0 & 0 \\
     \end{pmatrix}\end{small}\;, \,\,
     \bm{B}_3=\begin{small}\begin{pmatrix}
       0 & 1 & 0 & 0 \\
       -1 & 0 & 0 & 0 \\
       0 & 0 & 0 & 1 \\
       0 & 0 & -1 & 0 \\
     \end{pmatrix}\end{small}\;,
\end{equation}
satisfying the algebra $\bm{B}_j\bm{B}_k=-\delta_{jk}\bm{\mathbb{1}}-\bm{\epsilon}_{ijk}\bm{B}_i$, where $\bm{\epsilon}_{ijk}$ is the total antisymmetric or Levi-Civita tensor and summation over equal indices is intended. Furthermore, given a unit quaternion $\bar{\bm{\gamma}}$, $\lbrace \bm{B}_1\bar{\bm{\gamma}},\bm{B}_2\bar{\bm{\gamma}},\bm{B}_3\bar{\bm{\gamma}},\bar{\bm{\gamma}}\rbrace$ is an orthonormal basis of $\mathbb{\bm{R}}^4$ and each quadruple of Euler parameters $\bm{\gamma}$ (hence each rotation) can be expressed in coordinates with respect to the latter basis. In particular, for one hemisphere of $S^3$, we consider the new variable $\bm{\mathsf{b}}=(\mathsf{b}_1,\mathsf{b}_2,\mathsf{b}_3)\in {B}_1^3$ living in the open ball of $\mathbb{R}^3$ such that $\bm{\gamma}(\bm{\mathsf{b}})=\sum\limits_{i=1}^3{\mathsf{b}_i\bm{B}_i\bar{\bm{\gamma}}}+\sqrt{1-\Vert \bm{\mathsf{b}}\Vert^2}\bar{\bm{\gamma}}$.
Therefore, $\bm{\gamma}(\bm{\mathsf{b}})$ defines a $1$-to-$1$ parametrisation of $SO(3)$, adapted to the rotation expressed by the unit quaternion $\bar{\bm{\gamma}}$, meaning that $\bm{\gamma}(\bm{\mathsf{b}}=\bm{0})=\bar{\bm{\gamma}}$. To be precise, we should remark that the image of such a parametrisation does not include the elements lying on a maximal circle (which depends on $\bar{\bm{\gamma}}$) of the unit sphere in $\mathbb{R}^4$, since $SO(3)$ is not simply connected and rotations about a generic axis of a fixed angle are inevitably neglected.

For Euler parameters, the infinitesimal measure is given by $\dd \bm{q}_j=\delta\left(1-\Vert\bm{\gamma}_j\Vert^2\right)\dd \bm{\gamma}_j\,\dd \bm{r}_j$, so that the Haar volume measure on $SO(3)$ becomes a surface measure on $S^3$ \cite{BookTung}. Thus, the parametrisation $\bm{\phi}=\bm{\gamma}(\bm{\mathsf{b}}):{B}_1^3\subseteq\mathbb{R}^3\rightarrow\mathcal{M}\subseteq\mathbb{R}^4,$ with $\mathcal{M}$ an hemisphere of $S^3$, naturally induces a metric tensor $\bm{\mathsf{g}}$ on the tangent space at each point of $\mathcal{M}$. Denoting the coordinate vectors as $\bm{\phi}_i=\frac{\partial\bm{\phi}}{\partial\mathsf{b}_i}$, $i=1,2,3$, the components of the metric tensor are given by ${\mathsf{g}}_{i,k}=\bm{\phi}_i\cdot\bm{\phi}_k$, $i,k=1,2,3$, and we get $\bm{\mathsf{g}}(\bm{\mathsf{b}})=\bm{\mathbb{1}}+{\bm{\mathsf{b}}\otimes\bm{\mathsf{b}}}/({1-\Vert\bm{\mathsf{b}}\Vert^2})$, $\dd \bm{q}_j=\sqrt{\det{[\bm{\mathsf{g}}(\bm{\mathsf{b}}_j)]}}\,\dd \bm{\mathsf{b}}_j\,\dd \bm{r}_j$ with the metric correction being equal to ${1}/{\sqrt{1-\Vert\bm{\mathsf{b}}_j\Vert^2}}$.

Lastly, in order to deal with variables defined in the whole of $\mathbb{R}^3$, we introduce the Gibbs vector $\bm{\mathsf{c}}=\bm{\mathsf{b}}/(\sqrt{1-\Vert \bm{\mathsf{b}}\Vert^2})$. As a consequence, we have derived a $\bar{\bm{\gamma}}$-adapted parametrization of $SE(3)$ denoted by $\bm{\mathsf{q}}(s)=(\bm{\mathsf{c}}(s),\bm{\mathsf{t}}(s))\in\mathbb{R}^6$ as reported in Eq.~(\ref{par}). In particular, exploiting the Feynman discrete interpretation of the path integral measure \cite{BookFeynman}, we obtain Eq.~(\ref{par2}).

\section{Second variation matrices}
\begin{equation}\label{Cmat}
\begin{split}
\bm{\mathcal{C}}=\begin{pmatrix}
        \bm{\mathcal{K}}\bm{\mathsf{u}}^{\times}+\bm{\mathcal{B}}\bm{\mathsf{v}}^{\times}-\frac{1}{2}\bm{\mathsf{m}}^{\times}, &\,\,\,\, \bm{\mathcal{B}}\bm{\mathsf{u}}^{\times}  \\
       \\
       \bm{\mathcal{A}}\bm{\mathsf{v}}^{\times}+\bm{\mathcal{B}}^T\bm{\mathsf{u}}^{\times}-\bm{\mathsf{n}}^{\times}, &\,\,\,\, \bm{\mathcal{A}}\bm{\mathsf{u}}^{\times}
     \end{pmatrix}\;,\,\,\,
\bm{\mathcal{Q}}=\begin{pmatrix}
       \bm{\mathcal{Q}}_{1,1}, & \bm{\mathcal{Q}}_{1,2}  \\
       \\
       \bm{\mathcal{Q}}^T_{1,2}, & \bm{\mathcal{Q}}_{2,2}
     \end{pmatrix}\;,\,\,\,\text{with}
\end{split}
\end{equation}
\begin{equation}\label{Qmat}
\begin{split}
&\bm{\mathcal{Q}}_{1,1}=\frac{1}{2}(\bm{\mathsf{n}}^{\times}\bm{\mathsf{v}}^{\times}+\bm{\mathsf{v}}^{\times}\bm{\mathsf{n}}^{\times})+\frac{1}{2}(\bm{\mathsf{m}}^{\times}\bm{\mathsf{u}}^{\times}+\bm{\mathsf{u}}^{\times}\bm{\mathsf{m}}^{\times})-\bm{\mathsf{u}}^{\times}\bm{\mathcal{K}}\bm{\mathsf{u}}^{\times}-\bm{\mathsf{v}}^{\times}\bm{\mathcal{A}}\bm{\mathsf{v}}^{\times}-\bm{\mathsf{u}}^{\times}\bm{\mathcal{B}}\bm{\mathsf{v}}^{\times}-\bm{\mathsf{v}}^{\times}\bm{\mathcal{B}}^T\bm{\mathsf{u}}^{\times}\;,\\
&\bm{\mathcal{Q}}_{1,2}=-\bm{\mathsf{u}}^{\times}\bm{\mathcal{B}}\bm{\mathsf{u}}^{\times}-\bm{\mathsf{v}}^{\times}\bm{\mathcal{A}}\bm{\mathsf{u}}^{\times}+\bm{\mathsf{n}}^{\times}\bm{\mathsf{u}}^{\times}\;,\,\,\,\,\bm{\mathcal{Q}}_{2,2}=-\bm{\mathsf{u}}^{\times}\bm{\mathcal{A}}\bm{\mathsf{u}}^{\times}\;.
\end{split}
\end{equation}

\begin{equation}\label{E1}
\bm{\mathsf{E}}(s)=\begin{pmatrix}
     \bm{\mathsf{E}}_{1,1}={\bm{\mathsf{C}}}^T{\bm{\mathsf{P}}}^{-1}{\bm{\mathsf{C}}}-{\bm{\mathsf{Q}}}\,\,\,\,\, & \bm{\mathsf{E}}_{1,2}=-{\bm{\mathsf{C}}}^T{\bm{\mathsf{P}}}^{-1}\\
       \bm{\mathsf{E}}_{2,1}=\bm{\mathsf{E}}_{1,2}^{^T}\,\,\,\,\,\, & \bm{\mathsf{E}}_{2,2}={\bm{\mathsf{P}}}^{-1} 
       \end{pmatrix}\in\mathbb{R}^{12\times 12}\;.
\end{equation}
\begin{equation}\label{E2}
\bm{\mathsf{E}}_{1,1}=\bm{\mathcal{D}}\begin{pmatrix}
\frac{1}{2}(\bm{\mathsf{n}}^{\times}\bm{\mathsf{v}}^{\times}+\bm{\mathsf{v}}^{\times}\bm{\mathsf{n}}^{\times})-\frac{1}{4}\bm{\mathsf{m}}^{\times}\bm{\mathcal{R}}_{1,1}\bm{\mathsf{m}}^{\times}-\frac{1}{2}(\bm{\mathsf{m}}^{\times}\bm{\mathcal{R}}_{1,2}\bm{\mathsf{n}}^{\times}+\bm{\mathsf{n}}^{\times}\bm{\mathcal{R}}^T_{1,2}\bm{\mathsf{m}}^{\times})-\bm{\mathsf{n}}^{\times}\bm{\mathcal{R}}_{2,2}\bm{\mathsf{n}}^{\times}, \quad\quad \bm{\mathbb{0}}  \\
\\
\hspace{55mm}\bm{\mathbb{0}}, \hspace{69mm} \bm{\mathbb{0}}
\end{pmatrix}\bm{\mathcal{D}}\;,
\end{equation}
\begin{equation}\label{E3}
\bm{\mathsf{E}}_{1,2}=\bm{\mathcal{D}}\begin{pmatrix}
\bm{\mathsf{u}}^{\times}-\frac{1}{2}\bm{\mathsf{m}}^{\times}\bm{\mathcal{R}}_{1,1}-\bm{\mathsf{n}}^{\times}\bm{\mathcal{R}}_{1,2}^T, &\,\,\,\, \bm{\mathsf{v}}^{\times}-\frac{1}{2}\bm{\mathsf{m}}^{\times}\bm{\mathcal{R}}_{1,2}-\bm{\mathsf{n}}^{\times}\bm{\mathcal{R}}_{2,2}  \\
 \\ \bm{\mathbb{0}}, &\,\,\,\, \bm{\mathsf{u}}^{\times}
\end{pmatrix}\bm{\mathcal{D}}^{-1}\;,\,\,\,\bm{\mathsf{E}}_{2,2}=\bm{\mathcal{D}}^{-1}\bm{\mathcal{R}}\bm{\mathcal{D}}^{-1}\;.
\end{equation}

\vspace{3mm}
\section{A determinant identity}
\begin{equation}\label{Id}
\begin{split}
\text{det}&\left[\begin{pmatrix}
      \bm{\mathsf{X}}_{1,1} &   \bm{\mathsf{X}}_{1,2} \\
     \bm{\mathbb{0}} & \bm{\mathbb{0}}
     \end{pmatrix}\begin{pmatrix}
      \bm{\mathsf{A}} &   \bm{\mathsf{B}} \\
     \bm{\mathsf{C}} & \bm{\mathsf{D}}
     \end{pmatrix}-\begin{pmatrix}
      \bm{\mathsf{\alpha}} &   \bm{\mathsf{\beta}} \\
     \bm{\mathsf{\gamma}} & \bm{\mathsf{\delta}}
     \end{pmatrix}\begin{pmatrix}
      \bm{\mathsf{A}} &   \bm{\mathsf{B}} \\
     \bm{\mathsf{C}} & \bm{\mathsf{D}}
     \end{pmatrix}^{-1}\begin{pmatrix}
      \bm{\mathbb{0}} &   \bm{\mathbb{0}} \\
     \bm{\mathsf{X}}_{2,1} & \bm{\mathsf{X}}_{2,2}
     \end{pmatrix}\begin{pmatrix}
      \bm{\mathsf{A}} &   \bm{\mathsf{B}} \\
     \bm{\mathsf{C}} & \bm{\mathsf{D}}
     \end{pmatrix}\right]=\\
     &=(-1)^n\text{det}\left[\begin{pmatrix}
      \bm{\mathsf{X}}_{1,1} &   \bm{\mathsf{X}}_{1,2} \\
     \bm{\mathsf{X}}_{2,1} & \bm{\mathsf{X}}_{2,2}
     \end{pmatrix}\begin{pmatrix}
      \bm{\mathsf{A}} &   \bm{\mathsf{B}} \\
     \bm{\mathsf{\gamma}} & \bm{\mathsf{\delta}}
     \end{pmatrix}\right]\;,
     \end{split}
\end{equation}
for given matrices $\bm{\mathsf{X}}_{1,1}$, $\bm{\mathsf{X}}_{1,2}$, $\bm{\mathsf{X}}_{2,1}$, $\bm{\mathsf{X}}_{2,2}$, $\bm{\mathsf{A}}$, $\bm{\mathsf{B}}$, $\bm{\mathsf{C}}$, $\bm{\mathsf{D}}$, $\bm{\mathsf{\alpha}}$, $\bm{\mathsf{\beta}}$, $\bm{\mathsf{\gamma}}$, $\bm{\mathsf{\delta}}\in\mathbb{R}^{n\times n}$, which can be proven to be true by direct computation. 

\end{widetext}
\newpage

\bibliographystyle{unsrt}
\bibliography{biblio}

\end{document}